%% file: main.tex
\documentclass[preprint,12pt]{elsarticle}

\journal{Physics Reports}
\usepackage[a4paper, total={6in, 9in}]{geometry}
\usepackage{amsmath}
\usepackage{amssymb}
\usepackage[utf8]{inputenc}
\usepackage{graphicx}
\usepackage{bm}
\def\R{\mathbb{R}}

\def\P{\mathbb{P}}

\def\cA{\mathcal{A}}

\def\cC{\mathcal{C}}

\def\cG{\mathcal{G}}

\def\cJ{\mathcal{J}}

\def\cS{\mathcal{S}}

\def\cX{\mathcal{X}}	
\def\cY{\mathcal{Y}}

\def\txtd{{\textnormal{d}}}
\def\txte{{\textnormal{e}}}

\def\txtD{{\textnormal{D}}}

\def\ra{\rightarrow}
\def\I{\infty}

\newcommand{\be}{\begin{equation}}
\newcommand{\ee}{\end{equation}}
\newcommand{\benn}{\begin{equation*}}
\newcommand{\eenn}{\end{equation*}}
\newcommand{\bea}{\begin{eqnarray}}
\newcommand{\eea}{\end{eqnarray}}
\newcommand{\beann}{\begin{eqnarray*}}
\newcommand{\eeann}{\end{eqnarray*}}

\usepackage{xcolor}

\usepackage{hyperref}

\begin{document}

\begin{frontmatter}

\title{Adaptive Dynamical Networks}

\author[1,2]{Rico Berner}
\author[3,4,5]{Thilo Gross}
\author[6,7,8]{Christian Kuehn}
\author[1,9]{J\"urgen Kurths}
\author[9,10]{Serhiy Yanchuk\corref{cor1}}
\affiliation[1]{
organization={Department of Physics, Humboldt-Universität zu Berlin}, 
addressline={Newtonstr. 15},
city = {Berlin},
postcode = {12489},
country = {Germany}}
\affiliation[2]{
organization = {Institute for Theoretical Physics, TU Berlin},
addressline = {Hardenbergstr. 36}, 
city = {Berlin}, 
postcode = {10623},
country = {Germany}}
\affiliation[3]{
organizaiton = {Helmholtz Institute for Functional Marine Biodiversity (HIFMB)}, 
addressline = {Ammerländer Heerstr. 231}, 
city = {Oldenburg},
postcode = {26129}, 
country = {Germany}}
\affiliation[4]{
organizaiton = {Institute for Chemistry and Biology of the Marine Environment (ICBM), Carl-von-Ossietzky University Oldenburg}, 
addressline = {Carl-von-Ossietzky-Str. 9-11}, 
city = {Oldenburg},
postcode = {26129}, 
country = {Germany}}
\affiliation[5]{
organizaiton = {Alfred-Wegener Institute (AWI), Helmholtz Centre for Marine and Polar Research}, 
addressline = {Am Handelshafen 12},  
city = {Bremerhaven}, 
postcode = {27570},
country = {Germany}}
\affiliation[6]{
organization = {Department of Mathematics, School of Computation, Information and Technology, Technical University of Munich}, 
addressline = {Boltzmannstr. 3}, 
city = {Garching bei München}, 
postcode = {85748}, 
country = {Germany}}
\affiliation[7]{
organization = {Munich Data Science Institute (MDSI), Technical University of Munich}, 
addressline = {Walther-von-Dyck Str. 10}, 
city = {Garching bei München}, 
postcode = {85748},
country = {Germany}}
\affiliation[8]{
organization = {Complexity Science Hub Vienna}, 
addressline = {Josefstaedter Str. 39}, city = {Vienna}, postcode = {1080},  country = {Austria}}
\affiliation[9]{
organization = {Potsdam Institute for Climate Impact Research (PIK)}, addressline = {Telegrafenberg A 31}, city = {Potsdam}, postcode = {14473}, country = {Germany}}
\affiliation[10]{
organization = {Department of Mathematics, Humboldt-Universität zu Berlin}, adressline = {Rudower Ch. 25}, city = {Berlin}, postcode = {12489}, country = {Germany}}
\cortext[cor1]{Corresponding author: yanchuk@pik-potsdam.de}

\begin{abstract}
    It is a fundamental challenge to understand how the function of a network is related to its structural organization. 
    Adaptive dynamical networks represent a broad class of systems that can change their connectivity over time depending on their dynamical state. The most important feature of such
    systems is that their function depends on their structure and vice versa. While the properties of static networks have been extensively investigated in the past, the study of adaptive networks is much more challenging. Moreover, adaptive dynamical networks are of tremendous importance for various application fields, in particular, for the models for neuronal synaptic plasticity, adaptive networks in chemical, epidemic, biological, transport, and social systems, to name a few. 
    In this review, we provide a detailed description of adaptive dynamical networks, show their applications in various areas of research, highlight their dynamical features and describe the arising dynamical phenomena, and give an overview of the available mathematical methods developed for understanding  adaptive dynamical networks. 
\end{abstract}

\begin{keyword}
Adaptation \sep dynamics \sep networks \sep co-evolutionary \sep complexity \sep  neuronal plasticity \sep rewiring \sep  event-based adaptation \sep multiscale \sep cooperation \sep  opinion formation \sep  mean-field
\end{keyword}

\end{frontmatter}

\tableofcontents

%----------------------------------------------------------------------------------------
%	INTRODUCTION
%----------------------------------------------------------------------------------------
\section{Introduction}\label{sec:intro}
\input{parts/intro}
%----------------------------------------------------------------------------------------
%	COMPLEX DYNAMICAL NETWORKS
%----------------------------------------------------------------------------------------
\section{Complex dynamical networks}\label{sec:cplxDynNw}
\input{parts/cplxDynNw}
%----------------------------------------------------------------------------------------
%	CLASSIFICATION OF ADAPTIVE NETWORKS
%----------------------------------------------------------------------------------------
\section{Classification of adaptive networks}\label{sec:classification}
\input{parts/classification}
%----------------------------------------------------------------------------------------
%	Adaptive neuronal systems
%----------------------------------------------------------------------------------------
\section{Adaptive neuronal systems}\label{sec:neuro}
\input{parts/neuro}
%----------------------------------------------------------------------------------------
%	Adaptive physiological networks
%----------------------------------------------------------------------------------------
\section{Adaptive physiological networks}\label{sec:physiol}
\input{parts/physiol}
%--------------------------------------------------
% Machine learning applications
%--------------------------------------------------
\section{Machine learning applications}\label{sec:machineLearning}
\input{parts/machineLearning.tex}
%--------------------------------------------------
% Adaptive networks in control theory
%--------------------------------------------------
\section{Adaptive networks in control theory}\label{sec:control}
\input{parts/control}
%----------------------------------------------------------------------------------------
%	Adaptive networks and the relation to power grid models
%----------------------------------------------------------------------------------------
\section{Adaptive networks and the relation to power grid models}\label{sec:powerGrids}
\input{parts/powerGrids}

%--------------------------------------------------
% Human and animal behavior
%--------------------------------------------------
\section{Human and animal behavior}\label{sec:behaviour}
%Relationship between second-order consensus and adaptation
%Opinion formation, polarization dynamics

\input{parts/behaviour}
%--------------------------------------------------
% Epidemiological models
%--------------------------------------------------
\section{Epidemiological models}\label{sec:epidemics}
\input{parts/epidemics}
%--------------------------------------------------
% Adaptive transport networks
%--------------------------------------------------
\section{Adaptive transport networks}\label{sec:transport}
\input{parts/transport}

%--------------------------------------------------
% Machine learning applications
%--------------------------------------------------
\section{Climate modeling -- Adaptivity in networks of tipping elements}\label{sec:climate}
% TODO
% teleconnections adaptivity
% climate- socio-economic coupling
\input{parts/climate}

%----------------------------------------------------------------------------------------
%	Dynamical networks with adaptive time delays
%----------------------------------------------------------------------------------------
\section{Dynamical networks with adaptive time delays}\label{sec:delay}
\input{parts/delay}
%--------------------------------------------------
% 
%--------------------------------------------------
\section{Dynamical phenomena in adaptive networks}\label{sec:dynPhenomena}
\input{parts/dynPhenomena}
%----------------------------------------------------------------------------------------
% Mathematical Methods for adaptive networks
%----------------------------------------------------------------------------------------
   
%%%%%%%%%%%%%%%%%%%%%%%%%%%%%%%%%%%%%%%%%%%%%%%%%%%%%%%%%%%%%%%%%%%%%%%%%%%%%
\section{Mathematical methods for adaptive networks}\label{sec:mathmeth}
\input{parts/mathmeth}
%----------------------------------------------------------------------------------------
%	PERSPECTIVES AND CONCLUSIONS
%----------------------------------------------------------------------------------------
\section{Conclusions and Perspectives}\label{sec:conclusion}
\input{parts/conclusion}

\section*{Acknowledgements}
CK acknowledges partial support by the Volkswagen Stiftung via a Lichtenberg Professorship and by German Research Foundation DFG, Projects No. 444753754.
SY acknowledges funding from the German Research Foundation DFG,
Project No. 411803875.
%----------------------------------------------------------------------------------------
%	BIBLIOGRAPHY
%----------------------------------------------------------------------------------------
%\bibliographystyle{prwithtitle_aglabel}
%\bibliographystyle{alpha}
%\bibliographystyle{prwithtitle}
%\bibliographystyle{elsarticle-num} 
%\bibliography{references,refs_RB,sublib,epidemics,opinions}

\end{document}

%% file: parts/intro.tex
In nature and technology, complex networks have been already a long-standing framework with a broad range of applications from physics, chemistry, biology, neuroscience, socio-economic and others~\cite{NEW03}. 
Besides the paradigm of static networks, the analysis of interconnected dynamical systems on temporally evolving connectivity structures has gained more and more importance~\cite{POR20}. 
In this context, two basic modeling approaches can be distinguished: 
i) The first approach makes use of a prescribed temporal evolution of the network structure~\cite{HOL12,HOL15a}, called temporal dynamical networks. ii) In the second approach, the temporal evolution of the network depends on the dynamical state of the network and coevolves with the network nodes~\cite{GRO08a,GRO09}, called adaptive dynamical networks. 
In this review, we present the state-of-the-art for adaptive dynamical networks and provide perspectives for future research using this modeling framework.

Adaptive dynamical networks have been used to describe the dynamics of a variety of complex systems. 
The corresponding models are commonly used for understanding dynamical phenomena induced by synaptic plasticity~\cite{GER96,CAP08a}, which is a mechanisms that causes adaptation by leading to persistent changes in neural connections. 
An example is spike timing-dependent plasticity (STDP), which describes the change of the synaptic weight as a function of the difference of spiking times between pre- and post-synaptic neurons~\cite{GER96,MAR97a,BI98,ABB00,BI01,CAP08a,MEI09a,LUE16}. As a result of STDP, the network structure adaptively reorganizes  in response to neuronal dynamics. 
%Similarly, chemical systems have been investigated~\cite{JAI01}, where the reaction rates adapt dynamically depending on the variables of the system. 

Besides direct application to neuroscience, spike timing-dependent plasticity rules have been discussed for neurocomputing~\cite{HOP99} and have recently also been implemented into memristive devices~\cite{DU15,JOH18}. 
%For early work on electronic switching and S-shaped current-voltage characteristics see~\cite{SCH87,SHA92,SCH01}. 
Memristors and memristor arrays play an important role in the development of neuromorphic computing~\cite{PIC13,WAL13,IGN15,HAN17,Birkoben2020}, which is believed to be an important aspect in the future of artificial intelligence. 
An exhaustive survey on approaches to neuromorphic computing is presented in ~\cite{SCH17k}. 
%The huge interest of the scientific community in developing brain-inspired technologies is not least expressed by the flagship project "The human brain project"~\cite{MAR12c,AMU19} or the NIH BRAIN Initiative~\cite{KOR18a}. Therefore, there is clearly an urge for the analysis of adaptive dynamical networks and the development of novel methods to understand these systems.

Overall, adaptive networks have been reported for chemical~\cite{JAI01,JAI02,JAI02a,KUE19a}, epidemic~\cite{GRO06b}, biological~\cite{PRO05a}, physiological~\cite{SAW21b,BER22}, transport~\cite{GAU09a,TER10} and vascular systems~\cite{MAR17b}, social systems~\cite{GRO08a,ANT15a,BAU20,HOR20}, genomic organization~\cite{RAJ11}, or for technological systems as seen in artificial intelligence~\cite{Morales2021}, control schemes~\cite{LIU16} and power grids~\cite{BER21a}. A paradigmatic class of models of adaptively coupled phase oscillators has recently attracted much attention~\cite{GUT11,ZHA15a,Maslennikov2017,KAS17,ASL18a,KAS18,BER19,FEK20,BER21b}, and it appears to be useful for predicting and describing phenomena in more realistic and detailed models~\cite{POP15,LUE16,CHA17a,ROE19a}.

In this review, we cover many of these fields of applications and show how models of adaptively coupled dynamical systems are used to understand the dynamics in real-world systems. 
We further provide a comprehensive introduction to adaptive dynamical networks and classify these models with respect to certain common features. 
Additionally, we show which dynamical phenomena arise in systems with an adaptive network structure and introduce mathematical techniques to study these phenomena.

The article consists of four main parts concerning: (i) the definition and classification, (ii) applications of adaptive dynamical networks, (iii) effects and dynamical phenomena in systems with adaptive network structure, and (iv) mathematical methods to study these systems. In sections~\ref{sec:adaptiveNet} and~\ref{sec:classification} we provide a detailed introduction of adaptive dynamical networks and suggest several ways to classify them. 
These two sections serve as a reference point for the subsequent discussions. 
The following sections concern the application of adaptive dynamical networks to model neural (section~\ref{sec:neuro}) and physiological systems (section~\ref{sec:physiol}), to machine learning~(section~\ref{sec:machineLearning}), to build control mechanisms~(section~\ref{sec:control}), to investigate power grid systems~(section~\ref{sec:powerGrids}), and to study the behavior in social~(section~\ref{sec:behaviour}), epidemiological~(section~\ref{sec:epidemics}) as well as transport networks~(section~\ref{sec:transport}). 
In section \ref{sec:delay} we consider networks with time-delayed interactions and adaptive delays. 
Further in section~\ref{sec:dynPhenomena}, we discuss dynamical phenomena arising in the various adaptive dynamical network models outlined in the applications sections. In the last section~\ref{sec:mathmeth}, we summarize the currently available mathematical methods that have been introduced to study the effects of an adaptive network structure.

%% file: parts/cplxDynNw.tex
This section is devoted to the introduction of basic graph theoretical concepts for networks and the description of special classes of networks that will be considered in this review. Here, we follow standard textbooks and literature on graph theory and complex networks~\cite{GOD01,COS07,NEW10}.

%	A brief graph-theoretical primer
\subsection{A brief graph-theoretical primer}
\textit{Networks} are formally described by graphs~\cite{GOD01,COS07} which we will introduce in the following. However, we use the notion 'network' for consistency. A \textit{directed network} $\mathcal{N}$ is defined as a triple $\mathcal{N}=(V,E,\Psi)$ where $V$ is the set of all nodes, $E$ is the set of all links, and $\Psi:E\to\{(v,w)\in V\times V\}$ assigns each link to an out-going $v$ and in-going node $w$. The total number of nodes and links of a network are denoted by $N=|V|$ and $M=|E|$, respectively. In case of an \textit{undirected network}, we may restrict the link assignment to $\Psi:E\to\{X\subseteq V: |X|=2\}$. In particular, $\Psi$ maps the link $e\in E$ to a pair of elements $X=\{v,w\}$ meaning that $\Psi(e)$ refers to an undirected link between node $v$ and $w$. Note further if $v=w$, the link $e$ describes self-coupling. If all links can be uniquely identified with their images under the map $\Psi$, the network is called a \textit{simple network}. %However, by definition, it is possible that two links are mapped through $\Psi$ to the same element of $V\times V$, i.e., these links have the same out- and in-going nodes. In such a case, there exist two links $e,e' \in E$ such that $\Psi(e)=\Psi(e')$, and the network is called a \textit{multinetwork} or \textit{multigraph}.

Simple networks are the most commonly studied structures in the theory of complex dynamical networks. Therefore, we restrict our attention to this type of networks. For simple networks with $V=\{v_1,\dots,v_{N}\}$, every link $e$ can be uniquely assigned to the image $\Psi(e)=(v_j,v_i)$. Hence, we introduce the shorthand notation $e_{ij}$ meaning that the link $e$ connects two nodes going-out at node $v_j$ and going-in at node $v_i$ ($i,j\in \{1,\dots,N\}$). Due to the latter fact, the map $\Psi$ can be dropped in the definition of a network and we may define a simple network as $\mathcal{N}=(V,E)$ with $N$ nodes $V=\{v_1,\dots,v_{N}\}$ and $M$ links $E=\{e_{ij}: (i,j)\in \{1,\dots,N\}^2\}$. For the sake of simplicity, we refer to simple networks as networks unless stated differently. 
%Further, we associate an undirected network $\mathcal{N}'=(V',E')$ to a directed network $\mathcal{N}=(V,E)$ by $V'=V$, $E'=\{e_{ij} : e_{ij}\in E \text{ or } e_{ji}\in E, i,j=1,\dots,N\}$. Note that it is also possible to assign a directed network to an undirected network by introducing an orientation of the links~\cite{GOD01}.

We further introduce the notion of a \textit{subnetwork} $\mathcal{M}$ of a network $\mathcal{N}$. A network $\mathcal{M}=(V',E')$ is a subnetwork of a network $\mathcal{N}$ if $V'\subseteq V$ and $E'\subseteq E$. In addition, a subnetwork is denoted an induced subnetwork if $E'=\{e_{ij}\in E: v_i,v_j\in V'\}$.

Another way of describing the structure of a simple network is provided by the $N\times N$ \textit{adjacency matrix} $A$ with the entries
\begin{align}\label{eq:adjacencyMatrix}
	a_{ij} = \begin{cases}
		1, & \text{if } e_{ij}\in E,\\
		0, & \text{otherwise}.
	\end{cases}
\end{align}
The adjacency matrix provides an algebraic view on networks by representing their structure in form of a matrix. We will use this perspective frequently throughout this review. Moreover, the adjacency matrix $A$ allows us to define the following network quantities. The \textit{in-degree} $d(i)$ of node $v_i$ is given by the $i$th row sum of $A$, i.e.,
\begin{align}\label{eq:inDegree}
	d(i) = \sum_{j=1}^{N} a_{ij}.
\end{align}
Further, we define the \textit{Laplacian matrix} $L$ of a network as
\begin{align}\label{eq:laplacianMatrix}
	L = \begin{pmatrix}
		d(1) & 0 & \cdots & 0\\
		0 & \ddots & \ddots & \vdots\\
		\vdots & \ddots & \ddots & 0\\
		0 & \cdots & 0 & d(N)\\
	\end{pmatrix} - A.
\end{align}
This matrix is a discrete version of the well-known Laplacian operator known from the theory of partial differential equations. In particular, for certain dynamical networks such as nonlocally coupled ring networks, an explicit relation between the discrete and the continuum version can be derived~\cite{KOU14,ISE15,ISE15a}.

%A $v_i$-$v_j$ walk of length $L\in \mathbb{N}$ on a network is a sequence $v_i,e_{k_1 i},v_{k_1},\dots,e_{j k_{L-1}},v_j$ for $v_i,v_j\in V$ and $e_{k_m k_{m-1}}\in E$ for all $m=1,\dots,L$ where $k_0=i$, and $k_{L}=j$. With this, a network is said to be connected (or weakly connected~\cite{KOR18}) if for any two nodes $v_i,v_j\in V$ ($v_i\ne v_j$) exists at least one $v_i$-$v_j$ walk on the associated undirected network. Note that according to this definition, two nodes $v_i$ and $v_j$ are connected if either $e_{ij}\in E$ or $e_{ji}\in E$. We can utilize the adjacency matrix in order to analyse whether there is a $v_i-v_j$ walk of length $L$ between two nodes. For this, we consider the $L$th power of the adjacency matrix for the associated undirected network and find that there exists a $v_i-v_j$ walk if $(A^L)_{ji}\ne 0$. Therefore, a network is connected if the matrix $\sum_{l=1}^{N} A^l$ possesses no vanishing entry. The spectrum of the Laplacian matrix, also called graph spectrum, allows for another way to find whether a network is connected. Note that by definition any Laplacian matrix possesses a zero eigenvalue which belongs to the eigenvector $(1,\dots,1)^T$. Further, for undirected networks the Laplacian is symmetric and therefore its spectrum consist of eigenvalues such that $0=\lambda_1(L)\le \lambda_2(L) \le \dots \le \lambda_{|V|}(L)$. The second smallest eigenvalues is called algebraic connectivity (sometimes denoted as Fiedler number) and vanishes if and only if the network is unconnected~\cite{FIE73,FIE89}.

Lastly, for each network we can define a link weighting $\Xi:E\to \mathbb{R}$ which assigns a real number, the \textit{link weight}, to each link of the network. According to this map $\Xi$, the weight matrix $\kappa$ with entries is given by
\begin{align}\label{eq:weightMatrix}
	\kappa_{ij} = \begin{cases}
		\Xi(e_{ij}), & \text{if } e_{ij}\in E,\\
		0, & \text{otherwise}.
	\end{cases}
\end{align}
Consequently, a \textit{weighted} (simple) network $\mathcal{W}$ is defined by the triple $(V,E,K)$ where $K=\Xi(E)$ for a given weighting map $\Xi$. With regards to complex dynamical networks, the weight matrix is often called coupling matrix and its entries $\kappa_{ij}$ coupling weights. Note that each network can be also regarded as weighted dynamical network by considering $\Xi(e_{ij})=1$ for all $e_{ij}\in E$. Building on the definition of (weighted) networks, we establish the concepts of dynamical and adaptive dynamical networks in the next sections.

%	Dynamical networks
\subsection{Dynamical networks}\label{sec:dynNetwork}

A dynamical system $(\mathcal{X},\mathcal{T},\Phi)$ consists of a state space $\mathcal{X}$, a set of times $\mathcal{T}$ and a flow $\Phi:\mathcal{T}\times \mathcal{X}\rightarrow \mathcal{X}$. For the state space, we distinguish between discrete and continuous spaces. In the latter case, we usually consider $\mathcal{X}=\mathbb{R}^d$ or $\mathcal{X}=\mathbb{C}^d$ with dimension $d\in\mathbb{N}$. 
For the set of times, we consider either a continuous time $\mathcal{T}=\mathbb{R}$ or a discrete set of times $\mathcal{T}=\mathbb{Z}$. Frequently, the temporal evolution of a state space variable $\bm{x}$ for $\mathcal{T}=\mathbb{R}$ will be given by the ordinary differential equation (ODE)
\begin{align}\label{eq:dynSyst}
    \frac{\mathrm{d}}{\mathrm{d}t}\bm{x} = F(\bm{x},t),\qquad \bm{x}(0)=\bm{x}_0,
\end{align}
which directly generates a flow $\Phi(t,\bm{x}_0)=\bm{x}(t)$ if the vector field $F$ is sufficiently regular and autonomous $F(\bm{x},t)=F(\bm{x})$. The temporal derivative can be abbreviated by $\dot{\bm{x}}=\mathrm{d}\bm{x}/\mathrm{d}t$. We remark that upon appending $\dot{t}=1$ one can also generate a flow for the non-autonomous case at the cost of adding one additional dimension. In case of discrete time, we write the temporal evolution of the state space as the iterated map
\begin{align}\label{eq:dynSystDisc}
    \bm{x}_{n+1} = F(\bm{x}_n,n).
\end{align}
In the following, we write all equations in continuous time but often one can translate the setting directly to discrete time as well. 

In order to define a dynamical network, we first assign a state space $\mathcal{X}_i$ to each node $v_i\in V$ of a given network $\mathcal{N}$. The network state vector is then given by $\bm{x}=(\bm{x}_1,\dots,\bm{x}_N)\in \mathcal{X} = \mathcal{X}_1\times \cdots \times \mathcal{X}_N$. A \textit{dynamical network} is then generally defined by the differential equation
\begin{align}\label{eq:dynNetworkGen_global}
    \dot{\bm{x}}=f({\bm{x}},t; E)
\end{align}
with the vector field $f:\mathcal{X}_1\times\cdots\times \mathcal{X}_N \times \mathcal{T} \to \mathcal{X}_1\times\cdots\times \mathcal{X}_N$, where the set of edges $E$ (or $K$ in case of weighted networks) of a given network enters as a parameter (denoted by the separation with a semicolon) of the system. If the vector field is autonomous, we again get a dynamical system, which is the case we will be mostly interested in here.

In practice, a more structured form of a dynamical network compared to the general form given above is often used. This form reads 
\begin{align}\label{eq:dynNetworkGen}
    \dot{\bm{x}}_i=f_i({\bm{x}}_i,t) + \sum_{\{e_{ij}: j\in{1,\dots,N}, e_{ij} \in E\}} g_{ij}(\bm{x}_i,\bm{x}_j,t),
\end{align}
where $f_i:\mathcal{X}_i\times \mathcal{T}\to \mathcal{X}_i$ and $g_{ij}:\mathcal{X}_i\times \mathcal{X}_j \times \mathcal{T} \to \mathcal{X}_i$. 
Note that each state space $\mathcal{X}_i$ together with $f_i$ defines a single dynamical system for each node if $f_i$ and $g_{ij}$ are independent of time. This system is referred to as local (dynamical) system or as the local dynamics of node $i$. Hence, we refer to $f_i$ as the local vector field. The coupling functions $g_{ij}$ describe the directed interactions of two nodes for each link $e_{ij}\in E$. By the explicit use of the adjacency matrix we may write~\eqref{eq:dynNetworkGen} as
\begin{align}\label{eq:dynNetworkGenAdj}
    \dot{\bm{x}}_i = f_i({\bm{x}}_i,t) + \sum_{j=1}^N a_{ij} g_{ij}(\bm{x}_i,\bm{x}_j,t).
\end{align}
In most of the literature, however, simpler models are studied. It is commonly assumed that the same state space $\mathcal{Y}$ is assigned to each node and that the coupling function $g:\mathcal{Y}\times \mathcal{Y} \times T \to \mathcal{Y}$ is the same for all links. Hence, the ordinary differential equations (ODEs)~\eqref{eq:dynNetworkGenAdj} simplify to
\begin{align}\label{eq:dynNetwork}
    \dot{\bm{x}}_i=f_i({\bm{x}}_i,t) + \sum_{j=1}^N a_{ij} g(\bm{x}_i,\bm{x}_j,t)
\end{align}
with $\bm{x}_i\in \mathcal{Y}$.

It is important to mention that the structure of a network is used here in order to model the interaction of only two nodes/vertices, i.e., using pairwise interactions. One could think of using the network structures also to describe higher order interactions. However, in these cases usually other structures such as simplicial complexes and hyper-networks are introduced~\cite{MUL20,BIA21a,GAM21,BAT22}. 

Another frequently used generalization of dynamical networks are \textit{weighted dynamical networks}. In these cases, we consider a weighted network $\mathcal{W}=(V,E,K)$ and define the dynamics via
\begin{align}\label{eq:weightedDynNetwork}
    \dot{\bm{x}}_i = f_i({\bm{x}}_i) + \sum_{j=1}^N \bm{\kappa}_{ij} g(\bm{x}_i,\bm{x}_j,t)
\end{align}
with coupling weights $\bm{\kappa}_{ij}\in \mathbb{R}$. 
Note that by definition $\bm{\kappa}_{ij}=0$ whenever $a_{ij}=0$ and hence $a_{ij}$ could be dropped in~\eqref{eq:weightedDynNetwork}.

Further generalization are \textit{dynamical networks with delayed interactions}. In the weighted case, they are defined by $\mathcal{W}=(V,E,K,\mathfrak{T})$, where the matrix $\mathfrak{T}=(\tau_{ij})$, $\tau_{ij}\in\mathbb{R}\ge 0$ determines time-delays for the interactions along the links $e_{ij}$. The resulting set of delay differential equations is 
\begin{align}\label{eq:DelayweightedDynNetwork}
    \dot{\bm{x}}_i(t) = f_i({\bm{x}}_i(t)) + \sum_{j=1}^N \bm{\kappa}_{ij} g(\bm{x}_i(t),\bm{x}_j(t-\tau_{ij}), t),
\end{align}
where the time-dependence is explicitly shown for the state variables to distinguish between delayed and non-delayed terms. 
In contrast to the non-delayed networks, the phase space of system \eqref{eq:DelayweightedDynNetwork} is not just a direct product of the phase spaces of individual dynamical systems. 
Since solutions of  system \eqref{eq:DelayweightedDynNetwork} depend on their past history, the resulting system is infinite-dimensional and the natural phase space is the space of functions that includes the history
\cite{Hale1993,Yanchuk2017}. 

%	Adaptive dynamical networks
\subsection{Adaptive dynamical networks}\label{sec:adaptiveNet}

\begin{figure}[t]
\includegraphics[width=9cm]{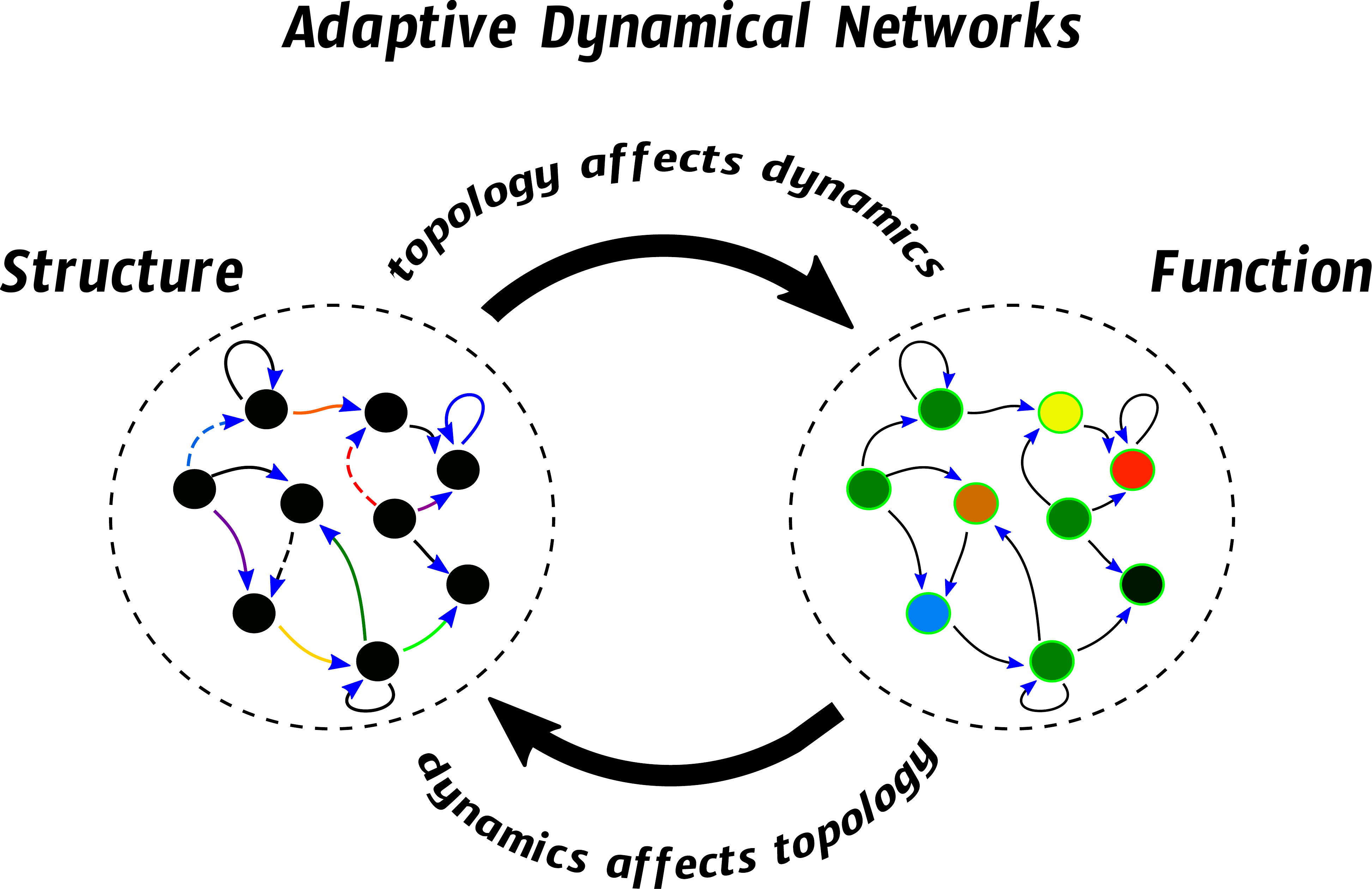}
\centering
\caption{Adaptive dynamical networks. \label{fig:an}}
\end{figure}

Here we introduce dynamical networks where the network structure is part of the temporal evolution and not static, see Fig.\ref{fig:an}.
Such systems go beyond the scope of dynamical network models with a static interaction structure introduced in the previous section. 
Based on the representation of the network structure via the adjacency matrix or  the coupling matrix, we extend systems~\eqref{eq:dynNetworkGen_global}.

Starting from~\eqref{eq:dynNetworkGen_global}, for a given set of edges $E$ (weighted edges $K$) we extend the state space by the set $\mathcal{L}=\{0,1\}^M$ (or $\mathcal{L}=\mathbb{R}^M$) where $M=|E|$ and $\bm{l}\in \mathcal{L}$ is the network state vector. A general adaptive dynamical network can then be written as
\begin{align}
    \dot{\bm{x}}&=f({\bm{x}}_1,\dots,{\bm{x}}_N,l_{1},\dots,l_{M},t),\label{eq:adapDynNetworkGen_global_x}\\
    \dot{\bm{l}}&=h({\bm{x}}_1,\dots,{\bm{x}}_N,l_{1},\dots,l_{M},t).
    \label{eq:adapDynNetworkGen_global_l}
\end{align}
with the nodal vector field $f:\mathcal{X} \times \mathcal{L} \times \mathcal{T} \to \mathcal{X}$ and the adaptation vector field $h:\mathcal{X} \times \mathcal{L} \times \mathcal{T} \to \mathcal{L}$.
Note that the dynamics of the network structure formally given by (\ref{eq:adapDynNetworkGen_global_l}) allows discontinuous changes in $\bm{l}$ by using generalised functions for $h$.

As in the case of dynamical networks, simpler models are often used in practice. 
To describe these models, let $a_{ij}\in\{0,1\}$ be the entries of the adjacency matrix and $\mathcal{Y}^N\times\{0,1\}^{N^2}$ be the extension of the state space of~\eqref{eq:dynNetwork}. We call a system \textit{an adaptive (or coevolutionary) dynamical network} if the corresponding vector field takes the form
\begin{align}\label{eq:adapDynNetwork_x}
    \dot{\bm{x}}_i&=f_i({\bm{x}}_i,t) + \sum_{j=1}^N a_{ij} g(\bm{x}_i,\bm{x}_j,t),\\
    \label{eq:adapDynNetwork_a}
    {a}_{ij}(t)&= H_{ij}^t[\bm{x}(\cdot),A(\cdot),t],
\end{align}
where with $H_{ij}^t$ we denote a general adaptation evolution operator, which depends on the states and the history of the nodes $\bm{x}$, as well as on the past states of the adjacency matrix $A=(a_{ij})$. The adjacency matrix becomes now the time-dependent variable $A(t)$. The evolution of $H_{ij}^t$ is usually discontinuous as it is associated with the appearance and disappearance of links, i.e., jumps of the values of $a_{ij}$ between $0$ and $1$. 

Analogously, we may define an \textit{adaptive (or coevolutionary) dynamical weighted network} on the state space $\mathcal{Y}^N\times \mathbb{R}^{N^2}$
\begin{align}\label{eq:adapWeightedDynNetwork_x}
    \dot{\bm{x}}_i&=f_i({\bm{x}}_i,t) + \sum_{j=1}^N \bm{\kappa}_{ij} g(\bm{x}_i,\bm{x}_j,t),\\
    \label{eq:adapWeightedDynNetwork_k}
    \dot{\bm{\kappa}}_{ij}&=h(\bm{x}_i, \bm{x}_j, t),
\end{align}
with the adaptation function $h: \mathcal{Y} \times \mathcal{Y} \times \mathcal{T} \to \mathbb{R}^{N\times N}$. We note that in both systems~\eqref{eq:adapDynNetwork_x}-\eqref{eq:adapDynNetwork_a} and~\eqref{eq:adapWeightedDynNetwork_x}--\eqref{eq:adapWeightedDynNetwork_k}, the adaptation depends explicitly on the state vectors of node $i$ and $j$. This feature of the adaptation function lies at the heart of adaptive networks as it allows the structure to rearrange according to the states of the nodes in the network. Therefore, we explicitly require this dependency of the adaptation function for our definition of adaptive dynamical networks. Analogously to \eqref{eq:adapWeightedDynNetwork_x}--\eqref{eq:adapWeightedDynNetwork_k}, dynamical networks with adaptive delays can be defined, for example, as follows
\begin{align}\label{eq:DelAdapWeightedDynNetwork_x}
    \dot{\bm{x}}_i(t)&=f_i({\bm{x}}_i(t),t) + \sum_{j=1}^N \bm{\kappa}_{ij} g(\bm{x}_i(t),\bm{x}_j(t-\tau_{ij}(t)),t),\\
    \label{eq:DelAdapWeightedDynNetwork_k}
    \dot{\tau}_{ij}&=h(\bm{x}_i(t), \bm{x}_j(t), t).
\end{align}
However, delay adaptation is much less studied until now, and we will only discuss this case in Sec.~\ref{sec:delay}. 

System \eqref{eq:adapWeightedDynNetwork_x}--\eqref{eq:adapWeightedDynNetwork_k} includes the adaptation rule where the weight of the link $\bm{\kappa}_{ij}$ depends only on the states of the in- and out-going nodes $\bm{x}_j$ and $\bm{x}_i$. Other higher-order adaptation rules are also possible, as we will show in this review. For example, the adaptation rule can depend on average ensemble quantities.

For dynamical systems such as~\eqref{eq:adapDynNetwork_x}-\eqref{eq:adapDynNetwork_a} and~\eqref{eq:adapWeightedDynNetwork_x}--\eqref{eq:adapWeightedDynNetwork_k} the notions coevolutionary and adaptive have been used interchangeably~\cite{GRO08}. In this review, we use the terminology of adaptive (weighted) dynamical networks. Adaptive networks of the form~\eqref{eq:adapDynNetwork_x}-\eqref{eq:adapDynNetwork_a} and~\eqref{eq:adapWeightedDynNetwork_x}--\eqref{eq:adapWeightedDynNetwork_k} have been studied extensively over the last years. Various forms of adaptation functions have been introduced in order to describe dynamical systems with a plethora of applications. In the next sections, we provide both, a systematic overview over frequently used adaptations rules, and an overview on the fields of applications for adaptive dynamical networks.

%% file: parts/classification.tex
% This section provides a systematic overview of several forms of adaptation functions used in the literature over the last years. 
%	Temporally evolving network structures
\subsection{Temporally evolving network structures that are not adaptive \label{sec:Temporally-evolving}}

In the previous section, we have described adaptive dynamical networks as a generalization of dynamical networks with temporally evolving network structure. In particular, we assumed that the change of the network structure depends on the states of the dynamical nodes. This feature of adaptive networks is different to other dynamical systems with temporally evolving network structures. Moreover, sometimes the notion of "adaptivity" is used in another context as in this review. In order to sort out the main unique modelling features of adaptive networks, we briefly discuss the differences to models of network evolution and adaptation.

A famous example of an evolved network structure are Barabasi-Albert networks. They  are the result of a network growth process that follows the rule of preferential attachment. By making the linking probability proportional to the node degree, preferential attachment guarantees that new nodes connect only to sites that are already well connected~\cite{BAR99}. The interplay of network growth and preferential attachment has been successfully used to provide a model explaining the scale-free structure of many real-world networks~\cite{ALB02a}. In Barabasi-Albert networks, the evolution of the network structure depends on the node degree. Similar evolution models of the network structure have also been used to investigate social and economic systems. In this context, models have been introduced where the network structure evolves based on the node feature "utility". In particular, starting from a given random network, links may be inserted or deleted depending on the utility value of the corresponding nodes~\cite{JAC02,KOE09}. All these models describe an evolution process of a network structure. However, the network nodes there are not dynamical, but only their topological properties determine the evolution of the network structure. Therefore, these classes of models are not captured by our definition of adaptive dynamical networks. 
%Moreover, we note that in case of Barabasi-Albert networks a growth process is considered while for adaptive dynamical networks the number of network nodes is kept fixed.

Another type of dynamical systems that includes a temporally changing network structure are temporal dynamical networks~\cite{HOL12,SCH14o,MAS16a,HOL19a,GHO22}. In contrast to adaptive dynamical networks, the networks structure evolves in time independently of the dynamics on the network, i.e., the state of the dynamical nodes. As pointed out in Section~\ref{sec:adaptiveNet}, however, the interdependence of network structure and node dynamics is a crucial feature of adaptive dynamical networks, that is encoded in the adaptation function. Temporal dynamical networks have served as models for many applications~\cite{HOL15a}. In these models, the temporal evolution of the topological structures are prescribed, e.g. by empirical data, and thus it sometimes does not explain the mechanism causing topological changes. Therefore, the interplay of the network structure with the node dynamics is still an open question. In the review article by Holme and Saramäki on temporal networks~\cite{HOL12}, they point out: "This question comes close to the goal of adaptive network studies [162] that model the feedback from network structure (and how it affects dynamics on the network) to the success of the agents forming the network (and how they seek to change their position in it). If one could include when contacts happen along an edge into adaptive network models and thereby explain some observed temporal–topological correlations, this would be a breakthrough (no matter what the objective system is)."

\subsection{Non-network-based adaptive systems \label{sec:noNetAdapt}}

As we describe in the next sections, adaptive dynamical networks are commonly used models in computational neuroscience. In particular, with relation to synaptic plasticity, adaptive networks have been used to get insights into the interaction of neural cells with their connectivity structure. An important mechanism leading to changes in the synaptic coupling strength is called synaptic short-term plasticity. Building on the work by Tsodyks et.al.~\cite{TSO98}, short-term plasticity has become important for modelling features of the working memory~\cite{MON08a} and even recently implemented in models for describing coarse-grained microscopic models of coupled neurons~\cite{TAH20,SCH20c}. Even though the models for short-term plasticity describe a change of the synaptic strength $\kappa_{ij}$ depending on the dynamics of the neurons $i$ and $j$, they are not adaptive dynamical networks in the sense considered here. In particular, the additional dynamical variables $u,x$ that describe the short-term plasticity \cite{TSO98} (denoting the fraction of resources that remain available after neurotransmitter depletion) can be included in the neuron model $\bm{x}_i$. 
%the adaptation function can be written as $h(\kappa_{ij}, \bm{x}_i, \bm{x}_j, t)=h_1(\bm{x}_i, t) h_2(\bm{x}_j, t)$ that allows for including the adaptivity as part of the local node dynamics.

It is also worth to mention that the notion "adaptive" is used in particular models of single neurons as well~\cite{TOU08,NAU08,AUG17}. More precisely, in this context adaptivity denotes a feedback to the membrane potential through internal processes in the neuronal cell.

In the following section, we discuss how adaptive dynamical networks of the form~\eqref{eq:adapDynNetwork_x}-\eqref{eq:adapDynNetwork_a} and~\eqref{eq:adapWeightedDynNetwork_x}--\eqref{eq:adapWeightedDynNetwork_k} can be classified with respect to their adaptation functions.

%	Event-based adaptation
\subsection{Event-based and continuous adaptation}

Adaptive dynamical networks, as treated here, can be distinguished by their adaptation rules ${a}_{ij}(t)= H_{ij}^t[\bm{x}(\cdot),A(\cdot),t]$. The adaptation functions determine how the dynamics of the nodes $\bm{x}_i(t)$ shape the network structure $A(t)=(a_{ij}(t))$ (or $\bm{\kappa}_{ij}(t)$). One can distinguish between event-based and continuous (in time) forms of adaptation.

\subsubsection{Event-based adaptation}

For the event-based adaptation, the network structure changes at certain discrete points in time. 
The triggers for the changes can be of various form and depend on different components of the dynamical system. 
A convenient way of writing this is by using event functions $e(\bm{x}, A, t)$. The adaptation function is split up into the form
\begin{align}
\left. \Delta A(t) \right|_{e(\bm{x}, A, t)=0} 
= \left. \left[ A(t_+)-A(t_-) \right] \right|_{e(\bm{x}, A, t)=0}
= \delta H(\bm{x}, A, t),
\end{align}
where  $\Delta A(t)$ is an instantaneous discontinuous change of the coupling structure, $A(t_+)=\lim_{\epsilon\downarrow 0} A(t+\epsilon)$ and $A(t_-)=\lim_{\epsilon\downarrow 0} A(t-\epsilon)$.  The amount of this change $\delta H(\bm{x}, A, t)$ can depend on the state of the dynamical network. Such a change can include, e.g., adding or removing certain links.
For all other time moment, the coupling structure is assumed to be constant. 
Thus, at any point in time, the event function evaluates the state of the adaptive dynamical network and activates the adaptation if the state meets the condition $e(\bm{x}, A, t)=0$. A typical example of such a condition is given by thresholding
\begin{align}
    e(\bm{x}, A, t) = \Tilde{e}(\bm{x}) = ((\bm{x}_i)_1-x_\mathrm{th})((\bm{x}_j)_1-x_\mathrm{th}),
\end{align}
where one of the components (here the first component) of the state vectors $\bm{x}_i$ or $\bm{x}_j$ hits a threshold value $x_\mathrm{th}$. 
Note that also the state dynamics can be subjected to the same event function. When the event function depends on the state vectors of the network, we call them \textit{spatial event functions}. This type is often used in neuroscience in the context of spike-timing-dependent plasticity, see Sec.~\ref{sec:STDP}.

Another type of event function can be found in models of epidemics or models of molecular species, see Sec.~\ref{sec:epidemics}. Here, the event depends on the current time of the dynamical system and we call them \textit{temporal event functions}. Temporal event functions may for instance have the following form
\begin{align}
    e(\bm{x}, A, t) = \Tilde{e}(t) = \prod_{k=1}^\infty (t-k T),
\end{align}
where $e(\bm{x}, A, t)$ activates the adaptation periodically with the period $T$.

All the above mentioned forms of event-based adaption can be found in a variety of models in the subsequent sections.

\subsubsection{Continuous adaptation}

In contrast to event-based adaptation rules, the adaptation function in \textit{continuous adaptive dynamical networks} $h(\bm{x}_i,\bm{x}_j,t)$ in  \eqref{eq:adapWeightedDynNetwork_k} (or $H^t_{ij}[\bm{x}(\cdot), A(\cdot), t]$ in~\eqref{eq:adapDynNetwork_a}) is continuous in time $t\in \mathcal{T}$. Note that the notion also applies to discrete time systems \eqref{eq:adapDynNetwork_a} with $\mathcal{T}=\mathbb{Z}$.

As well as the event-based adaptation rules, continuous adaptive dynamical networks have been studied extensively in the literature. Examples are provided in the subsequent sections of this article. In the following section, we introduce a widely studied model that belongs to the class of continuous adaptive dynamical networks.

%	Adaptively coupled phase oscillators models
\subsection{Adaptively coupled phase oscillator models
\label{sec:phaseOscModel}}

For the theory of synchronization phenomena, models of phase oscillators such as the Kuramoto model~\cite{KUR84} are of great  importance~\cite{KUR84,Strogatz1,ACE05,OME12b,PIK08}. A particularly important feature of coupled nonlinear oscillator systems is their reducibility to phase oscillator networks in the case of weak interactions~\cite{WIN80,HOP97,PIK01,PIE19a}. The reduction of complex dynamical systems to networks of coupled phase oscillators is well-known and there exist exhaustive reviews highlighting the importance of phase oscillator models~\cite{PIK01,ASH16,PIE19a}. Recent studies, in addition, aim for making phase oscillators models even more powerful by lifting conditions under which reductions to phase oscillator models can be achieved~\cite{KLI17a,ROS19a,ROS19b,ERM19}. Therefore, it is not surprising that also adaptive networks have been studied over the years based on the phase oscillator modeling paradigm.

In accordance with~\eqref{eq:adapWeightedDynNetwork_x}--\eqref{eq:adapWeightedDynNetwork_k}, an adaptive dynamical weighted network of $N$ coupled phase oscillators is written as
\begin{align}
    \dot{\phi}_i &= \omega_i + \sum_{j=1}^N a_{ij}\kappa_{ij} g(\phi_i,\phi_j,t), \label{eq:APO_phi_gen}\\
    \dot{\kappa}_{ij} & = h(\kappa_{ij},\phi_i,\phi_j,t), \label{eq:APO_kappa_gen}
\end{align}
where $\phi_{i}\in[0,2\pi)$ represents the phase of the $i$th oscillator ($i=1,\dots,N$), $\omega_i$ is its natural frequency, and $\kappa_{ij}$ is the coupling weight of the connection from node $j$ to $i$. 
%Note that the dynamics of an unweighted adaptive dynamical network could be written analogously, see also~\eqref{eq:adapDynNetwork_x}--\eqref{eq:adapDynNetwork_a}.

As discussed in the previous sections, the form of the coupling vector field $g$ and the adaptation function $h$ can have various forms. One class of adaptive phase oscillator models that has recently gained a lot of attention takes the following form~\cite{KAS17,BER19,GKO22} 
\begin{align}
\dot{\phi}_i &= \omega_i + \sum_{j=1}^N a_{ij}\kappa_{ij} g(\phi_i-\phi_j), \label{eq:APO_phi}\\
\dot{\kappa}_{ij} & = -\epsilon\left(\kappa_{ij} + h(\phi_i - \phi_j)\right). \label{eq:APO_kappa}
\end{align}
The functions $g$ and $h$ are $2\pi$-periodic functions and $\epsilon$ is the adaptation time constant which is often considered to be small $\epsilon\ll 1$~\cite{BER19,GKO22}.

A rather simple model for the oscillators coupling dynamics, known as Kuramoto-Sakaguchi type model~\cite{SAK86}, describes a dynamical network of $N$ coupled phase oscillators. Equipped with an adaptation function similar to the phase interaction function, it reads
\begin{align}
	\dot{\phi}_{i} &=\omega-\frac{1}{N}\sum_{j=1}^{N}\kappa_{ij}\sin(\phi_{i}-\phi_{j}+\alpha),\label{eq:AdaptiveKS_phi} \\
	\dot{\kappa}_{ij}&=-\epsilon\left(\kappa_{ij}+\sin(\phi_{i}-\phi_{j}+\beta)\right).\label{eq:AdaptiveKS_kappa}
\end{align}
The interaction as well as the adaptation function are each homogeneously chosen, i.e., the same for each pair of phase oscillators, and given by a sinusoidal coupling kernel. The parameter $\alpha$ and $\beta$ can be considered as phase-lags of the interaction~\cite{SAK86} and the adaptation function~\cite{AOK11,KAS17}, respectively. 
Systems similar to~\eqref{eq:AdaptiveKS_phi}--\eqref{eq:AdaptiveKS_kappa} have been extensively studied \cite{REN07,AOK09,AOK11,PIC11a,TIM14,GUS15a,KAS16a,NEK16,KAS17,AVA18,BER19}.

%	Slow and fast adaption
\subsection{Slow and fast adaption limits}

As already introduced in the previous section, the dynamics of the nodal states $\bm{x}_i$ does not necessarily take place on the same time scale with the adaptation dynamics of the network's links. In the following sections, we describe the plethora of adaptation rules found and studied in various fields of research. In many examples, the role of different time scales is also discussed and empirical evidence for the time scale splitting is provided. 

A simple way to include different time scales for the nodal and the link dynamics is to explicitly introduce a time scale separation parameter $\epsilon$ in the adaptation function, i.e., $h(\bm{x}_i,\bm{x}_j,\kappa_{ij},t)=\epsilon \hat{h}(\bm{x}_i,\bm{x}_j,\kappa_{ij},t)$. Many works utilize a strict separation of the time scales for slow ($\epsilon\to 0$) or fast ($1/\epsilon \to 0$) adaptation, e.g. to derive analytic conditions for the emergence of observed phenomena or mean-field models. The differences between slow and fast adaptation rates have been analyzed, for example, in the context of slow and fast learning in~\cite{NIY09}. A detailed mathematical description of multiple time scale techniques will be discussed in sections~\ref{sec:recurrent} and~\ref{sec:multiscale}; here we discuss the scale separation formally from the perspective of applications.

Let us for simplicity consider a weighted adaptive dynamical network of the form~\eqref{eq:adapWeightedDynNetwork_x}--\eqref{eq:adapWeightedDynNetwork_k} with $h(\bm{x}_i,\bm{x}_j,t)=\epsilon \hat{h}(\bm{x}_i,\bm{x}_j,t)$. Then, in the formal slow adaptation limit ($\epsilon\to 0$), the nodal dynamics~\eqref{eq:adapWeightedDynNetwork_x} is much faster than the dynamics of the network structure. Hence, on a time scale of $\mathcal{O}(1)$, one can attempt to approximate the network as static and thus \eqref{eq:adapWeightedDynNetwork_x} can be treated as a dynamical network, where the network structures enters as a parameter. On the other hand, the network structure adapts to the nodal dynamics slowly. 

%The network dynamics can be approximated using a temporal average 
%\begin{align*}
%    \frac{d}{dt'}\kappa_{ij}(t') =  \frac1T \int_{t'-T}^{t'} h(\bm{x}_i(t), \bm{x}_j(t),\kappa_{ij}, t)\, \mathrm{d}t,
%\end{align*}
%where $t'=\epsilon t$ is the slow timescale, see~\cite{SAN07c} for details on averaging methods.

In the case of the fast adaptation $(1/\epsilon\to 0)$, the situation is inverse. Then, the network structure is on the fast time scale $t$ determined by the equation
\begin{align*}
    \dot{\kappa}_{ij} = h(\bm{x}_i, \bm{x}_j, \kappa_{ij}, t),
\end{align*}
where the nodal states $\bm{x}_i$ enter as parameters. In case of model~\eqref{eq:AdaptiveKS_kappa}, this would lead to $\kappa_{ij}=-\sin(\phi_i -\phi_j+\beta)$ for the network configuration. 

%The slow motion of the nodal dynamics can be determined here also by considering temporal averaging.

Thus, a separation of the time scales between the nodal and network dynamics can have favorable analytic consequences. In some realistic systems such a separation is indeed present and we will discuss these models in the next sections.
%\begin{itemize}
%\item local interaction induced adaptation (as in STDP)
%\item Fixed structure with weighted adaptive couplings
%\begin{itemize}
%    \item Differential (iterative for maps) rules for adaptation
%    \item Discontinuous
%\end{itemize}
%\item Evolving topology (as in structural plasticity)
%\item classification depending on the underlying structure (ring, all-to-all, multilayer, etc.)
%\item self-organized criticality Bornholt  Rolf
%\end{itemize}

%% file: parts/neuro.tex
Neural and neuronal systems have been a major research objects over the last decades. In this section, we introduce two forms of synaptic plasticity that have been studied in order to understand the function of the human brain. We note, however, adaptive dynamical networks have been also developed to understand e.g. the neuronal activity in Alzheimer's disease~\cite{GOR20,alexandersen2023multi} and other forms of adaptive mechanisms have been introduced to study the voltage dependence of synaptic plasticity~\cite{CLO10,MEI20a}, activity-dependent rewiring rules~\cite{RUB09}, short-term synaptic plasticity~\cite{TSO98,TAH20,SCH20c} or to incorporate resource constrains~\cite{VIR16c,KRO21,FRA22}.

\subsection{Spike-timing-dependent plasticity}\label{sec:STDP}

In neuronal systems, spike-timing-dependent plasticity (STDP) is an adaptation mechanism describing changes in coupling weights (synaptic efficacy), which are caused by the relative differences between firing times of pre- and postsynaptic neurons \cite{GER96,MAR97a,BI98,CLO10}. If a presynaptic neuron fires at time $t_{j,\text{pre}}$ and postsynaptic at time $t_{\text{i,post}}$, then the change of the coupling weight $\kappa_{ij}$ is a function of $\Delta t_{ij} = t_{i,\text{post}} - t_{j,\text{pre}}$. In experiments, such changes of synaptic efficiency were measured by forcing neurons to fire repetitively with a fixed inter-spike interval $\Delta t_{ij}$ \cite{BI98}. In modeling approaches, the coupling weights $\kappa_{ij}$ are updated in a point process-like manner as 
\begin{equation}
\label{eq:STDP-update}
\kappa_{ij}(t_+) = \kappa_{ij}(t_-) + \epsilon W(\Delta t_{ij})
\end{equation}
at each time-moment $t$ when either neuron $i$ or neuron $j$ fires.
At each individual spike, the update is small, which is reflected by the smallness of $\epsilon$. This also implies that the STDP adaptation takes place on a slower timescale than the neuronal spiking dynamics. 

\begin{figure}[t]
\includegraphics[width=8cm]{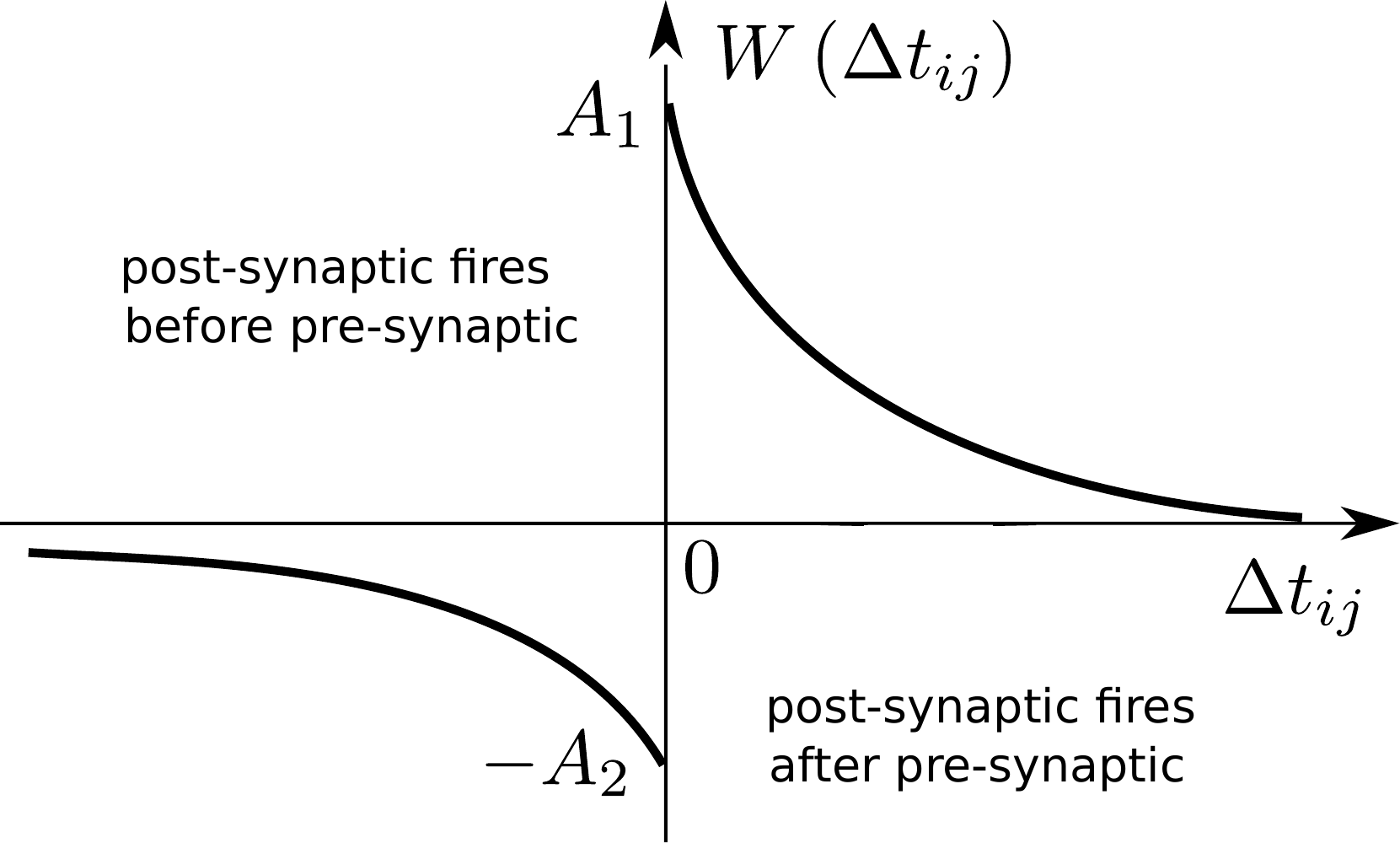}
\centering
\caption{A possible STDP adaptation function $W(\Delta t_{ij})$ given by Eq.~\eqref{eq:STDP} and measured experimentally \cite{BI98}. 
\label{fig:STDP}}
\end{figure}

The STDP update function $W(\Delta t_{ij})$ can be of the following form
\begin{equation}
\label{eq:STDP}
W\left(\Delta t_{i j}\right)=\left\{\begin{array}{ll}
A_{1} \mathrm{e}^{-\Delta t_{i j} / \tau_{1}}, & \Delta t_{i j} \geq 0 \\
-A_{2} \mathrm{e}^{\Delta t_{i j} / \tau_{2}}, & \Delta t_{i j}<0
\end{array}\right. ,
\end{equation}
which is an approximation of experimentally measured potentiation and depression of glutamatergic synapses induced by correlated spiking of presynaptic and postsynaptic neurons \cite{BI98}; see also \cite{GER14a} and Fig.~\ref{fig:STDP}. Here, $A_1,A_2,\tau_1$, and $\tau_2$ are real valued parameters that determine the shape of $W$. The STDP function can, however, have a form different from \eqref{eq:STDP}. 
It may depend, for instance, on the dendritic location of the connections \cite{Froemke2005,Froemke2010,Sjostrom2006,Letzkus2006,MEI20a}. 
As a result, the update rule may become even of the opposite sign. 
The work \cite{ROE19a} considers a symmetric update function in the form of a ''Mexican hat''. 

A typical general form of the adaptive dynamical network with STDP can be written as follows
\begin{eqnarray}
     \bm{\dot x}_i(t) & = & F_i(\bm{x}_i) + \sum_j \kappa_{ij} 
    G(\bm{x}_i,\bm{x}_j), 
    \label{eq:STDP-network-1}\\
    \left.\Delta  \kappa_{ij}(t)\right|_{t\in U_{i}\cup U_j} & = &
    \epsilon W(\Delta t_{ij}),
    \label{eq:STDP-network-2}
\end{eqnarray}
where $\bm{\dot x}_i = F_i(\bm{x}_i)$ determines the dynamics of an uncoupled neuron, $G(\bm{x}_i,\bm{x}_j)$  is an (often nonlinear) coupling term, $\Delta \kappa_{ij}(t) = \kappa_{ij}(t_+)-\kappa_{ij}(t_-)$ denotes a discontinuous jump of the coupling weight $\kappa_{ij}$ at time moment $t$ where one of the neurons $i$ or $j$ fires; we denote the set of such time moments as $U_i$ and $U_j$, respectively. Depending on the model for the individual neuron, such a spiking event can be determined differently. For example, it can be a moment where the voltage variable crosses the zero level. $\Delta t_{ij}$ is the interval between the current spike time $t$ and the previous spike time of the reciprocal neuron. The models of individual neurons (nodes) in the networks with STDP can be of different complexity, ranging from phase oscillators \cite{LUE16,MAI07,ROE19a,POP13} to more realistic conductance-based neurons \cite{POP15,POP13,ROE19a}. When using the update rule \eqref{eq:STDP-update}, the neuronal models determine the spiking times $t_i$ and $t_j$, while the STDP updates \eqref{eq:STDP-network-2} change the coupling weights $\kappa_{ij}$ affecting the neuronal dynamics. We note that the point process-like, discontinuous, update rule \eqref{eq:STDP-update} introduces additional challenges in the numerics and, especially, in theoretical analysis of systems with STDP. One of the possible approaches for the complexity reduction is to approximate the discontinuous update process by a continuous update \cite{LUE16,ROE19a} using an averaging technique.

We briefly summarize the main properties of the STDP adaptation: 
\begin{itemize}
\item
The update of the coupling weight occurs on a much longer timescale than the dynamics of individual nodes.
\item
The precise timing of individual neurons spiking matters. Also, the order of spiking plays an important role. 
\item
The adaptation of the coupling weight $\kappa_{ij}$ depends directly only on the relative dynamics of the adjacent nodes $i$ and $j$. 
\end{itemize}
Motivated by STDP, simplified phase-oscillator models were proposed (see Sec.~\ref{sec:phaseOscModel}), which possess the above features. It was shown that such models are much simpler for theoretical and numerical studies, while at the same time they possess a predictive power for STDP \cite{ROE19a,Thiele2023,LUE16,POP13}. 

\subsection{Structural plasticity}

Contrary to STDP, the structural plasticity mechanism allows connections between neurons to be deleted and created, and not only alter their weights \cite{Butz2009,Butz2013,VanOoyen}. 
It serves a homeostatic purpose to reach and maintain the target firing rate of the network. The model for structural plasticity contains one-compartment neurons that carry sets of synaptic (axonal and dendritic) elements, so-called contact points. Synapses are formed by merging corresponding synaptic elements or are deleted when synaptic elements are lost. In current models \cite{Nowke2018,Diaz-Pier2016,VanOoyen,Butz2013,Butz2009,MAN21}, the averaged neuronal activity is effectively represented by the neuron’s intracellular calcium concentration, which drives changes in the number of synaptic elements per neuron. The following experimental observation sets the ground for structural plasticity:  When the average activity of a neuron exceeds some level, the neuron withdraws dendritic spines, and retracts neurite branches, thereby reducing connectivity and hence activity. If the activity becomes lower, the neuron generates synaptic elements, increasing connectivity and activity. Accordingly to structural plasticity, connectivity in the network is updated on a much slower timescale than the electrical activity of neurons. The model works at the single neuron level, so that each neuron follows this rule independently. 

Here we sketch the main ideas of a structural plasticity model; more details can be found in  \cite{Butz2009,Butz2013,VanOoyen}. 
Important variables are $A_i$ the number of axonal elements of neuron $i$, $D_i^{\text{ex}}$ and $D_i^{\text{in}}$ are excitatory and inhibitory dendritic elements of neuron $i$.  
Axonal elements can be excitatory or inhibitory, for excitatory or inhibitory neurons, respectively.
$A_i$, $D_i^{\text{ex}}$, and  $D_i^{\text{in}}$ are continuous variables for integration, but they are rounded down to integer values for manipulating with synapse formation.

\begin{figure}[t]
\includegraphics[width=8cm]{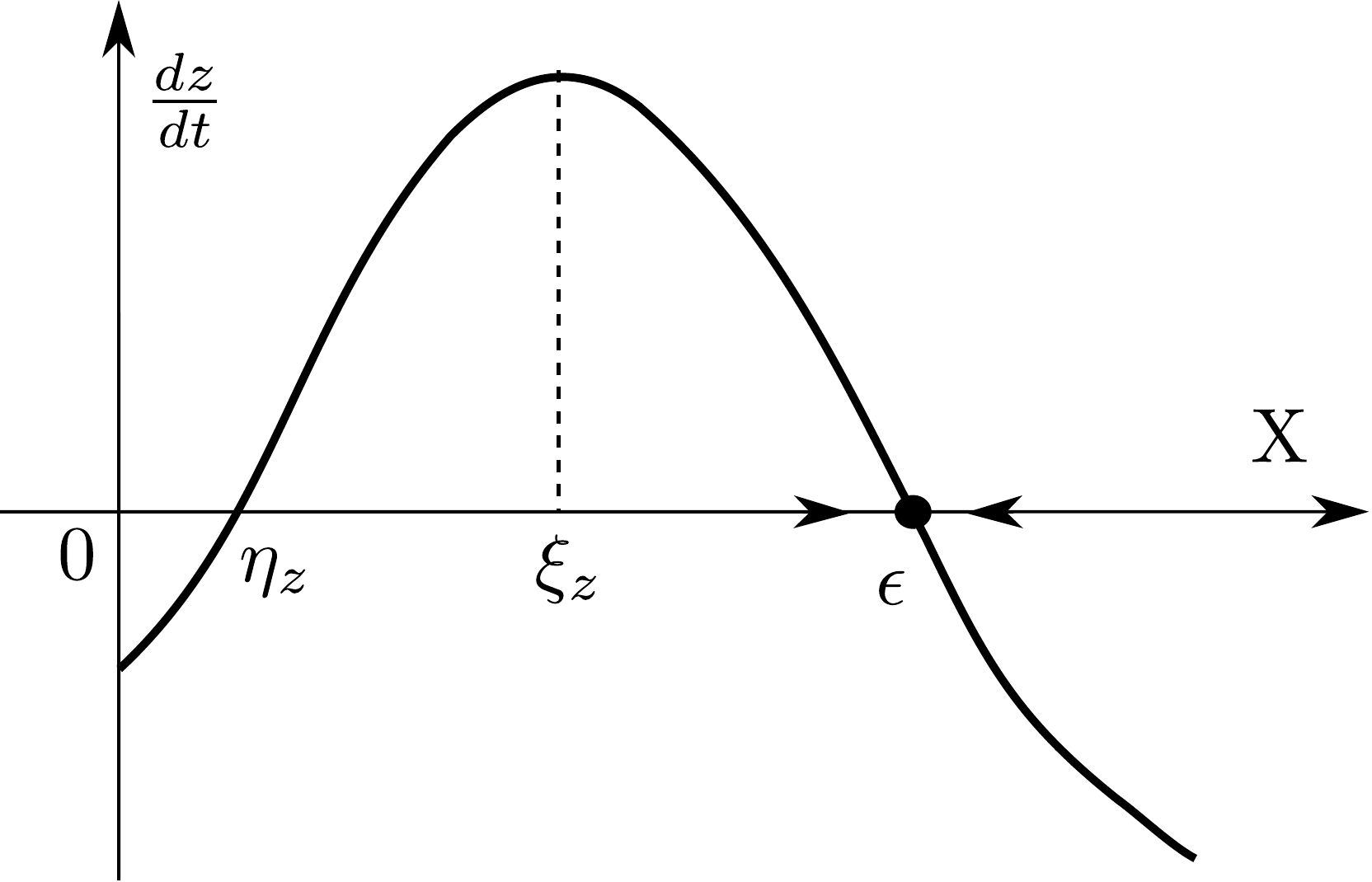}
\centering
\caption{Growth rate curve for synaptic elements $z_i$ used in structural plasticity; see Eq.~\eqref{eq:syn-el}. The curve is schematic and holds qualitatively for $\epsilon>\eta_z$ and $\nu>0$.
\label{fig:structural}}
\end{figure}

The dynamics of dendritic and axonal elements $z_i \in\left\{ A_i, D^{\text{ex}}_i, D^{\text{in}}_i\right\}$ are described by the following system \cite{Butz2013} 
\begin{equation}
\label{eq:syn-el}
\frac{d z_{i}}{d t}=\nu \left( 2 \, \mathrm{exp}{\left( -\left(\frac{X_i-\xi_{z}}{\zeta_{z}}\right)^{2}\right)}-1\right),
\end{equation}
where $X_i$ is the averaged activity of neuron $i$. The parameters  $\xi_{z}=(\eta_{z}+X_H)/2$ and $\zeta_{z}=(\eta_{z}-X_H)/(2 \sqrt{\ln 2})$ are such that the right-hand side of Eq.~\eqref{eq:syn-el} has a Gaussian shape and crosses zero at $\eta_z$ and $X_H>\eta_z$, see Fig.~\ref{fig:structural}. 
Such a shape of the growth curve guarantees the stabilization of the activity level at the target value $X_H$.
The neuronal activity can be measured by the level of calcium concentration  $X_i = \left[\mathrm{Ca}^{2+}\right]_{i}$, see \cite{Butz2013}.
The slowness of the structural update is determined by the maximum rate $\nu$, which is chosen to be $\nu=0.0001$ms$^{-1}$ in \cite{Butz2013}.

The level of calcium concentration (activity level) of neuron $i$ can be determined \cite{Butz2013} as
\begin{equation}
    \label{eq:Ca}
    \frac{dX_i}{dt}  = 
    \left\{
    \begin{matrix}
    -\frac{X_i}{\tau_\text{Ca}} + \beta,\quad V\ge 30\text{mV} , \\
    -\frac{X_i}{\tau_\text{Ca}} \quad \text{otherwise},
    \end{matrix}
    \right. 
\end{equation}
where $V$ is the membrane voltage variable.

Algorithmically, the update of the synaptic elements is performed discretely in a periodic manner (every 100ms in \cite{Butz2013}). The  synaptic elements are either randomly deleted if the corresponding value of $z_i$ given by Eq.~\eqref{eq:syn-el} decreases, or inserted otherwise. 

Above we described the dynamics of synaptic elements (connection points). 
For synapse formation, all vacant synaptic elements from all neurons are simultaneously  randomly assigned to a complementary synaptic element, i.e. excitatory axonal elements to excitatory dendritic elements and inhibitory axonal elements to inhibitory dendritic elements.
Whether or not the assigned pairs of complementary synaptic elements actually form synapses depends on the Euclidean distance between the neurons. For this, a distance-dependent kernel is introduced determining the distance-dependent likelihood for synapse formation between any pair of neurons. 

Summarising, the structure of the dynamical adaptive network system corresponding to the structural plasticity algorithm from \cite{Butz2013}, can be written in the following form
%\rem{SY}{add vacant elements dynamics}
\begin{eqnarray}
\dot x_i & = & F(x_i) + \sum_j \kappa_{ij} G(x_i,x_j), \label{eq:SP-1}
\\
\label{eq:SP-2}
\dot  z_{i} & = & \nu \left( 2 \, \mathrm{exp}{\left( -\left(\frac{X_i-\xi_{z}}{\zeta_{z}}\right)^{2}\right)}-1\right),\quad z \in\left\{ A, D^{\text{ex}}, D^{\text{in}}\right\}
\\
\label{eq:SP-3}
\dot X_i &  = & \left\{
\begin{matrix}
    -\frac{X_i}{\tau_\text{Ca}} + \beta,\quad V_i=x_i^1\ge 30\text{mV},  \\
    -\frac{X_i}{\tau_\text{Ca}}, \quad \text{otherwise}.
    \end{matrix}
\right. 
\\
\label{eq:SP-4}
\left. \Delta \kappa_{ij} \right|_{t=kT} 
& = &  P(z_i,z_j,i,j),\quad T=100\text{ms}.
\end{eqnarray}
Here the function $P(z_i,z_j,i,j)$ denotes the change of the connectivity between the neurons $i$ and $j$, which depends on the number of corresponding dendritic and axonal elements $z_i,z_j$ of these neurons. As determined in \cite{Butz2013}, the dependence $P$ is not deterministic, since connections may appear or disappear randomly, with the probabilities depending not only on $z_i,z_j$ but also on other geometric properties such as the distance between the neurons. Such a dependence is expressed by the arguments $i,j$ of this function. Note that the range of $\kappa_{ij}$ values in \cite{Butz2013} are discrete and include $0$ for the case when there are no connections between a given pair of neurons. In system \eqref{eq:SP-1}-\eqref{eq:SP-4}, the dynamics of the single node $i$ is described by the collection of variables $(x_i,A_i,D_i^\text{ex},D_i^\text{in},X_i)$, and the adaptation rule is discrete and event-based. Since $T$ is much larger than the dynamics of the single neuron, the structural plasticity adaptation is slow. Manos et al.~\cite{MAN21} show that by taking structural plasticity into account, clinically important phenomena, such as the increase of desynchronizing effects of coordinated reset neuromodulation can be reproduced computationally.

% \bigskip

% \remg{remark}{short-term, long-term
% the role of phenomenological models, comparisons (e.g. our PLOS paper comparing HH and phase models, Tinnitus und Parkinson works by Tass)
% check Gerstner’s book} 

%% file: parts/physiol.tex
Network physiology is an interdisciplinary research area bridging between physiological modeling approaches from the micro to the macro scale. Bringing together network science, dynamical system theory and physiological modeling, network physiology aims for getting insights into systemic diseases such as cancer, sepsis and others~\cite{IVA21,SCH22a}. Except from neural systems, modelling approaches using an adaptive network structure are rare. Recently little steps into this direction have been taken in the works~\cite{SAW21b,BER22}, where the focus lies on the functional modeling of interactions between different systems in the living organism, not on a detailed biochemical modeling of a single organ or system, see e.g.~\cite{schuurman2023embracing} for a recent perspective on the modeling of sepsis. In this section, we describe briefly the functional model (unified disease model) recently proposed to study the emergence of tumor disease and sepsis.

\begin{figure}
	\centering
	\includegraphics[width = \textwidth]{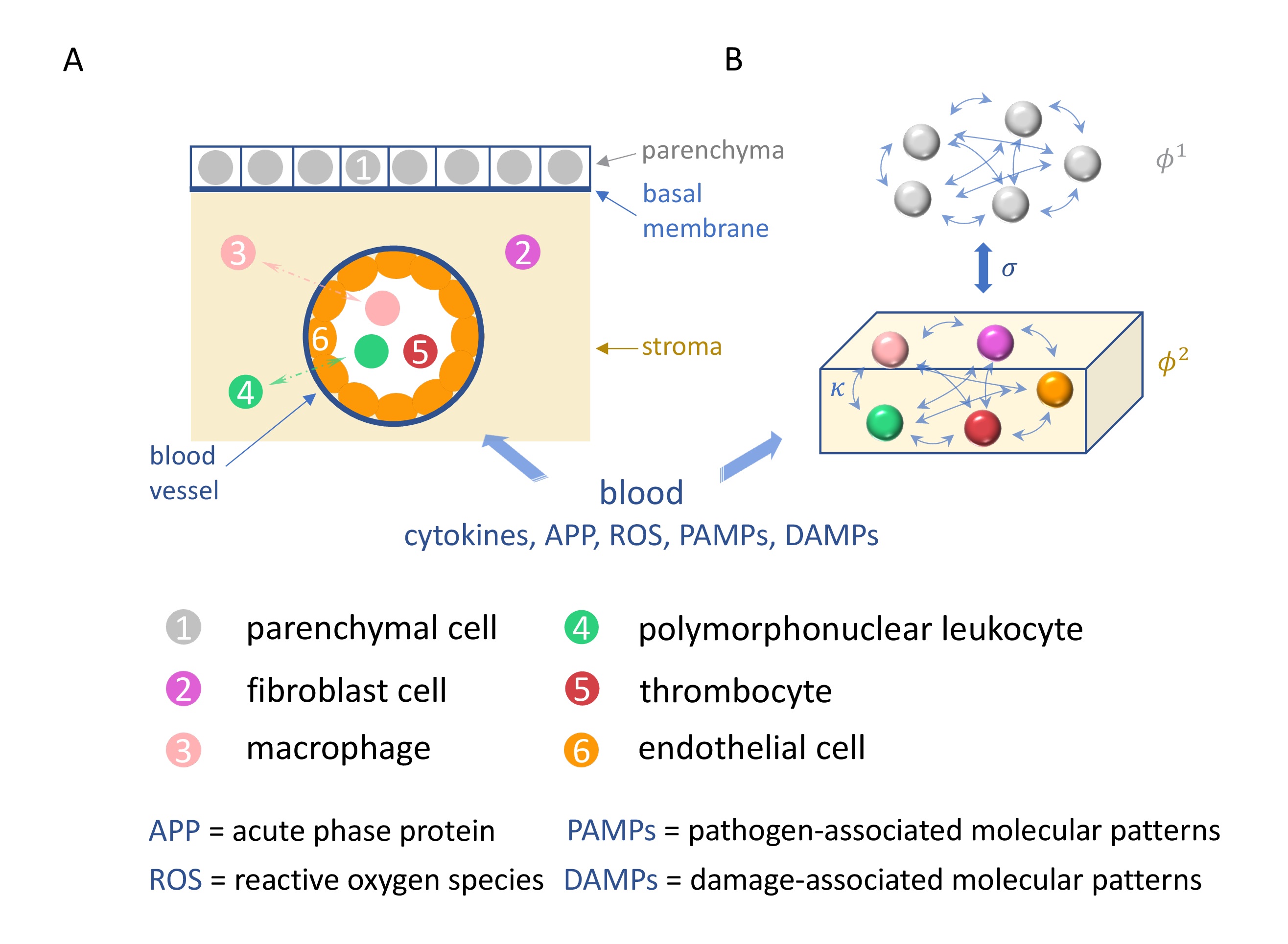}
	\caption{Schematic illustration of a unified disease model. (A) A tissue element is depicted, in which the basic processes take place: shown are the cells (colored) involved such as parenchymal, fibroblast, endothelial cells, and macrophages, polymorphonuclear leukocytes and thrombocytes in the parenchyma (grey), the stroma (yellow), and the capillary blood vessel. (B) depicts the functional interactions within and between the two corresponding network layers in our model, the parenchyma and the stroma (immune layer). Figure taken from~\cite{BER22}.}
	\label{fig:physModel}
\end{figure}

\subsection{Functional two-layer network model}

The unified disease model is centered around the \textit{nonspecific immune system}, which includes disease-specific initial conditions, parameters and infection-driven cytokine dysregulation. For the analysis of tumor disease or sepsis, we consider a volume element of tissue consisting of parenchyma, basal membrane and stroma, see Fig.~\ref{fig:physModel} A. The network layer of parenchymal cells (superscript 1) are represented by $N$ phase oscillators $\phi_i^{P}$, $i=1,\ldots, N$ and the network layer of immune cells (superscript 2) are presented by $N$ phase oscillators $\phi_i^{S}$. The coupling weights in the parenchymal layer are considered to be partly fixed and partly adaptive, while in the immune layer the coupling weights are completely adaptive. Coupling weights model here the communication through cytokines which mediate the interaction between the parenchymal cells by the coupling weights $\kappa_{ij}^{P}$, and those between the immune cells by coupling weights $\kappa_{ij}^{S}$. Note that $\phi_i^S$ and $\kappa_{ij}^{S}$ represent the collective dynamics of all dynamical units of the stroma, see Fig.~\ref{fig:physModel} B. The use of phase oscillators for the functional modeling of the interacting parenchymal cells and immune cells is motivated by the fact that phase oscillator networks are a paradigmatic model for collective coherent and incoherent dynamics. The healthy state is assumed to be characterized by regular periodic, fully synchronized dynamics of the phase oscillators. Healthy and pathological cells differ by their metabolic activity, i.e., pathological cells shut down their mitochondrial cellular respiration and switch to aerobic glycolysis. Therefore they are less energy-efficient and thus have a modified cellular metabolism and reduced function, which is reflected in the phase oscillator model by a different frequency.

A multiplex network with two layers each consisting of $N$ identical adaptively coupled phase oscillators of the following form has been introduced:
\begin{align}
\label{eq:DGL_somatic}
	\dot{\phi}_{i}^{P} &=\omega_i^{P}-\frac{1}{N}\sum_{j=1}^{N}(a_{ij}^P+\kappa_{ij}^P)\sin(\phi_{i}^P-\phi_{j}^P+\alpha^{PP}) %\nonumber \\[-.15cm]
	-\sigma \sin(\phi_{i}^P-\phi_{i}^S), \\
		\dot{\kappa}_{ij}^P&=-\epsilon^P \left (\kappa_{ij}^P+\sin(\phi_{i}^P-\phi_{j}^P-\beta)\right), \nonumber
\end{align}
\begin{align}
\label{eq:DGL_immune}
	\dot{\phi}_{i}^{S} &=\omega^{S}-\frac{1}{N}\sum_{j=1}^{N}\kappa_{ij}^S\sin(\phi_{i}^S-\phi_{j}^S+\alpha^{SS}) %\nonumber \\[-.15cm]
	-\sigma \sin(\phi_{i}^S-\phi_{i}^S), \\
	\dot{\kappa}_{ij}^S&=-\epsilon^S \left (\kappa_{ij}^S+\sin(\phi_{i}^S-\phi_{j}^S-\beta)\right), \nonumber
\end{align}
where $\phi_i^{\mu}\in [0,2\pi)$ represents the phase of the $i$th oscillator ($i=1,\dots,N$) in the $\mu$th layer ($\mu=P,S$), $\omega_i^P\equiv \omega_i$ are the natural oscillator frequencies of the parenchymal cells which are distributed according to a probability distribution $\rho(\omega^P)= (1-r)\delta(\omega^P-\omega^h) + r\delta(\omega^P-\omega^{pat})$ where $r$ is the fraction of pathological parenchymal cells (tumor cells) relative to the number of all parenchymal cells $N$, $\delta$ is the Dirac delta function, and $\omega^{pat}$ and $\omega^h$ are the natural frequencies of pathological and healthy parenchymal cells, respectively. The value of $\omega^S \equiv \omega$ is the natural frequency of the immune cells. The interaction between the oscillators within each layer is determined by the intralayer connectivity weights $a_{ij}^P\in[0,1]$ (fixed interaction within an organ) and $\kappa_{ij}^{\mu}\in[-1,1]$ (adaptive interaction mediated by cytokines). Between the layers the interlayer coupling weights $\sigma \ge 0$ are fixed and symmetric for both directions of interaction. Further the interactions within the layer depend on the phase lag parameters $\alpha^{PP}$ and $\alpha^{SS}$. The adaptation rates $0<\epsilon^\mu \ll 1$ separate the time scales of the slow dynamics of the coupling weights and the fast dynamics of the oscillatory system. The adaptation rate of the parenchymal layer $\epsilon^P$ is assumed to be slow compared to the adaptation rate of the immune layer $\epsilon^S$, i.e., $\epsilon^P \ll \epsilon^S$ to account for the faster reaction of the immune cells~\cite{MOR13a,ALT19}. Thus there are two classes of adaptive coupling weights modeling two different cytokine mechanisms on two different timescales.

The parameter $\beta$ plays an essential role in the model because it governs the adaptivity rule of the cytokines. It is called age parameter as it mimics a systemic sum parameter which accounts for different influences, such as physiological changes due to age, inflammaging, systemic and local inflammatory baseline, adiposity, pre-existing illness, physical inactivity, nutritional influence, etc.

\subsection{Predicting hospitalization incidences for sepsis}

\begin{figure}
	\centering
	\includegraphics[width = 1.0\textwidth]{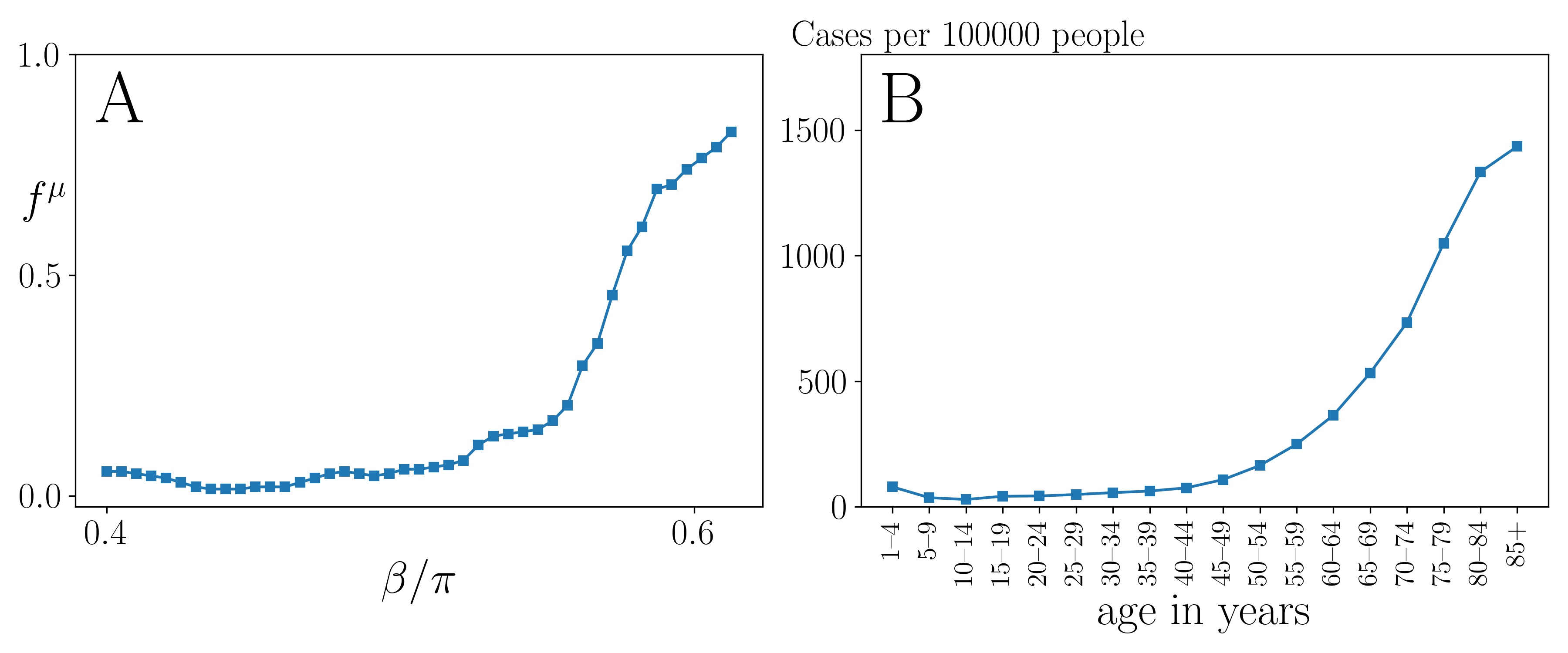}
	\caption{Qualitative comparison of model prediction with empirical data. (A) Pathological states ratio for $\sigma=1$ vs the age parameter $\beta$ where all data points were averaged over a sliding window of $4$ neighboring data points. (B) Empirical data taken from~\cite{FLE16} showing the hospitalization incidence of sepsis per 100,000 inhabitants in Germany by age group for the years from 2007 to 2013. Figure taken from~\cite{BER22} where also further details on the simulations and the model can be found.}
	\label{fig:sepsisInc}
\end{figure}

A first remarkable result achieved by this functional modeling approach  (Eqs.~\eqref{eq:DGL_somatic}--\eqref{eq:DGL_immune}) can be seen in Fig.~\ref{fig:sepsisInc}. Figure~\ref{fig:sepsisInc}A presents the ratio of observed pathological states compared to the number of considered initial conditions for the simulations depending on the age parameter $\beta$, see~\cite{BER22} for details. It shows that the probability of a pathological sepsis state sharply rises for the age parameter $\beta$ above approximately $\beta>0.5\pi$. This curve compares favorably with empirical data of patients which gives the number of cases of sepsis per 100,000 inhabitants in Germany as a function of age, presented in Fig.~\ref{fig:sepsisInc} B. This comparison shows a striking similarity that needs to be investigated in further studies, however, providing first evidence for the potential of the proposed functional modeling approach.

%% file: parts/machineLearning.tex
A machine learning implementation should ideally possess self-adaptation capabilities. 
That means it should adapt to changes in the real-world that were not anticipated during the initial development and learn with new data in both the runtime and development environments. 
Such machine learning methods should be able to dynamically adjust learning and goals based on real-time feedback, making them suitable for operations with changing external environments or operations that require optimized response.
In this section, we demonstrate the relevance of adaptive networks in machine learning applications. 

\subsection{Deep neural networks as an adaptive dynamical network during training phase}

\begin{figure}
\includegraphics[width=8cm]{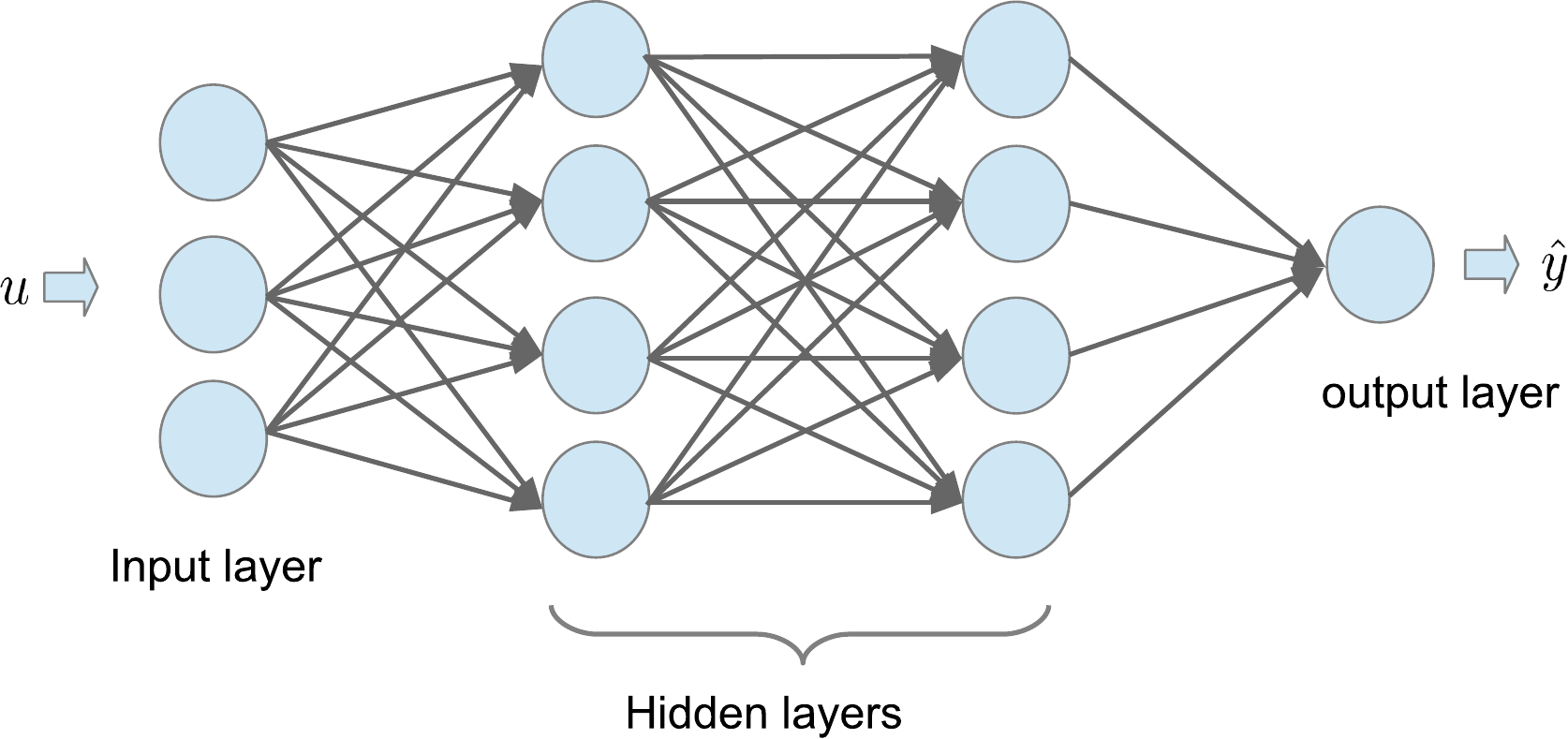}
\centering
\caption{Schematic representation of the multilayer perceptron. The layers are connected in a feed-forward manner. Each connection is trained during the training phase. 
\label{fig:DNN}}
\end{figure}

Many machine learning applications employ artificial neural networks with adaptive weights. The multilayer perceptron, for example, consists of layers of artificial neurons coupled in a feed-forward manner \cite{LeCun2015}, see Fig.~\ref{fig:DNN}. The corresponding coupling weights are adapted during the training phase to optimally perform a specific task. After training is completed, the coupling weights are fixed and a network with a static coupling structure is used for inference. Thus, the adaptive features of the network are used only during the training phase. 

%In standard deep neural networks (DNN), node dynamics is implemented with discrete nonlinear maps, and training is performed with the gradient descend method. To find the necessary gradients of a loss function with respect to the coupling weights, the back-propagation algorithm is used \cite{Goodfellow2016,Rumelhart1986}. 

Let us show that a training process for deep neural networks  (DNNs) is based on an adaptive dynamical network with discrete time. 
Let $\boldsymbol{\kappa}$ be the vector containing the coupling weights in a DNN, and $(\mathbf{u}(n),\mathbf{y}(n))$ is a sequence of input-output pairs used for the training. 
A successively trained neural network is expected to reliably reproduce not only the outputs  $\mathbf{y}(n)$ for the inputs $\mathbf{u}(n)$ of the training sequence with possibly small error, but also the input-output relationships for a test data set $(\mathbf{u}_\text{test}(n),\mathbf{y}_\text{test}(n))$.

Consider the training step $n$ with the training data $(\mathbf{u}(n),\mathbf{y}(n))$ and coupling weights $\boldsymbol{\kappa}({n-1})$ resulting from the training step $n-1$. 
Denote the output of the DNN as 
\begin{equation}
\label{eq:DNN}
\mathbf{\hat y}({n}) = F_{\text{DNN}}(\mathbf{u}(n),\boldsymbol{\kappa}({n-1}),\mathbf{\hat y}({n-1})). 
\end{equation}
In most cases, the dependence \eqref{eq:DNN} does not include the output of the previous training $\mathbf{\hat y}({n-1})$, i.e., Eq.~\eqref{eq:DNN} reads $\mathbf{\hat y}({n}) = F_{\text{DNN}}(\mathbf{u}(n),\boldsymbol{\kappa}({n-1}))$. 
However, for generality, we add such a dependency to incorporate possible applications to temporal data, where the output of the neural network is further used as an input.
Equation \eqref{eq:DNN} represents a dynamical relation determining the functional response of the neural network to the input $\mathbf{u}(n)$. 
The calculation of \eqref{eq:DNN} implicitly includes the updates for the states of the network nodes $\mathbf{x}(n)$ at discrete times $n$ as
$$
\mathbf{\hat y}(n)= F^{\text{out}}(\mathbf{x}(n)); \quad
\mathbf{x}(n) = X(\mathbf{u}(n),\boldsymbol{\kappa}(n-1),\mathbf{\hat y}(n-1)),
$$
where the function $X$ includes the recurrent applications of certain activation functions accordingly to the scheme in Fig.~\ref{fig:DNN}. In fact, the output $\mathbf{\hat y}(n)= F^{\text{out}}(\mathbf{x}(n))$ is a function of the nodes in the output layer only. For our purpose, we do not specify the functions $F_{\text{DNN}}$, $F^{\text{out}}$, or $X$ explicitly.

To compute the updated coupling weights $\boldsymbol{\kappa}_{n}$, one usually considers a loss function $L(\mathbf{\hat y}(n),\mathbf{y}(n))$, which could be, for example, $L=\| \mathbf{\hat y}(n) - \mathbf{y}(n)\|^2$. 
The simplest realization of the gradient descent algorithm  leads to the update rule 
\begin{equation}
\label{eq:DNN-2}
\boldsymbol{\kappa}(n) = \boldsymbol{\kappa}(n-1) - \varepsilon \nabla_{\boldsymbol{\kappa}} L (\boldsymbol{\kappa}(n-1),\mathbf{u}(n),\mathbf{y}(n),\mathbf{\hat y}(n-1)),
\end{equation}
where the parameter $\varepsilon$ controls the update rate (it can depend on the discrete time $n$). 
A common way for computation of the gradient $\nabla_{\boldsymbol{\kappa}} L$ in DNNs is the back-propagation algorithm, see details in \cite{Goodfellow2016,Rumelhart1986}. 

Finally, the obtained set of dynamical equations during the training phase is summarized as
\begin{eqnarray}
\label{eq:DNN-f2}
\mathbf{x}(n) & = & X(\mathbf{u}(n),\boldsymbol{\kappa}(n-1),\mathbf{\hat y}(n-1)), \\
\label{eq:DNN-f1}
\mathbf{\hat y}(n) & = & F^{\text{out}}(\mathbf{x}(n)), \\
\label{eq:DNN-f3}
\boldsymbol{\kappa}(n) & = & \boldsymbol{\kappa}(n-1) - \varepsilon \nabla_{\boldsymbol{\kappa}} L (\mathbf{u}(n),\mathbf{y}(n),\boldsymbol{\kappa}(n-1),\mathbf{\hat y}(n-1)),
\end{eqnarray}
and it has the form of an adaptive network with the nodal dynamics \eqref{eq:DNN-f2}, some ''macroscopic'' network characteristic given by the output \eqref{eq:DNN-f1}, and the adaptation rule \eqref{eq:DNN-f3} for the coupling weights. The adaptation rule depends on the output \eqref{eq:DNN-f1}. Since the gradient descent parameter $\varepsilon$ is usually small, the adaptation dynamics is slow. 

We note that the adaptive dynamical network \eqref{eq:DNN-f2}--\eqref{eq:DNN-f3} is \textit{non-autonomous}, i.e., it is driven by the time-dependent inputs $(\mathbf{u}(n),\mathbf{y}(n))$.

% Recently, it has been shown that DNNs and adaptive dynamical networks of the form \eqref{eq:DNN-f2}--\eqref{eq:DNN-f3} can be constructed from a single delay system with multiple-delays feedback \cite{Stelzer2021a,Stelzer2020e}. Such a scheme was applied successfully to several machine learning classification and denoising tasks \cite{Stelzer2021a}.

\subsection{Adaptive networks in reservoir computers}

Reservoir Computing (RC) combines the computational capabilities of neural networks with a fast and rather simple training  \cite{Jaeger2001,Maass2002}. In RC, only the linear weights of the output layer are trained using the relatively simple ridge regression method. Most reservoir computing approaches consider a reservoir with fixed internal connection weights. However, plasticity as an unsupervised and biologically inspired adaptation seems to be beneficial for the performance of RC \cite{Morales2021,Babinec2007,Triesch2005,Schrauwen2008,Steil2007,Wang2019}. 

The work \cite{Morales2021} considers a combination of several learning rules in RC. First, the anti-Hebbian learning for synaptic plasticity can be exemplified by the following adaptation rule for the coupling weights $w_{kj}(t)$ (the time $t$ in this system is discrete)
\begin{equation}
\label{eq:RC}
w_{kj}(t+1) = \frac{ w_{kj}(t) + \eta x_k(t+1) x_j(t)}
{\sqrt{\sum_j \left( w_{kj}(t) + \eta x_k(t+1) x_j(t) \right)^2}},
\end{equation}
where $\eta$ is a parameter controlling the learning rate, 
 $x_j(t)$ and $x_k(t+1)$ represent the activity of the pre- and post-synaptic neurons, respectively. 
The work \cite{Morales2021} also uses another type of plasticity, called intrinsic plasticity, which alters the excitability of individual neurons to match a Gaussian target distribution function. This is a type of homeostatic plasticity. It has been shown that the best performance is usually achieved by a combination of synaptic and homeostatic plasticity.

In \cite{Babinec2007}, Hebbian-type synaptic plasticity was employed for modifying the reservoir weights dynamically. In their model, the so-called anti-Oja’s update rule is used
$$
w_{kj}(t+1) = w_{kj}(t) - \eta x_k(t+1) \left[ x_j(t) - x_k(t+1)w_{kj}(t) \right],
$$
which approximates Eq.~\eqref{eq:RC} for small updates $\eta$, see \cite{Morales2021}. The prediction error achieved by the RC with adaptation was substantially smaller compared to the prediction error achieved by a standard algorithm.

In \cite{Triesch2005}, the intrinsic plasticity rule is developed as a gradient adaptation dynamics based on information theory. Such a rule allows the neuron to bring its firing rate distribution into an approximately exponential regime, as observed in visual cortical neurons. In \cite{Schrauwen2008}, the intrinsic plasticity was applied to RC. The authors demonstrate that the intrinsic plasticity is able to make RC more robust: the internal dynamics can autonomously adjust to the dynamic regime that is optimal for a given task, independent of the initial weights or input scaling.
Other plasticity rules in RC networks can be found in \cite{Steil2007,Yusoff2016,Wang2019}.

Another related research direction is the prediction of the behavior of adaptive networks with machine learning tools.  For example, in \cite{Andreev2022}, the authors have predicted the macroscopic signal of an adaptive network of Kuramoto phase oscillators using RC. To improve the prediction, an additional time-delay embedding of the input was performed. 

%% file: parts/control.tex
%In the previous section, we have elaborated on the relevance of adaptive dynamical networks for modelling real-world dynamical systems. In addition, 
Adaptive systems and in particular adaptive dynamical networks have been studied with regards to various control problems and several methods have been developed~\cite{yakubovich1969,Annaswamy2021AHP,LIU16}. In this context, adaptive dynamical networks are also called state-dependent networks~\cite{MES05,RAJ11}. In the following, we provide a short overview on selected control schemes that have been developed using adaptive network structures.Note that there are also other control approaches using adaptive dynamical networks, e.g. control through rewiring~\cite{XUA11} or adaptive gradient networks~\cite{CHE08c}.

\subsection{Continuous adaptive control of full synchronization in complex networks}

Of particular interest in control theory is the control of full synchronization in a dynamical system. For this, consider a dynamical network of the form
\begin{align}
    \frac{\mathrm{d}\bm{x}_i}{\mathrm{d}t}=f(\bm{x}_i)+\sum_{j=1}g(a_{ij}, \kappa_{ij}, \bm{x}_i, \bm{x}_j),
\end{align}
see our more general definition of adaptive dynamical networks in Sec.~\ref{sec:adaptiveNet}. Note that the coupling function can take the forms e.g. $a_{ij}\kappa_{ij}g(\bm{x}_i, \bm{x}_j)$ (Eq.~(\ref{eq:dynNetworkGenAdj})) or $l_{ij}g(\bm{x}_i, \bm{x}_j)$ where $l_{ij}$ denote the entries of the (weighted) Laplacian matrix. For such a system, we assume that the fully synchronized state, i.e., the solution for which $\bm{x}_i(t)=\bm{s}(t)$ for all $i=1,\dots,N$, exists. In simple words, the goal of control theory is to guarantee the stability of this fully synchronized state against perturbations and, moreover, to make the size of its basin of attraction as large as possible, desirably covering the whole phase space, i.e., global stability. A couple of different approaches to control these states based on an adaptive change of the network structure have been developed. In this section, we briefly introduce some of them. For a more detailed review, we refer to~\cite{LEL10b}.

The weights $\kappa_{ij}$ are also called coupling gains and hence, the type of control is often called \textit{gain control decentralized adaptive strategy}. In its general form the gain control scheme can be written as
\begin{align}\label{eq:gainControl}
    \dot{\kappa}_{ij} = h_{ij}(\bm{x}_1,\dots,\bm{x_N}).
\end{align}
At this point, the distinction between vertex-based and edge-based adaptive strategies can be made. For the vertex-based strategy, for each vertex (node) of the network a single coupling gain $\kappa_i$ is considered and hence the dynamical equations~\eqref{eq:gainControl} reduces to $\dot{\kappa}_{i} = h_{i}(\bm{x}_1,\dots,\bm{x}_N)$. Control schemes like this have been proposed e.g. by Kurths and Zhou~\cite{ZHO06f}, DeLellis et. al.~\cite{LEL08}, and Sorrentino and collaborators~\cite{SOR08,SOR10a,RAV09}. As outlined in the preliminaries of this review, these vertex-based control strategies do not belong to the group of adaptive dynamical networks. For the edge-based adaptive control strategies, the control mechanism is implemented in the individual coupling weights by considering a strategy of the form $\dot{\kappa}_{ij} = h_{ij}(\bm{x}_1,\dots,\bm{x}_N)$. Hence, the dynamical network together with its control strategy forms an adaptive dynamical network.

Edge-based strategies have been considered depending on the local synchronization error $\bm{e}_{ij}=\bm{x}_i-\bm{x}_j$. DeLellis et al.~\cite{LEL10b} considered two classes of strategies $h_{ij}(\bm{x}_1,\dots,\bm{x}_N)=h(\bm{e}_{ij})$ with (i) $h(\bm{e}_{ij})=\alpha \|\bm{e}_{ij}\|^p$ where $\alpha\in\mathbb{R}$, $\|\cdot\|$ indicating the Euclidean norm and $0<p\le 2$, or (ii) a monotonously increasing function $h(\bm{e}_{ij})$ with $h(0)=0$ and $0\le h(\bm{e}_{ij})<\infty$. Examples are $h(\bm{e}_{ij})=\alpha \|\bm{e}_{ij}\|$ and $h(\bm{e}_{ij})=\frac{\|\bm{e}_{ij}\|}{\|1+\bm{e}_{ij}\|}$ for class (i) or (ii), respectively. For dynamical networks for which the corresponding vector field is of QUAD-type, see~\cite{LEL11} for a definition, it was shown that an edge-based adaptive control strategy of the first or second class guarantees global stability of the fully synchronized state~\cite{LEL09a}. These results have been also confirmed by numerical results~\cite{LEL08,LEL08a,LEL09} for several dynamical systems including the Kuramoto model, consensus models, Chua's circuits and Rössler oscillators. Another control scheme has been analyzed in~\cite{YU12} where instead of the coupling weights, the entries of the weighted Laplacian matrix are updated, i.e., the authors investigated an adaptation rule of the form $h_{ij} (\bm{e}_{ij}) = -\alpha_{ij}\bm{e}_{ij}^T \Gamma \bm{e}_{ij} $ where $\Gamma$ is a positive semi-definite diagonal matrix and $\alpha_{ij}=\alpha_{ji}\in\mathbb{R}$. Also for these systems global stability could be proved and numerically confirmed.

\subsection{Edge snapping control}

While in the previous section the control strategies build on a prescribed network structure, edge snapping describes a control scheme which allows for reorganization of the network topology~\cite{LEL10b}. Note in this regard that the control strategies lead to a constantly increasing coupling weight $\kappa_{ij}$ until full local synchronization $\bm{e}_{ij}$ is reached. Hence, whenever a link is active, i.e., $a_{ij}=1$, the corresponding coupling weight is positive when the asymptotic state is reached. In order to allow also for vanishing coupling weights, in~\cite{LEL10} a second order differential equation for the coupling weights of the following form has been introduced:
\begin{align*}
    \ddot{\kappa}_{ij}+d\dot{\kappa}_{ij} + \frac{\partial}{\partial \kappa_{ij}}V(\kappa_{ij}) = h(\bm{e}_{ij}),
\end{align*}
where $d$ is a damping constant. The dynamics of the network topology is governed by the (bistable) potential $V$, which is considered to be sufficiently smooth and to posses two local minima at $\kappa_{ij}=0$ (inactive link) and at $\kappa_{ij}=1$ (active link). It has been shown that also for this control strategy full synchronization is asymptotically achieved. Moreover, depending on the form of the potential $V$ the resulting network supporting full synchronization can vary strongly. See~\cite{LEL10b} for examples.

For edge snapping also event-based strategies have been considered. The so-called hybrid adaptive coupling weights are given by 
\begin{align*}
    \kappa_{ij} = \begin{cases} \hat{\kappa}_{ij}, & \text{if } 
    % \Theta(\hat{\kappa}_{ij},\bm{e}_{ij})>1,\\
    \hat{\kappa}_{ij} >1,\\
    0, & \text{otherwise},
    \end{cases}
\end{align*}
where the reference coupling weight $\hat{\kappa}_{ij}$ is controlled by $\mathrm{d}\hat{\kappa}_{ij}/\mathrm{d}t=\alpha \|\bm{e}_{ij}\|$ and the event function 
%$\Theta$ 
is chosen to be $\Theta(\hat{\kappa}_{ij},\bm{e}_{ij})=\hat{\kappa}_{ij}$. 
To render the resulting network structures unweighted and similar to those obtained by the edge snapping described above, the following rule could be considered: $\kappa_{ij} = 1$ if $\Theta(\hat{\kappa}_{ij},\bm{e}_{ij})>1$ and $\kappa_{ij}=0$ otherwise.

\subsection{Speed-gradient method and cluster synchronization}

Another control scheme that have been used to control the synchronization dynamics with adaptive coupling weights in complex dynamical networks is the speed-gradient method~\cite{FRA07}. In~\cite{LEH14,LEH15}, the authors apply the speed-gradient method to a coupled system of Stuart-Landau oscillators where the coupling weights were used as control variables. The control scheme is obtained by first choosing an appropriate goal function for the desired $M$-cluster state, i.e., a state where the system splits up into $M$ groups of completely synchronized oscillators. Second the dynamical equation
\begin{align*}
    \dot{\bm{u}}=-\Gamma\nabla \frac{\partial Q(\bm{z}_1,\dots,\bm{z}_N,u,t)}{\partial t}
\end{align*}
determines the state of the control variables $\bm{u}$, where $\Gamma$ is a positive definite gain matrix and $\bm{z}_i$ is the state vector of the $i$th Stuart-Landau oscillator. Choosing the coupling weights $\kappa_{ij}$ as control variables and $\Gamma_{ij}=\gamma_G \delta_{ij}$ ($\gamma_G>0$), a continuous adaptive coupling strategy can be derived that possesses the form
\begin{align*}
    \dot{\kappa}_{ij}=h(\bm{z}_i,\bm{z}_j, \hat{h}_i(\bm{z}_1,\dots,\bm{z}_N), \tilde{h}_i(\kappa_{11},\dots,\kappa_{NN}))
\end{align*}
with smooth functions $\hat{h}_i$ and $\tilde{h}_i$ that describe special local mean values for the oscillatory system and the coupling topology, respectively. For more details we refer to~\cite{LEH14,LEH15b}.

%% file: parts/powerGrids.tex
Network models describing power systems as well as micro and macro power grids have been analyzed intensively~\cite{BER81,SAL84,SAU98a,FIL08a,SCH16o}. It was shown that rather simple low-dimensional models capture certain aspects of the short-time dynamics of power grids very well~\cite{DOE13,WEC13,NIS15,AUE16}. In particular, the model of phase oscillators with inertia has been widely used to understand synchronization phenomena of complex networks and as a paradigm for the dynamics of modern power grids~\cite{DOE12,ROH12,COR13,MOT13a,DOE14,MEN14,ROD16,WIT16,AUE17,MEH18,SCH18c,TAH19,HEL20,KUE19,MOL20,TOT20,ZHA20c}.

Adaptive dynamical networks emerge naturally in power grid systems as a result of the inductances and capacitances to the ground of the power grid lines \cite{Gros2019}. Hence, the lines possess their own dynamics that depend on the state of the neighboring nodes. 
However, in this review, we do not cover the topic of the power grids with transmission-line dynamics. Instead, we consider a simpler class of models consisting of $N$ coupled phase oscillators with inertia is given by
\begin{align}\label{eq:KwI_2order}
    M\ddot{\phi}_i +\gamma\dot{\phi}_i & = P_i + \sum_{j=1}^N a_{ij}h(\phi_i-\phi_j),
\end{align}
where $M$ is the inertia coefficient, $\gamma$ is the damping constant, $P_i$ is the power of the $i$th oscillator (related to the natural frequency $\omega_i = {P_i}/{\gamma}$), $h$ is the coupling function, and $a_{ij}$ is the adjacency matrix as defined in Eq.~\eqref{eq:APO_phi}. We note that the phase space of~\eqref{eq:KwI_2order} is $2N$-dimensional, i.e., of lower dimension than that of the adaptive network model Eqs.~\eqref{eq:APO_phi}--\eqref{eq:APO_kappa}.

Over the last years, studies on phase oscillator models such as~\eqref{eq:KwI_2order} and oscillators on adaptive networks such as Eq.~\eqref{eq:APO_phi}, revealed a plethora of common dynamical scenarios including solitary states~\cite{JAR18,TAH19,HEL20,BER20c}, frequency clusters~\cite{BEL16a,BER19,BER19a,TUM19}, chimera states~\cite{OLM15a,KAS17,KAS18}, hysteretic behavior and non-smooth synchronization transitions~\cite{OLM14a,ZHA15a,BAR16a,TUM18}. Also hybrid systems with phase dynamics combining inertia with adaptive coupling weights have been investigated, for example, to account for a changing network topology due to line failures~\cite{YAN17a}, to include voltage dynamics~\cite{SCH14m} or to study the emergence of collective excitability and bursting~\cite{CIS20}.

Building on the previously described observations, it has been shown that dynamical power grid models have a deep relations with adaptive networks~\cite{BER21a}. In particular, a mathematical relation between these two models have been found. In the following, we sketch the main idea of this relation, discuss its dynamical implications and describe two routes for its generalization.
%--------------------------------------------------
% Dynamical relation between the phase oscillator models
%--------------------------------------------------
\subsection{Dynamical relation between the phase oscillator models}\label{sec:dynRel}

In order to find a relation, first write Eq.~\eqref{eq:KwI_2order} in the form
\begin{align}
\dot{\phi}_i &= \omega_i + \psi_i, \label{eq:KwI_1order_phi}\\
\dot{\psi}_i & = -\frac{\gamma}{M}\left({\psi}_i - \frac{1}{\gamma}\sum_{j=1}^N a_{ij}h(\phi_i - \phi_j)\right), \label{eq:KwI_1order_psi}
\end{align}
where $\psi_i$ is the deviation of the instantaneous phase velocity from the natural frequency $\omega_i$. We observe that this is a system of $N$ phase oscillators~\eqref{eq:KwI_1order_phi} augmented by the adaptation~\eqref{eq:KwI_1order_psi} of the frequency deviation $\psi_i$. Similar systems with a direct frequency adaptation have been studied in~\cite{ACE98,ACE05,TAY10,SKA13a}.  
Note that the coupling between the phase oscillators is realized in the frequency adaptation which is different from the classical Kuramoto system~\cite{KUR84}. As we know from the theory of adaptively coupled phase oscillators~\cite{KAS17,BER19}, a frequency adaptation can also be achieved indirectly by an adaptation of the coupling matrix. 

\begin{figure}
    \centering
    \includegraphics{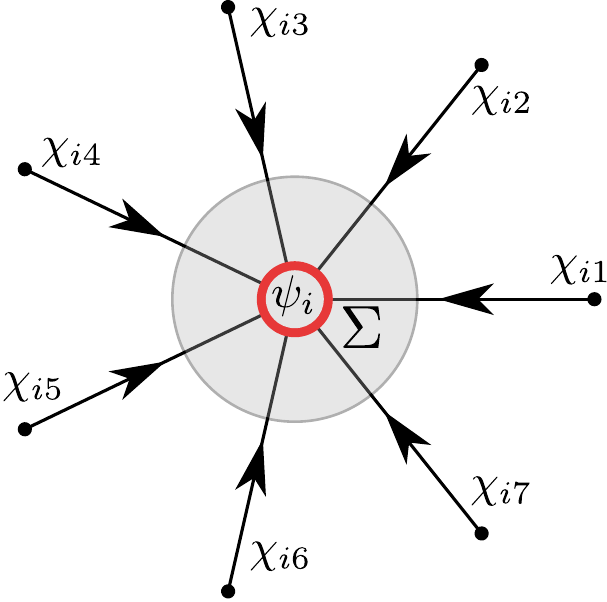}
    \caption{Illustration for the definition and meaning of pseudo coupling weights $\chi_{ij}$ and their relation to the mean phase velocity}
    \label{fig:pseudoCoupling}
\end{figure}
In order to introduce coupling weights into system~\eqref{eq:KwI_1order_phi}--\eqref{eq:KwI_1order_psi}, we express the frequency deviation $\psi_i$ as the sum $\psi_i = \sum_{j=1}^N a_{ij}\chi_{ij}$ of the dynamical power flows $\chi_{ij}$ from the nodes $j$ that are coupled with node $i$. Figure~\ref{fig:pseudoCoupling} visualizes the relation between $\psi_i$ and $\chi_{ij}$. The power flows are governed by the equation $\dot{\chi}_{ij}  = -\epsilon\left({\chi}_{ij} + g(\phi_i - \phi_j)\right)$, where $g(\phi_i - \phi_j)\equiv-h(\phi_i - \phi_j)/\gamma$ are their stationary values~\cite{SCH18i} and $\epsilon =\gamma/M$. It is straightforward to check that $\psi_i$, defined in such a way, satisfies the dynamical equation~\eqref{eq:KwI_1order_psi}. In order to discuss a physical meaning of the coupling weights $\chi_{ij}$, we consider the power flows $F_{ij}$ from node $j$ to node $i$ given by $F_{ij}=-g(\phi_i - \phi_j)$~\cite{SCH18i}. Then each $\chi_{ij}$ is driven by the power flow from $j$ to $i$. In particular, for constant $F_{ij}$, $\chi_{ij}\to F_{ij}$ asymptotically as $t\to \infty$. Therefore, $\chi_{ij}$ acquires the meaning of a dynamic power flow.

As a result, we have shown that the swing equation~\eqref{eq:KwI_1order_phi}--\eqref{eq:KwI_1order_psi} can be  written as the following system of adaptively coupled phase oscillators
\begin{align}
    \dot{\phi}_i &= \omega_i+\sum_{j=1}^N a_{ij}\chi_{ij}, \label{eq:KwI_pseudo_phi}\\
    \dot{\chi}_{ij} & = -\epsilon\left({\chi}_{ij} + g(\phi_i - \phi_j)\right). \label{eq:KwI_pseudo_kappa}
\end{align}
The obtained system corresponds to~\eqref{eq:APO_phi}--\eqref{eq:APO_kappa} with coupling weights $\chi_{ij}$ and coupling function $f(\phi_i-\phi_j) \equiv 1$. The coupling weights form a pseudo coupling matrix $\chi$. Note that the base network topology $a_{ij}$ of the phase oscillator system with inertia Eq.~\eqref{eq:KwI_2order} is unaffected by this transformation.

With the introduction of the pseudo coupling weights $\chi_{ij}$, we embed the $2N$ dimensional system~\eqref{eq:KwI_1order_phi}--\eqref{eq:KwI_1order_psi} into a higher dimensional phase space. In~\cite{BER21a}, it was further shown that the dynamics of the higher dimensional system \eqref{eq:KwI_pseudo_phi}--\eqref{eq:KwI_pseudo_kappa} is completely governed by the system \eqref{eq:KwI_1order_phi}--\eqref{eq:KwI_1order_psi} on a $2N$ dimensional invariant submanifold. All together we yield a dynamical equivalence between \eqref{eq:KwI_1order_phi}--\eqref{eq:KwI_1order_psi} and \eqref{eq:KwI_pseudo_phi}--\eqref{eq:KwI_pseudo_kappa}. 

The obtained result suggests that the power grid model is a specific realization of adaptive neuronal networks. In the following, we briefly outline some implications of the found relation between both dynamical systems.
%--------------------------------------------------
% Consequences of the relation between network adaptivity and inertia
%--------------------------------------------------
\subsection{Consequences of the relation between network adaptivity and inertia}
The new relation between adaptive networks and power grid models provides an abstract mapping between two dynamical systems on the one hand. On the other, and probably even more important, this relation brings different research communities closer together and may trigger knowledge transfer among them. 
Below we describe some examples where such transfer has been very fruitful.

% Mixed frequency cluster states in phase oscillator models with inertia
The first example concerns multi-frequency-cluster states. Here, the relation between a phase oscillator with inertia and adaptively coupled phase oscillators has been successfully utilized. In fact, a novel type of multi-frequency-cluster state could be reported for the phase oscillator model with inertia, for details on multi-frequency-clusters we refer to section~\ref{sec:multicluster}. Various types of multicluster states including the special subclass of solitary states have been extensively described for adaptively coupled phase oscillators~\cite{KAS17,BER19a,BER20c}. For phase oscillator models with inertia, however, only one type of multicluster state, the in-phase multicluster, was known~\cite{JAR18,TUM18,TUM19}.

In~\cite{BER21a} a novel hierarchical mixed-type multicluster on a nonlocally coupled ring of phase oscillators with inertia was presented. This state consists of one large splay cluster with wavenumber $k=2$ and a small in-phase cluster and was found by using the experience from a system of adaptively coupled phase oscillators. The emergence of such a multicluster state breaks the dihedral symmetry of the nonlocally coupled ring network. In another work~\cite{BRE21}, similar symmetry breaking states referred to as Chimera states have been observed also in small globally coupled networks. Another novel observation for multicluster states in networks of phase oscillators with inertia is their hierarchical emergence. As reported in~\cite{KAS17} for adaptive networks, the clusters emerge in a temporal sequence from the largest to the smallest. By visualizing the pseudo coupling strengths at different points in time, the same hierarchical temporal formation of multicluster states could be also observed for phase oscillators with inertia.

% Solitary states in the German ultra-high voltage power grid network
A particular subclass of multicluster states are solitary states that are a common dynamical phenomenon observed in simulations of real power grid networks~\cite{TAH19,HEL20}. Using an approach similar to~\cite{BER19} which was developed for adaptive networks, a multiscale ansatz was used in~\cite{BER21a} to understand dynamical features of solitary states in the German ultra-high voltage power grid. By the transfer of the method, it was possible to approximate the temporal behavior of the solitary node as well as to quantify the temporal variation of the coupling weights between the solitary node and any other mode from another cluster (coherent background cluster). Note that due to the relation described in the previous section, the coupling weights are directly connected to the power flow on each line. Hence, the pseudo coupling approach further allows for a description of the power flow for each line. In~\cite{BER21a} it was shown how high power fluctuations emerge at the solitary nodes and how these fluctuations spread over the power grid.

% What power grids teach us about neural breakdowns
So far methods and ideas from adaptive networks have been transferred to understand dynamical phenomena in power grid models. The next example concerns the other direction. In particular, in~\cite{BER21a} the authors reinterpreted the dynamical cascading of line failures in power grid models as cascades leading to neural breakdown in the context of adaptive networks. For this the power grid setup from~\cite{SCH18i} was employed and the following interpretation was given. Let us regard the power flow on a line in a power grid as the (localized) synaptic input $F_{ij}(t)=\kappa_{ij}(t)f(\phi_i(t)-\phi_j(t))$ from oscillator $j$ to oscillator $i$. Then, we say that a line fails if the corresponding synaptic input exceeds a certain threshold $K\in\mathbb{R}$, i.e., $|F_{ij}(t)|> K$ at some time $t$. Correspondingly, the link is cut off, i.e., $a_{ij}=a_{ji}=0$.
A possible neuronal interpretation of such a temporal cut-off may be related to the presence of short term synaptic plasticity \cite{HEN13a,GER14a}. Indeed, when the signal between neurons or neuronal regions exceeds a certain critical level, then the corresponding connections can be affected by short term activity-dependent depression. As a result, such an activity implies an effective cut-off, at least temporarily. With this interpretation and the insights from the dynamical cascading of line failures in power grid models, dynamical cascading of synaptic failures in an adaptive network could be found and explained.

%--------------------------------------------------
% Generalization to the equation with voltage dynamics and second-order consensus models
%--------------------------------------------------
\subsection{Generalizations of the pseudo coupling approach\label{sec:generalization}}
In Sec.~\ref{sec:dynRel}, we have shown how adaptive phase oscillator networks and networks of coupled phase oscillators with inertia are related. In the following, we show two generalizations to this relation that have been reported in~\cite{BER21a}.

%Swing equation with voltage dynamics
The obtained results in Sec.~\ref{sec:dynRel} suggest that the power grid model is a specific realization of adaptive neuronal networks. Now, we proceed one step further and consider the swing equation with additional inclusion of the voltage dynamics~\cite{SCH14m,TAH19}. By using the technique developed in the Sec.~\ref{sec:dynRel}, these dynamical systems can be written as
\begin{align}
	\dot{\phi}_i &= \omega_i+\sum_{j=1}^N a_{ij}\chi_{ij},\\
	\dot{\chi}_{ij} & = -\frac{1}{M_i}\left(\gamma{\chi}_{ij} - E_i E_jh(\phi_i - \phi_j)\right),\label{eq:extended_KwI_pseudoCoup}\\
	m_i \dot{E}_i & = - E_i + E_{f,i} + \sum_{j=1}^N a_{ij}E_j v(\phi_i - \phi_j),\label{eq:extended_voltagedyn_E}
\end{align}
where the additional dynamical variable $E_i$ is the voltage amplitude. The functions $h$ and $v$ are $2\pi$-periodic, and $m_i$ and $E_{f,i}$ are machine parameters~\cite{SCH14m,TAH19}. Due to the voltage dynamics~\eqref{eq:extended_voltagedyn_E}, the adaptation function $g(\phi)=E_i(t) E_j(t)h(\phi)$ in~\eqref{eq:extended_KwI_pseudoCoup} possesses additional adaptivity. This kind of meta-adaptivity (meta-plasticity) has been shown to be of importance in neuronal networks~\cite{ABR96,ABR08a} as well as for neuromorphic devices~\cite{JOH18}.

%Second-order consensus models
As another example for a generalization, we consider a second-order consensus model. Consensus describes the result of a decision making process of autonomous mobile agents with positions $\bm{x}_i$ and velocities $\bm{v}_i$. The decision making process is described by the consensus protocol that is given as a dynamical system on a complex network structure. Consensus is achieved if the agents synchronize as time tends to infinity. Consensus models have a wide range of applications and are of particular importance in social science and engineering~\cite{REN05c}.

Let us consider the following second-order consensus model~\cite{YU10}
\begin{align}
	\dot{\bm{x}}_i &= \bm{v}_i,\label{eq:consensusX}\\
	\dot{\bm{v}}_i &= \rho\sum_{j=1}^N l_{ij} \bm{v}_j + \sigma \sum_{j=1}^N a_{ij} \bm{h}(\bm{x}_i-\bm{x}_j), \label{eq:consensusV}
\end{align}
where the dynamical variables $\bm{x}_i,\bm{v}_i\in\mathbb{R}^d$, $a_{ij}$ are the entries of the adjacency matrix of the network, $l_{ij}$ the entries of the Laplacian matrix of the network, i.e., $l_{ij}= a_{ij}$ for $i\ne j$ , $l_{ii}=-\sum_{j=1,j\ne i}^N a_{ij}$, and $\rho,\sigma\in\mathbb{R}$ are coupling constants. Let us introduce the vector-valued pseudo coupling matrix $\bm{\chi}_{ij}\in\mathbb{R}^d$ by $\bm{v}_i=\sum_{j=1}^N a_{ij}\bm{\chi}_{ij}$. Then the model~\eqref{eq:consensusX}--\eqref{eq:consensusV} can be written as
\begin{align}
	\dot{\bm{x}}_i &= \sum_{j=1}^N a_{ij}\bm{\chi}_{ij},\label{eq:consensus_pseudoX}\\
	\dot{\bm{\chi}}_{ij} &= -\rho l_{ii}\bm{\chi}_{ij} + \rho\sum_{k=1}^N a_{jk} \bm{\chi}_{jk} + \sigma\bm{h}(\bm{x}_i-\bm{x}_j). \label{eq:consensus_pseudoChi}
\end{align}
By using the same arguments in section~\ref{sec:dynRel}, the dynamical equivalence between both models~\eqref{eq:consensusX}--\eqref{eq:consensusV} and~\eqref{eq:consensus_pseudoX}--\eqref{eq:consensus_pseudoChi} can be proved. With this, we have shown that a second-order consensus model can be written as a dynamical network with a complex adaptive coupling scheme rather than a fixed coupling matrix. Note that the elements of the complex dynamical coupling scheme $\bm{\chi}_{ij}$ are not uniquely defined, but might be chosen according to their physical meaning.

%% file: parts/behaviour.tex
A pressing current question is how we can effectively make decisions together to address major societal problems. In the age of post-truth, social media echo chambers, and widespread willful disinformation, our ability to determine what is real as a society is eroding. This makes it hard to reach a broad consensus even in areas where urgent action is needed. These concerns lead to the question how humans form their opinions and how this process is affected by social network structure, norms, and hierarchy.    

Studying the processes that form these networks, norms, and hierarchies naturally leads to adaptive network models. Networks (or higher order structures) offer a suitable framework to describe human social organization. In these social networks the nodes represent agents, whereas the links represent friendships, acquaintance, or professional relationships. Through their links the agents are exposed to different opinions, bits of information and types of behavior. In time these influences may shape internal properties of the agent, such as political affiliation, ideology, informational state, wealth,  health, fitness or social status. These changes in the agent may in turn induce changes in the network structure, for example cuts links to followers of competing ideologies or seeks to connect to individuals of higher status. This completes the adaptive feedback loop. 

The origin of our social networks, norms and hierarchies can be traced back to biological evolution where fundamental behaviours were shaped in an ecological 
context. It is therefore interesting to study adaptive social networks in animals as well, not only because these system are fascinating in their own right, but also they provide a new angle at human behaviour. In contrast to their human counterparts, social networks between animals are often easier to study. In these networks there are typically no privacy and data protection concerns, social behavior is less biased by the observation, networks can often be identified more easily from visual cues, and when rewards are offered as part of the experiments the costs are much cheaper than they would be for rewarding humans.   

Surveying the publications in this field of adaptive social networks is daunting because of the breadth of academic disciplines involved, which including notable contributions from Economics, Anthropology, Psychology, Sociology, Biology and Philosophy, besides Physics, Mathematics and Computer Science. In the following we resist the temptation to group the literature by discipline and instead organize it according to a set of central questions, each of which is addressed by multiple disciplines. Due to the volume of work that has been done, the overview presented here will be far from comprehensive and instead focuses on some important advances and illustrative examples.   

%%%%%
\subsection{Information flow in adaptive networks}
One of the core questions that is investigated in adaptive social networks is how self-interested choices by agents shape large scale network properties. An interesting precursor of this line of work is a paper by Paczuski et al.~\cite{paczuski2000selforganized}. This paper studies a set of agents who participate in a minority game. In each round, each agent selects one of two possible actions, with the aim of selecting the action that is chosen by the minority of agents in the system. The agents make their decision dependent on the previous actions of their neighbors using a Kauffman-style~\cite{kauffman1969homeostasis} look-up table. These look-up tables are then evolved as the agents try to optimize their success in the game. 

Although the model of Paczuski is played on a static network, it contains a degree of adaptivity as the lookup tables can evolve to ignore imput from certain neighbors. Thus the edges can be cut from the effective network of realized information flow. The authors show that this link-cutting is very common and continues to a point at which the network reaches a critical state. The model thus provides an example of (surprising) topological adaptation to a specific dynamical state. 

Interestingly the model by Paczuski et al.~appeared at the same time as the first adaptive network model of neural criticality \cite{BornholdtRohlf} which also showed self-organization to a state at or close to criticality. The study of the critical brain subsequently led to several other models that showed adaptive self-organized criticality. %(see Sec.~XXX). 
Meanwhile, papers in engineering demonstrated that adaptive networks that locally adapt links to optimize information processing could have have technical applications as well \cite{jiang2013distributed,NIPS}. 
 
\subsection{Adaptive cooperation}
Other parts of the literature are concerned with the impact of  adaptivity on the outcome of the dynamics on the network. A central question in this area is the emergence of cooperative behavior. This is typically studied in models where nodes are agents and links represent social or professional interactions, which are modelled as cooperation games. The agents often have the choice between cooperating and defecting, i.e.~making investments to produce social benefits or selfishly free-riding off the investments of others. The agents can then change their behavior or their local topology to maximize payoffs received in the game.     

An important early work in this direction are papers by Zimmermann et al.~\cite{zimmermann2011cooperation}, and Skyrms and Pemantle \cite{skyrms2000dynamic}. Similar models were then subsequently analyzed in a large number of publications which elucidated various aspects of the dynamics and topological evolution and refined analysis methods, including~\cite{ebel2002evolutionary,goyal2005network,graser2009disconnected,szolnoki2009emergence,vansegbroeck2009reacting,zhang2014phase,pacheco2006coevolution,hojman2006endogenous,biely2007prisoners,poncela2008complex} and many more. These works and several others were reviewed in \cite{perc2010coevolutionary}. Notably many of the predictions from these models can be confirmed in experiments \cite{santos2006cooperation,rand2011dynamic}.

One of the central results in this field is that adaptivity generally benefits 
the emergence of costly cooperation. In most of the games studied links confer a benefit, as links produce a positive payoff in average. Hence cutting links to defectors, which some earlier models (e.g.~\cite{zimmermann2011cooperation,pacheco2006coevolution}) implement constitutes a direct punishment of defection, which benefits the emergence of cooperation in a fairly direct way. However, some papers avoid this direct punishment of defectors by implementing rewiring rules that are agnostic of node strategies, such as rewiring selectively to successful players regardless of their strategy or direct benefit (e.g.~\cite{poncela2008complex,ZschalerTraulsenGross}). Even in these scenarios the adaptivity can benefit cooperation by leading to the emergence of dense clusters of highly successful cooperators (safe havens of cooperation) or by a complex dynamical mechanism that leads to a collective push to full cooperation \cite{ZschalerTraulsenGross}. The latter mechanism seems surprisingly generic and can also be observed in other games on adaptive networks \cite{demirel2011cyclic}. 

Another focus of game-theoretic work in adaptive networks is to explore the formation of hierarchies that develop in the course of adaptive self-organization
\cite{do2010patterns,skyrms2000dynamic,zimmermann2011cooperation,szolnoki2009emergence,mogielski2009mechanism,do2010patterns}. One of the cleanest models in this class was prosed by Holme and Ghoshal \cite{holme2006dynamics} where agents try to achieve a position of high centrality but low degree in the network. In principle the adaptivity of this model could be disputed as it does not feature dynamics on the network, however instead we have the computation of centralities which can be seen as a (fast) dynamical process on the network. In this sense the model of Holme and Ghoshal is similar to other adaptive network models with timescale separation and exhibits a similar degree of topological self-organization, including fragmentation transition and global hierarchies.     

The formation of large scale hierarchies was already observed in some of the earliest games on adaptive networks, however these analysis of these hierarchies was complicated due to the choice of discrete strategies.
In later models continuous time adaptation, weighted link dynamics (e.g.~\cite{mogielski2009mechanism}) and finally continuous time, links, and strategies \cite{do2010patterns} are explored. The paper by Do et al.~\cite{do2010patterns} demonstrates that large scale hierarchies and resource flows can emerge deterministically from an almost perfectly homogeneous initial condition.  

After 2010, the activity in this adaptivity in games sharply declined as many of the important mechanisms have been understood at this point. However, adaptive interplay and the phenomena it causes live on in many current more complex models as adaptivity arises indirectly or as one feature in a more complex framework, e.g.~\cite{deng2012network,yuan2018interpretable,fahimipour2021sharp}. One consequence of the shift of interest that has occurred in 2010 is that many of the adaptive network models in this area have not been analyzed with the powerful methods that were introduced subsequently. Hence reviving this area in the light of these developments could be very promising and profitable. 

\subsection{Opinion formation and Fragementation}
Potentially the most widely investigated phenomenon in social adaptive networks is social fragmentation.  Even before the post-truth crisis researchers have explored 
the possibility that adaptivity in networks of agents may lead to echo chambers and a fragmentation of society. 

In the context of adaptive networks researchers already noted in 2012 ``It thus seems likely that situations develop where a given
subset of the society (and the media by which it is represented)
pay attention only to information sources with the same belief
system, thus reinforcing and perpetuating myths that are never
confronted with opposing views. In this light, one may ask
whether we are heading for a society that is fractionated into
groups adhering to internally consistent but mutually exclusive
belief systems.''\cite{Zschaleretal}. 

Research in the social science and humanities has a longer tradition of analysing opinion formation and the potential of social fragmentation, dating back to the 1950s \cite{french1956formal,harary1959criterion}. A central idea that emerged from with work is \emph{bounded confidence} \cite{deffuant2000mixing,sznajdweron2001opinion,HegselmannKrause}, the idea that people are willing to take different opinions into account as long as these opinions are relatively closely aligned with their own, but completely disregard opinions that are too different. 

Models of bounded confidence naturally lead to adaptive networks, where nodes are agents and links denote social interactions. In time the agents update their opinions and links, taking the opinions of interacting partners into account but also shifting links toward agents with more similar opinion \cite{deffuant2000mixing,sznajdweron2001opinion,zanette2006opinion,gil2006coevolution,HolmeNewman,centola2007homophily,biely2008socioeconomical,sobkowicz2009studies,kozma2008consensus,NardiniKozmaBarrat,herrera2011general,durrett2012graph,KimuraHayakawa,wiedermann2015macroscopic,min2017fragmentation}. 

The most stylized model of opinion formation in adaptive networks is the adaptive voter model (aVM) \cite{vazquez2008generic}. In this model agent's opinion are binary variables and the links are unweighted. Nodes adopt their neighbors opinion at a certain rate, while also rewiring links from nodes with the opposed opinion to nodes that hold their own opinion. One of these rate parameters can be eliminated by means of time-scale renormalization, such that the only remaining parameters are the size of the network, the mean degree of the nodes, and the relative rate of rewiring compared to the total rate of rewiring and opinion copying events. This last parameter is simply called rewiring rate in the context of the aVM. 

The aVM can be considered a fair model, in the sense that it does not give any reproductive advantage to either of the two opinions, even if one of the opinions is in the numerical majority. The model therefore undergoes a random walk in the opinion space. If the rewiring rate is low this random walk in the number of supporters of either opinion can be observed for an extended period until the model eventually hits one of the absorbing states where one of the opinions vanishes. At low rewiring rates, one thus observes an extended dynamical phase followed by eventual consensus. 

If the rewiring rate is increased, a phase transition occurs, in which the meta-stable dynamical phase vanishes. After the transition links are rewired quickly enough such that all links that connect different opinions are quickly eliminated from the system \cite{vazquez2008generic}. The result is a fragmented state in which the number of agents believing in the two opinions is still very close to their initial values, but all links connecting agents of different opinions have been cut, which freezes the system in a state with partisan divide. 

Like the adaptive aSIS model (see Sec.~\ref{sec:epidemics}) the aVM can lay claim to being the simplest adaptive network model. Both of these models use two node states and a conserved number of unweighted, bidirectional links. In both cases the link dynamics is implemented as homophilic rewiring and the node dynamics are a simple contagion processes. 

Despite the many similarities, the two models respond very differently to mathematical analysis with moment expansions~\cite{DemirelVazquezBoehmeGross}. The aSIS model is extremely well-behaved and almost every moment expansion scheme that has been proposed in the literature yields very accurate results. By contrast in case of the aVM, all common 
moment expansions yield poor estimates of the transition point. 

For example, it was shown that an unbounded active-neighborhood approximation can be solved analytically~\cite{SilkDemirelHomerGross}. In this approximation the dynamics is described by an infinite system of differential equations that capture the opinion-colored joint degree distribution of the system. This infinite ordinary differential equation system can then be transformed into a two-dimensional partial differential equation system. Using the method of characteristics, these partial differential equations are then transformed into four ODEs which can be integrated analytically. However, the accuracy of this very extensive approximation is only marginally better than a simple homogeneous expansion and can overestimates the transition point by approximately~20\%.  

The difficulties with the aVM arise because moment expansions generically assume that the system is well-mixed beyond a certain correlation length~\cite{DemirelVazquezBoehmeGross}. However, when the fragmentation transition occurs in the aVM, the system is at the point where the network breaks into two internally homogeneous clusters, and hence the correlation length of node states is comparable to the network diameter. 

To estimate the aVM's fragmentation point accurately, one must avoid the mixing assumption, which normally enters in the moment closure approximation. By deriving a bespoke closure approximation that replaces the assumption of mixing with an assumption of opinion separation, one arrives at an approximation scheme that allows precise estimates of the fragmentation point using a comparatively simple calculation~\cite{BoehmeGross}.   

Since its inception the aVM has become an important toy model, which has inspired a large number of subsequent developments e.g.~\cite{BoehmeGross, rogers2013consensus,SilkDemirelHomerGross,carro2016noisy,klamser2017zealotry,raducha2020emergence,papanikolaou2022consensus}. These include multi-opinion versions as well as models on directed networks \cite{Zschaleretal}, multi-layer networks \cite{diakonova2014absorbing}, simplicial complexes \cite{HOR20} and hypergraphs \cite{papanikolaou2022consensus}. These extensions show a variety of new phenomena. For example in hypergraphs small initial imbalances in opinions are amplified leading to a faster consensus~\cite{papanikolaou2022consensus,HOR20}. By contrast in 'twitter-like' directed networks arbitrarily small rewiring rates can be sufficient to cause fragmentation \cite{Zschaleretal}

The aVM and its variants have been very successful in physics, revealing a range of new phenomena and contributing to a deeper understanding of the adaptive interplay between state and topology. An important challenge for the future will be to translate these insights back into the application domains that inspired these models. Recent developments including the post truth-crisis, Covid-19 response, and also global change highlight the need for a deeper understanding of real world social dynamics and particularly opinion formation processes \cite{mullerhansen2017towards}. 

For humans it is presently still difficult to relate aVM results directly to the real world. Data protection concerns makes working with real-life social network data difficult. Moreover the need for an informed consent makes it nearly impossible to conduct unbiased laboratory experiments with human participants. An interesting alternative is to explore animal models of decision making. Behavioral experiments with animals can be informed without consent and allow the researcher to study decision-making in animals in a highly controlled way. 

It was pointed out in \cite{huepe2011adaptivenetwork} that common swarming experiments with animals can also be interpreted as opinion formation experiments, as the individuals in the swarm make the decision where to go collectively. Using a variant of the aVM and an analysis based on moment expansions, the paper showed that experimental results can be understood with adaptive network models. A subsequent  paper \cite{couzin2011uninformed} then predicted a new phenomenon in the collective behavior that was verified in experiments with fish. These and other results  \cite{chen2016adaptive} suggest adaptive networks as a powerful tool to explore, understand and to some extent predict opinion formation in animals.   

%% file: parts/epidemics.tex
Infectious diseases have plagued humanity since its earliest days, from the killers of the ancient world, via medieval plagues, to the COVID-19 pandemic. New diseases typically arrive in the human population, when we change our behavior or environment and allowing new pathogens to spread \cite{karlen1996man}: Intravenous drug use contributed to HIV, the use of air conditioning caused legionellosis, climate change gave us ESME a resurgence of malaria and many others. 

Today we are changing our environment at an unprecedented pace. The human population is larger than ever. Our global transport systems, both for people and livestock, enable new and emerging diseases to rapidly spread around the world. The black death of the 14th century travelled across Europe as a wave, proceeding at the pace of an ox cart, and taking years to spread across the continent \cite{belik2011natural}. By contrast SARS traveled at the speed of a jetliner, spreading around the globe within days \cite{worldhealthorganization2006sars}. As a result new and reemerging pathogens arrive in the human population at an unprecedented rate. 

Over the past years, epidemic models have become rapidly more important as a planning tool, with model predictions providing a quantitative basis for policy decisions (and a way to assess the damage done when this quantitative information is ignored) \cite{pastorsatorras2015epidemic,wang2016statistical}. The vast majority of epidemic models follow a \emph{compartmental approach}~\cite{KissMillerSimon}, where different variables are used to track the number of individuals who have a certain epidemic status such as being susceptible to the disease or being currently infected and infectious \cite{kermack1997contribution}. In the simplest case a mean field approximation is used to derive differential equations for the size of the infected populations in the thermodynamic limit. 

Epidemic models are typically named to suggest the sequence of epidemic statuses that they consider \cite{anderson1979population}. In particular, the SIS model describes a situation in which susceptible individuals (S) can become infected (I) and eventually recover, becoming immediately susceptible (S) again. This immediate return to the susceptible status, without a period of immunity is typical for macro-parasite infections (e.g.~lice). The dynamics can then be described by 
\begin{eqnarray}
[\dot{I}] &=& -r[I] + p[SI], \label{eqaSISfirst}\\{}
[\dot{S}] &=& p[SI] - r[I],
\end{eqnarray}
where $[S]$ is the proportion of individuals that are susceptible, $[I]$ is the proportion of infected (and infectious) individuals, $[SI]$ denotes the effective rate of contact between infectious and susceptible individuals, and $r$ and $p$ are rate constants for recovery and contagion. 

To close the model, a mean field approximation is used to write the encounter density as 
\begin{equation}
[SI]=[S][I],    
\end{equation}
where the constant of proportionality has been set to 1, as any such constant that may exist in the real world system can be absorbed into the parameter $p$.
Furthermore, the proportions of infected and susceptible individuals obey
\begin{equation}
[I]+[S] = 1,
\end{equation}
which allows us to eliminate the variable $[S]$, leading to
\begin{eqnarray}
[\dot{I}] &=& -r[I] + p[I](1-[I]), \\{}
[S] &=& 1 - [I].
\end{eqnarray}
These equations are exact for a well-mixed population in the thermodynamic limit, but offer also a good approximation to epidemics on low-diameter (i.e.~small world) networks, where nodes represent individuals and links represent contacts; for a more detailed look at the mathematical challenges for moment methods applied to adaptive networks, we refer to Section \ref{sec:mathmeth}.

Many real world diseases have a more complex epidemiology. Hence, models need to  consider additional epidemic statuses such as recovered or removed individuals (R), who cannot be reinfected, asymptomatic infectious individuals (A), and individuals who have been exposed and become infected but have not yet reached the infectious phase (E).  

\subsection{Adaptive SIS model}
Epidemic diseases that are threatening or stigmatised trigger a variety of social responses ranging from the individual avoidance of risky activities, via physical distancing, increased hygiene, and vaccination to self-imposed or mandatory quarantine and large-scale preemptive lock downs \cite{funk2009spread,funk2010modelling,wang2016statistical}. 

To understand the phenomena that are triggered by the adaptive response of the network to the disease, it is useful to explore conceptual models. For illustration we consider the so-called adaptive SIS (aSIS) model, as it is the first and perhaps simplest of a class of similar models \cite{ShawSchwartz,zanette2008infection,risaugusman2009contact,vansegbroeck2010adaptive,wang2010epidemic,zhong2011time,ShawSchwartz1,jolad2012epidemic,zhou2012epidemic,valdez2012intermittent,shai2013coupled,youssef2013mitigation,tunc2014effects,rattana2014impact,zhang2014suppression,zhou2014epidemic,yang2015large,szabosolticzky2015oscillating,britton2016network,ogura2016epidemic,demirel2017dynamics,ball2019stochastic}. 

The model considers SIS on a network, where susceptible individuals try to avoid becoming infected, by breaking off links with infected \cite{GRO06b}. For every link that is broken in this way the susceptible individual establishes a new link to another susceptible individual such that the number of links in the system is conserved. 

The adaptive rewiring of links can have a significant impact on the dynamics of the epidemic, but is not captured in the mean field model, since rewiring itself does not change the number of infected nodes. To incorporate the effect of rewiring, we have to go back to Eq.~(\ref{eqaSISfirst}) and instead of closing at the mean field level write additional equations for the change in the number of links of given type, leading to
\begin{eqnarray}
 {[\dot{I}]} &=& -r[I] + p[SI], \label{eq:aSIS-1} \\ 
 {[\dot{SS}]} &=& -p[SSI] + r [SI],  \label{eq:aSIS-2} \\ 
 {[\dot{II}]} &=& -2r[I]+p[SI]+p[ISI], \label{eq:aSIS-3}
\end{eqnarray}
where we now interpret $[SI]$, $[SS]$, and $[II]$ as the number of links of the respective type per node, while $[ISI]$ is the number of ISI-chains, divided by the number of nodes. This somewhat peculiar normalization generally pays off as it often leads to more compact equations. The exact reasoning by which these equations are derived is explained in detail in \cite{GRO06b} and the basic procedure is also covered in Sec.~\ref{sec:mathmeth}.

\subsection{State-topology interplay: complex dynamics and explosive transitions}

The simple equations of the aSIS model \eqref{eq:aSIS-1}-\eqref{eq:aSIS-3} are interesting because they capture a crucial feature of adaptive networks: the interplay of node-state with topological degrees of freedom \cite{GRO08a}. Consider that discrete state dynamical systems on networks have typically low-dimensional dynamics. For example the dynamics of the non-adaptive SIS model can already be captured faithfully by a single dynamical variable. By contrast in adaptive networks the dynamic state of nodes impacts the topology. The network topology (with its very many degrees of freedom) thus becomes a memory that can be written and read by the node dynamics. How much of this memory is accessed depends on the specific dynamical process. In the aSIS model this interaction with the topology is limited and hence the dynamics can still be faithfully captured with three dynamical variables.     

Coupling node dynamics to topological degrees of freedom can lead to new phenomena as the topological memory allows higher-dimensional dynamics to occur.
For example in the aSIS self-sustained oscillations are possible at high rewiring rates. By contrast such oscillations cannot occur in the effectively one-dimensional dynamics of the non-adaptive SIS model. In the original aSIS model the basin of attraction of the oscillatory dynamics is small and hence very large simulations are necessary to observe such oscillations. However, variants of the model exhibit stable large scale-oscillations over a broad parameter range, e.g. \cite{gross2008robust,zhou2012epidemic,szabosolticzky2015oscillating}.  

The aSIS model was intentionally set up to model prudent~\cite{scarpino2016effect} behavior, i.e.~the susceptible individuals protect themselves by rewiring links in a way that always reduces the links that are available for disease transmission. However even this prudent behaviour has a detrimental side effect: It leads to the accumulation of links between susceptible individuals, which can fuel the future growth of the epidemic if these susceptible nodes should become infected.  

Rewiring impedes epidemic spreading effectively if the proportion of infected in the population is low. In this case the rewired links are diluted in the large pool of susceptibles. By contrast rewiring is less effective in states where a large proportion of the population is infected. In this case links quickly accumulate in a smaller set of susceptibles, which in turn can be quickly reinfected.   

\begin{figure}[htb]
    \centering
    \includegraphics[width=0.6\textwidth]{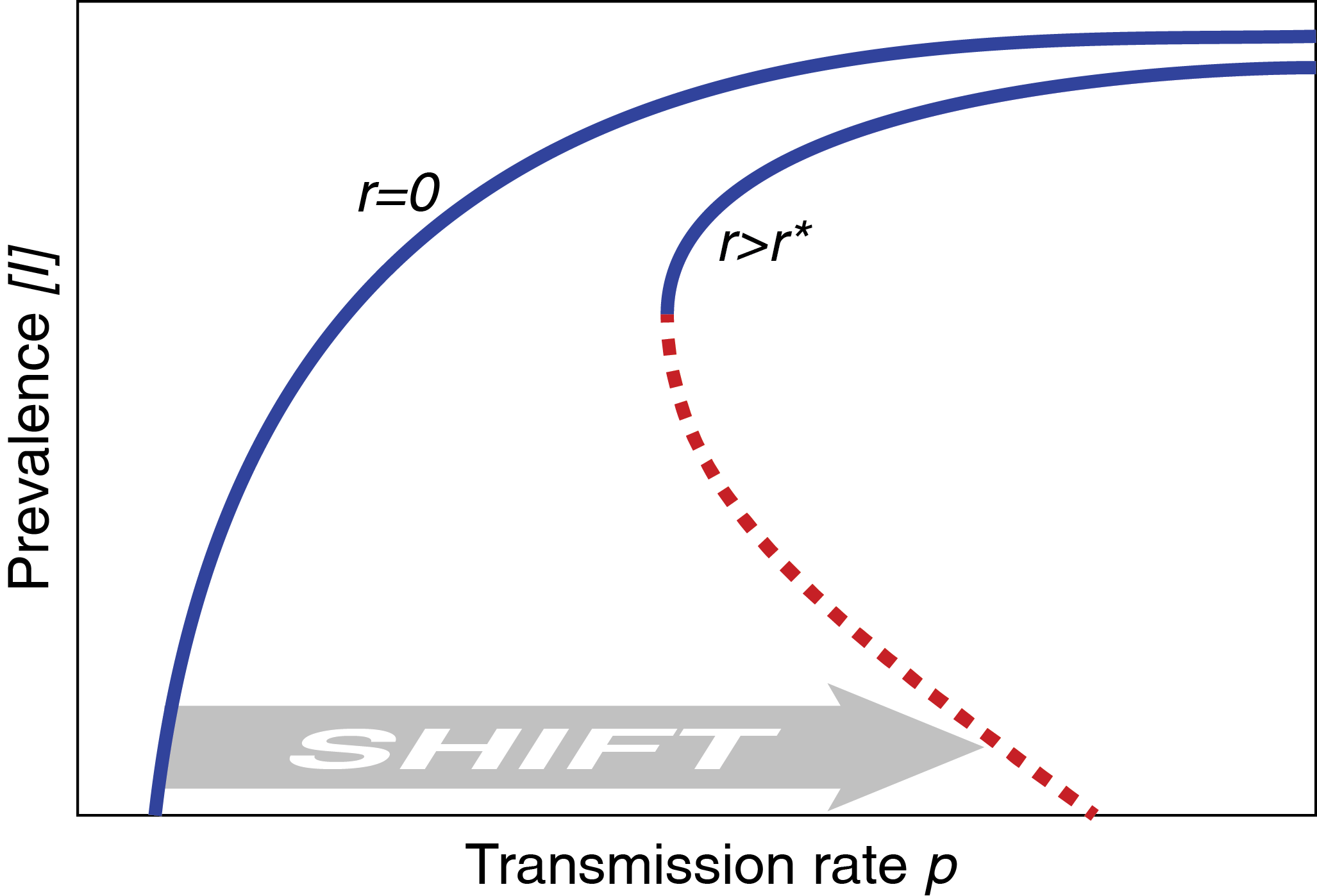}
    \caption{Explosive transitions in adaptive epidemic models. Shown is the sketch of a stylized bifurcation with stable (blue) and unstable (red, dashed) branches of steady states. Without adaptive rewiring ($r=0$) the onset of the epidemic occurs in a continuous transition as the infection rate is increased above the epidemic threshold. If the agents try to avoid contact with infected, this epidemic threshold can be moved to much higher infection rates ('shift'). However, with sufficiently strong rewiring ($r>r^*$) an established epidemic can persist even below the epidemic threshold, leading to hysteresis and explosive transitions.}
    \label{anshift}
\end{figure}

The higher effectiveness of rewiring at low numbers of infected leads to a specific phenomenon: Rewiring pushes the epidemic threshold that marks the critical infection rate at which the epidemic can invade the population. 
As a result this threshold is moved to much higher values of $p$ (Fig.~\ref{anshift}) \cite{GRO06b,funk2009spread}. However, due to lesser effect in the case when a large infected population exists, the persistence threshold for an established epidemic is pushed back significantly less. This leads to the formation of a region where an established epidemic can persist but a newly arriving epidemic cannot invade a healthy population. This region is then bounded by discontinuous phase transitions. The aSIS model thus provided one of the first examples for explosive transitions in network dynamics \cite{BOC16,SOU19}. 

The adaptive interplay also allows the topology to react to the dynamics in other ways. It is well known that preferential attachment (a topological process) creates scale-free structures that have vanishing epidemic threshold, because the second moment of the degree distribution becomes infinite. By contrast networks with finite second moment typically have a finite epidemic threshold. However if network growth by preferential attachment is coupled with SIR dynamics that removes nodes, the system approaches a state of finite second moment, while the epidemic threshold still vanishes \cite{demirel2017dynamics}. This is possible because the adaptive interplay pushed the second moment to a value that is just high enough for the disease to survive.   

\subsection{From concept and benchmark to real world epidemics}

The aSIS model \eqref{eq:aSIS-1}-\eqref{eq:aSIS-3} is a among a group of conceptual models which studied similar systems. For example \cite{ShawSchwartz} considers a susceptible-infected-recovered (SIR) model in which links to infected can be broken. The main purpose of these early models was to draw attention to the new phenomena that can be observed in adaptive networks, including higher-dimensional dynamics and bifurcations, new transition pathways and strongly shifted transition points. 

Moreover, the aSIS model was subsequently widely used as a well-behaved benchmark system that was later reanalysed and used for the development of refined methods, e.g. \cite{gross2008robust,marceau2010adaptive,vansegbroeck2010adaptive,graser2011separatrices,kuehn2012mathematical,yang2012efficient,juher2012outbreak,rogers2012stochastic,wieland2012structure,zhou2013linkbased,guo2013epidemic,trajanovski2015from,KuehnZschalerGross,yang2016network,kattis2016modeling,HorstmeyerKuehnThurner,sahneh2019contact,zhang2019complex,KUE21}. In particular it served as an example in the development of the active neighborhood approximation \cite{marceau2010adaptive} and in the construction of early warnings signals for saddle-escape transitions \cite{KuehnZschalerGross}. Kattis et al.~used it as an example system to develop a data-driven equation free modelling approach to network dynamics \cite{kattis2016modeling} and Kuehn and Bick showed that the transition between explosive and non-explosive behavior observed in this model is a generic pathway \cite{KUE21}. 

Besides these conceptual uses, it quickly became apparent that already such fairly simple models can be related to real world data and yield useful insights. An interesting work in this context is the paper by Scarpino et al.~\cite{scarpino2016effect}. The model is a variant of aSIS with a rewiring rule that captures the effect of relational exchange, the effect of the replacement of infected workers in key roles with susceptible substitutes. The resulting dynamics feature an aSIS-like hysteresis loop but are overall more favorable to the disease. In particular the dynamics explain an initial acceleration of cases at the onset of the disease that is observed in some epidemics, but cannot be explained by common non-adaptive models.  

Another interesting extension is the study of adaptive SIS and SIR models in which the network topology is created in more complex algorithms that more closely mimics real world social network formation processes \cite{mancastroppa2020active}, which adds further realism and allows a detailed exploration of the effects of quarantine.

\subsection{Epidemics and Information}
The Covid-19 pandemic has shown forcefully that human behaviour is an essential factor in the dynamics of infectious diseases. Particularly adaptive changes of the topology are often the result of active choices, such as avoiding contacts, increased hygiene, use of protective equipment or adoptions of vaccinations. The spreading of epidemics is thus coupled to the spreading of information (including mis- and dis-information) which happens on a related but different social network \cite{funk2009spread,funk2010modelling,meloni2011modeling,wang2015coupled,wang2016suppressing}.

From a network perspective the propagation of epidemics and information through a population are very similar processes. Both processes can be described as copying processes, where the payload is duplicated when it spreads. Moreover in both cases the content of the message, or properties of the pathogen can evolve between transmissions. 

When one considers epidemic spreading together with the spreading of information that triggers an adaptive response, a race between the epidemic and the information can ensue. Kiss et al.~\cite{kiss2010impact} derive the conditions under which the information can win this race, leading to the eradication of the epidemic. Further analysis, often with a stronger emphasis on the multi-layer nature of this race, was presented in subsequent papers  \cite{granell2013dynamical, wang2014asymmetrically,andrews2016impacts,bottcher2017critical,pires2018sudden,zhan2018coupling,evans2020infected}, recently reviewed in~\cite{wang2019coevolution}. In most of these models adaptivity exists only because becoming informed about the epidemic fully or partially removes the node from the disease propagation layer. However, already this limited adaptivity is sufficient to cause typical phenomena associated with adaptive networks auch as explosive transitions and complex dynamics.   

\subsection{Epidemiological modeling outlook}
A second important lesson from Covid-19 is that epidemic models play an essential role in forecasting and responding to pandemics \cite{mohamadou2020review}. As none of the factors that contributed to the worldwide spread of Covid-19 can be expected to change in the future (dense-connectivity, super-spreaders, environmental change, etc.) future pandemics of similar size and impact will appear highly likely.    

Covid-19 has also shown that the adaptive response of the population is an essential factor that has a strong impact on the time-course of the pandemic. 
However, in future predictive models of epidemics adaptivity will likely only be one aspect besides others, such as realistic social mixing patterns, demographic structure, geographic migration, and temporal forcing of mixing \cite{silk2021role}. Adaptivity and the explosive transitions and complex dynamics that it causes will thus become a small, but important puzzle piece in much larger modelling efforts. 

We hope that in addition to their use in public health, there will also be continued interest in adaptive epidemics in the complex systems community. The SIS model is one of the nicest and most well-behaved model of a continuous phase transition. Likewise its adaptive extensions such as aSIS provide simple and tractable models of state-topology interplay in network dynamics. This interplay remains relatively purely understood: We still do not have a comprehensive theory regarding the topological features a dynamical process senses from its network substrate, neither is there a good understanding what a given topology tells us about the adaptive changes that have shaped it. Gaining such an understanding could impact a wide variety of areas from neuroscience to computing. The universality of the simple propagation and rewiring mechanisms captured by aSIS and its cousins highlights these models as useful tools that can aid in making this progress.  

%% file: parts/transport.tex
Transport processes are crucial for the functioning of natural and technological systems. In most cases, such processes can be described by transport networks. One of the most advanced examples of a transport network is the mammalian vascular system. An effective adaptation mechanism in such networks is an important feature allowing to adapt to different operating conditions. 
For example, the mammalian vasculature is highly adaptive in that the diameter of the vessels dynamically adjusts to changes in flow properties, such as pressure or shear stress, through a variety of vascular response mechanisms. 

Several publications have studied such adaptation mechanisms in flow networks, and here we shortly mention the modeling approaches from \cite{Martens2019,MAR17b}. 
The phenomenological models used in \cite{Martens2019,MAR17b} can predict the emergence and partition of the flow network into tree-like and cyclic structures, thus, rendering the supply in the network more robust. 

The state of node $i$ is described by the pressure $p_i$, while the flow between the nodes $i$ and $j$ is given by $f_{ij} = C_{ij}(p_i - p_j)$, where the nodal accumulation rate is neglected (e.g. inelastic vessels, see \cite{Reichold2009}). Here $C_{ij}=C_{ji}>0$ is the conductance along the network edge $(i,j)$, while $C_{ij}=0$ if no such edge exists. External nodal flows are given as temporal functions $h_i(t)$ that can model sources ($h_i>0$) or sinks (for $h_i<0$). 
The mass balance then becomes 
\begin{equation} 
	\label{eq:mass-balance}
	h_i(t) = \sum_j C_{ij}(p_i(t) - p_j(t)).
\end{equation}
The main variables of the system are: 
\begin{itemize}
	\item 
	The state of each node $i$, $i=1,\dots,N$ is described by the pressure $p_i(t)$. We denote $\mathbf{p}=\left(p_1,\dots,p_n\right)^T$ the vector of all pressures. 
	\item 
	The weights are described by the conductances $C_{ij}$. In the case of an adaptive network, the conductances $C_{ij}$ can adapt their values, depending on the network states $\mathbf{p}(t)$, and become time-dependent $C_{ij}(t)$.
\end{itemize} 
The role of the node equations in this case is played by the linear algebraic system \eqref{eq:mass-balance}, which can also be written in the vector form 
$$
\mathbf{h} = \mathbf{K} \mathbf{p},
$$
where 
$K_{ij} = (\delta_{ij} \sum_j C_{ij}) - C_{ij}$ \cite{MAR17b}.
Taking into account a finite accumulation rate for the nodes would lead to a system of ODEs for the node dynamics instead of the algebraic system. However, we limit the description here to the algebraic case as in \cite{Martens2019,MAR17b}.  

The adaptation rule for the weights $C_{ij}$ is given as 
\begin{equation}
	\label{eq:transport-adaptation}
	\frac{d}{dt} C_{ij} = C_{ij}\left( \alpha_1  \left| p_j - p_i \right|^\gamma - \alpha_2 \right),
\end{equation}
where $\alpha_{1,2}>0$, and the first term in the right-hand side induces growth proportional to the power dissipated along the edge. This mitigates rising pressure differences by increasing the conductance along the edge. The exponent $\gamma$ was chosen to be $2$ in \cite{MAR17b}, and the effect of varying positive $\gamma$  was studied in \cite{Martens2019}.  The last term in Eq.~\eqref{eq:transport-adaptation} prevents an unlimited growth of the conductance. This adaptation rule acts towards minimizing the power consumption. 

Summarizing the modeling approach in \cite{Martens2019,MAR17b}, the complete adaptive system is given by Eqs.~\eqref{eq:mass-balance}--\eqref{eq:transport-adaptation} with the additional condition for the balance of the nodal flows $\sum_i h_i(t)=0$. It is demonstrated that such a model explains the intricate balance between cyclic and tree-like structures that emerge self-consistently and render the supply more robust in flow networks with (in-) or outflow fluctuations. 

\begin{figure}
    \centering
    \includegraphics{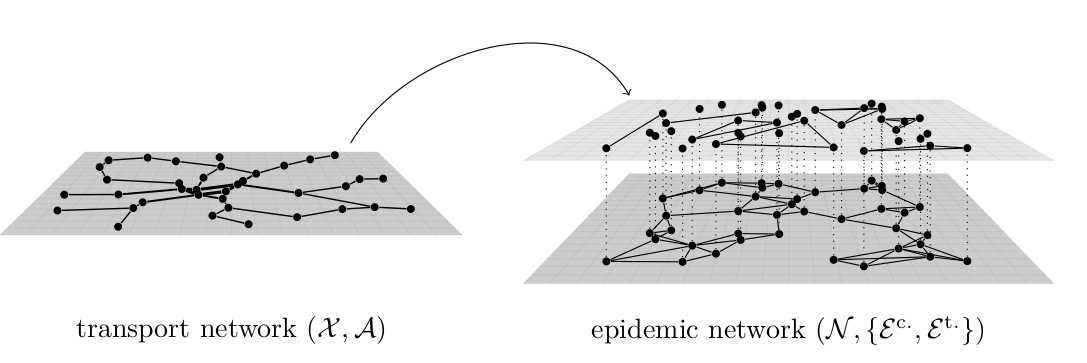}
    \caption{A model from \cite{KUE22a} combining epidemic dynamics on a network with a transport process. Reprinted from  \cite{KUE22a}.}
    \label{fig:transport}
\end{figure}

In more realistic situations for epidemic modeling, transportation networks must be coupled to the epidemic network leading to a challenging dynamical processes. An example for such a modeling approach is given in \cite{KUE22a}, where a multiplex network model emerge. The  main ingredients of the model from \cite{KUE22a} are as follows, see Fig.~\ref{fig:transport}: \\
(i) The transport process takes place on a given static network as a random walk of individuals. In particular, these individuals can occupy the same nodes.\\
(ii) The epidemic network consists of two layers: a static base "community" layer, and a dynamic layer, which includes links between the individuals only in the case when they occupy the same location in the transport network. As a result, the connectivity of the second epidemic layer is time-dependent and is governed by the transport dynamics.   
According to our classification, the model from \cite{KUE22a} is an example of a temporally-evolving network, since the temporal evolution of the second layer of the epidemic network is prescribed by the transport network dynamics, and there is no feedback from the epidemic state to the network topology. Such a situation would be, for example, when the transport process becomes adaptive, e.g. by restricting the mobility of individuals in a certain state of health or by avoiding sites with a high local prevalence of
the epidemic.

%% file: parts/climate.tex
The Earth is a complex system of interacting elements at different spatial and temporal scales. 
Obviously, there are a multitude of different adaptive mechanisms in this system that need to be investigated. 
At a present time, the authors are not aware of any modeling approaches that would emphasize adaptive (not merely time-varying) network structure there. In this section, we propose a possible approach that could be pursued in the future.  

%\subsection{Adaptivity in networks of tipping elements}

The approach relies on the so-called tipping elements, important large-scale parts of the Earth system that can have a significant impact on climate \cite{Lenton2008}.
Tipping elements can undergo large and even qualitative changes in response to natural or anthropogenic perturbations. 
In particular, such dangerous large-scale changes can represent "tipping points", such as the potential collapse of the Atlantic Meridional Overturning Circulation (AMOC), the dieback of the Amazon rainforest, or the disintegration of the Greenland ice sheet.
Tipping elements include, for example, Arctic sea-ice, Greenland ice sheet, West Antarctic ice sheet, AMOC, Indian Summer Monsoon, and others. 

Tipping elements interact with each other and form a complex dynamic network. In \cite{Wunderling2021}, for example, the authors examined the effects of interactions between the Greenland and West Antarctic ice sheets, the AMOC, and the Amazon Rainforest using a conceptual network approach. They showed that there is a high probability that the ice sheets will be the initiators of tipping cascades, while the AMOC acts as a mediator transmitting cascades. 

\begin{figure}
    \centering
    \includegraphics[width=0.7\textwidth]{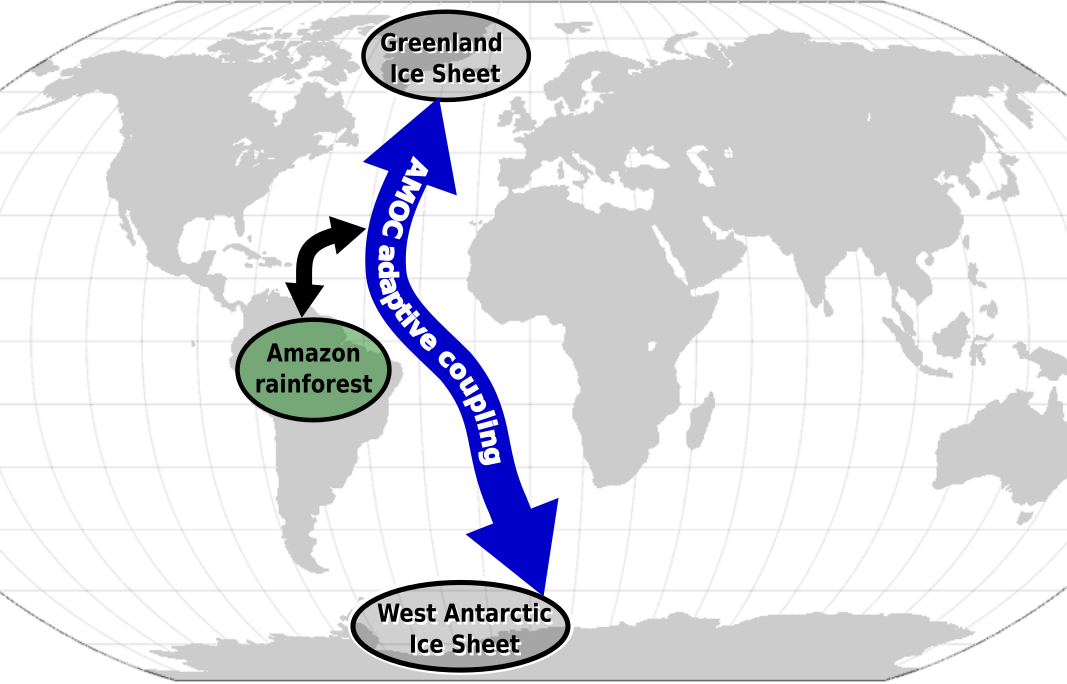}
    \caption{Adaptive coupling between the tipping elements Greenland Ice Sheet and the West Antarctic Ice Sheet realized through the Atlantic Meridional Overturning Circulation (AMOC). The dynamic state (weakening or strengthening) of the AMOC is influenced by another tipping element: the Amazon rainforest. The AMOC coupling is dynamic (or active) because of its interaction with the Amzon rainforest and also because of its complex dynamic nature.}
    \label{fig:climate}
\end{figure}

The results from \cite{Wunderling2021} imply that AMOC can be considered as an adaptive coupling between the ice sheets. Figure \ref{fig:climate} illustrates the interaction between the Greenland and the West Antarctic ice sheets, which is mediated by the AMOC. In addition, the Amazon rainforest influences the AMOC, leading to an additional active degree of freedom  for the adaptive coupling between the ice sheets. 

%% file: parts/delay.tex
Time delays occur in applications where signal propagation between the interacting elements cannot be neglected. This is often the case for neuronal networks \cite{Deco2009a,Wu2001} or semiconductor lasers \cite{Soriano2013,Giacomelli2020}. One of the possible forms of the dynamical networks with time delayed interactions is as follows 
\begin{equation}
     \label{eq:delay-network}
     \dot{\bm{x}}_i(t)=f_i({\bm{x}}_i(t)) + \sum_{j=1}^N \kappa_{ij} g(\bm{x}_i(t),\bm{x}_j(t-\tau_{ij})),
\end{equation}
where $\tau_{ij}$ is the time-delay for the interaction between the $j$-th and $i$-th system. Other variants are possible, for example, the time delays may appear for the self-feedback terms $g(\bm{x}_i(t-\tau),\bm{x}_j(t-\tau))$ or even in the individual dynamics. In the case of the networks with time delays, the properties of the edges of the corresponding graphs are described not only by the coupling weights $\kappa_{ij}$ but also by the time-delays $\tau_{ij}$.
 
\begin{figure}
    \centering
    \includegraphics[width=0.8\textwidth]{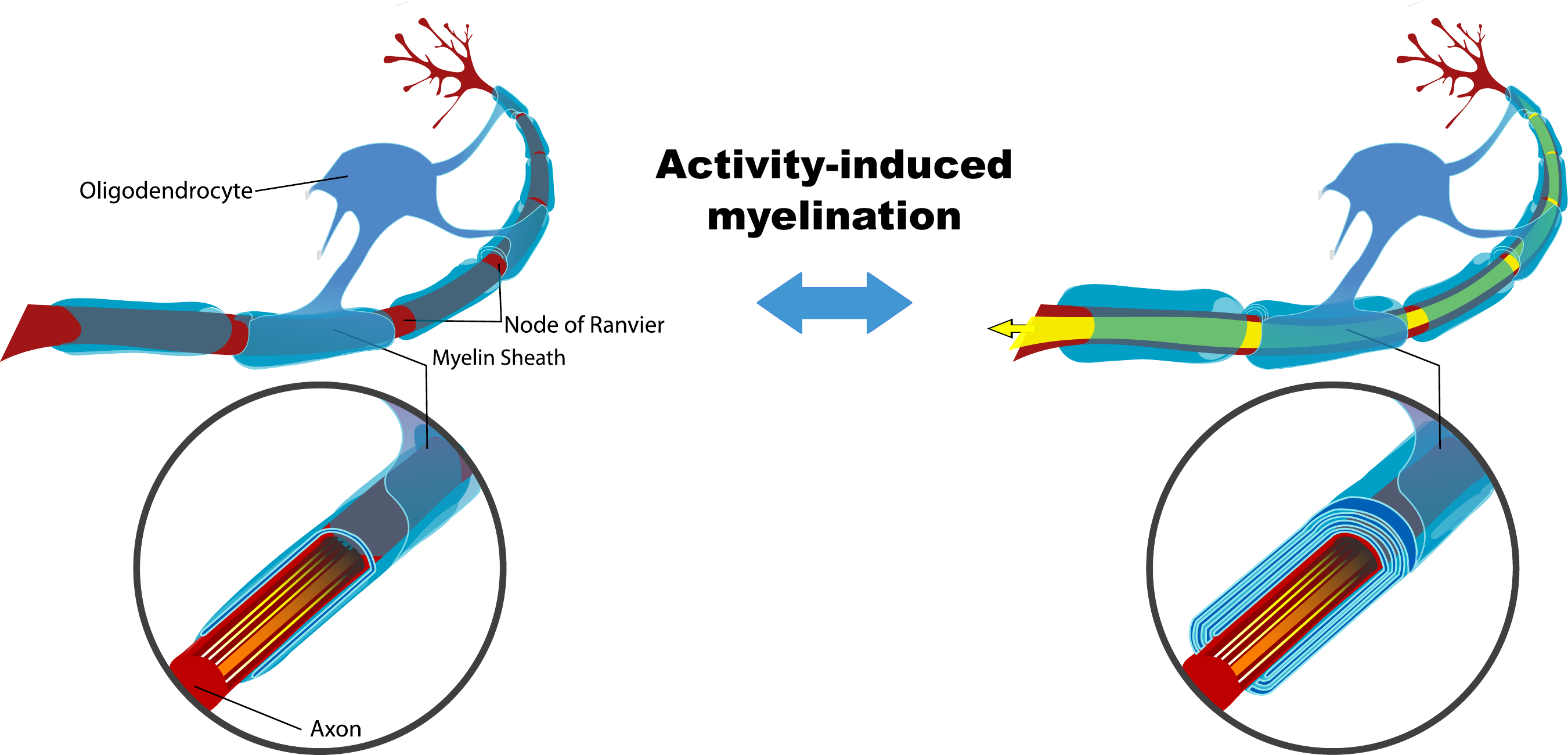}
    \caption{The extent of myelination may depend on neuronal activity. Myelination affects the propagation time delay along the axon. This results in an adaptive mechanism between the neuronal dynamics and the time delays along the axons.
    }
    \label{fig:myolinated}
\end{figure}

In contrast to the adaptive networks introduced in the previous sections, where the coupling weights $\kappa_{ij}$ were changing, the adaptation mechanism can exist for the time delays $\tau_{ij}$ as well \cite{Eurich1999,LueckenRosinWorlitzerEtAl2017,Paugam-Moisy2008,DBLP:conf/nips/KempterGHW95,Park2020,Fields2015,Pajevic2014,Almeida2017,Bechler2018,monje2018myelin,Senn2002,Fields2010}, i.e., time-delays become dependent on the activity of the nodes $\tau_{ij}=\tau_{ij}(\bm{x})$. The reason for the appearance of adaptive time-delays depends on the physical setup. For example, in neuronal systems, myelinated axons (white matter) regulate signal transmission velocity. Moreover, recent studies suggest that the level of myelination undergoes continuous changes, rather than remaining static \cite{monje2018myelin}, and these changes are activity-dependent \cite{Almeida2017,Bechler2018}. As a result, interaction delays continuously adjust in order to regulate the timing of neural signals propagating between different brain areas. 
In particular, active neurons become more myelinated \cite{Bechler2018}, see. Fig.~\ref{fig:myolinated}. In networks of interacting active particles, adaptive time-delays may occur due to the changes in the positions of the particles and, hence, the distant-dependent interaction delays. For machine learning applications, time-delay adaptation can serve as an additional learning mechanism \cite{Paugam-Moisy2008}.

The existing research with adaptive time delays is scarce. The reasons are twofold: firstly, the evidences of delay adaptation from neuroscience is more recent \cite{monje2018myelin,Almeida2017,Bechler2018} than the  weight adaptations, secondly, dealing with state-dependent time-delay is a much harder theoretical problem \cite{Hartung2006a}. 
 
Here we shortly mention two models from \cite{Park2020} and \cite{LueckenRosinWorlitzerEtAl2017}. In \cite{Park2020}, the coupled phase oscillators 
\begin{equation}
    \label{eq:Kuramoto-delay}
\frac{d}{d t} \theta_{i}(t)=\omega_{i}+ \sum_{j=1}^{N} \kappa_{i j} \sin \left(\theta_{j}\left(t-\tau_{i j}(t)\right)-\theta_{i}(t)\right), \quad 1 \leq i \leq N
\end{equation}
with adaptive time-delays of the following form 
\begin{equation}
\label{eq:Kuramoto-delay-adaptation}
 \frac{d}{d t} \tau_{i j}(t)= \varepsilon H\left(\tau_{i j}(t)\right) \left[-\left(\tau_{i j}(t)-\tau_{i j}^{0}\right)+\rho 
 \sin \left(\theta_{j}(t)-\theta_{i}(t)\right)\right]
\end{equation}
are considered. This form is similar to the adaptation rule \eqref{eq:AdaptiveKS_kappa} for the coupling weights. Here $H(\tau)$ is a "smoothed" Heaviside function intended to keep the time-delay non-negative. 
 
In \cite{LueckenRosinWorlitzerEtAl2017}, the event-based adaptation rule is used 
\begin{equation}
    \label{eq:delay-adaptation-STDP}
    \tau_{ij} \mapsto \tau_{ij} + \varepsilon W(\Delta t), 
    \text{ where } W(\Delta t) = \sin (\Delta t) e^{-|\Delta t|}
\end{equation}
and $\Delta t = t_i -t_j -\tau_{ij}$ the inter-spike distance that takes into account the propagation delay. The discrete adaptation \eqref{eq:delay-adaptation-STDP} is applied each time the neuron $i$ or neuron $j$ fires. Such an adaption rule us similar to the STDP plasticity discussed in  Sec.~\ref{sec:STDP} with the difference that the STDP leads to the updates in the coupling weights. As a nodal system for determining the state $\bm{x}(t)$, excitable coincidence detectors are used. this wikipedia picture

%% file: parts/dynPhenomena.tex
This section given an overview of dynamical phenomena discovered in adaptive dynamical networks. Some of these phenomena appear the same as in non-adaptive networks, such as complete synchronization (Sec.~\ref{sec:CS}), multistability (Sec.~\ref{sec:multistability}), explosive synchronization (Sec.~\ref{sec:explosive}), or chimera states (Sec.~\ref{sec:chimeras}). However, adaptivity either induces new specific features or requires a more sophisticated theoretical and numerical treatment of the adaptive systems involved. For example, the classical master-stability-function method for the study of complete synchronization \cite{Pecora1998} cannot be applied to adaptive networks, and an extended approach has been developed \cite{BER21b}. Also, multistability in adaptive networks appears to be much richer compared to dynamical networks without adaptivity. 

Of particular interest are phenomena that exclusively occur in networks with adaptive couplings. These are frequency clusters (Sec.~\ref{sec:multicluster}), solitary states (Sec.~\ref{sec:solitary}), recurrent synchronization (Sec.~\ref{sec:recurrent}), self-organized noise resistance (Sec.~\ref{sec:noise-resistance}) or heterogeneous nucleation (Sec.~\ref{sec:nucleation}).  
We note that while we report on the adaptation-induced phenomena, it is not excluded that some phenomenologically similar effects may occur in other classes of systems.
For example, solitary states were reported in \cite{Jaros2015} in the Kuramoto model with inertia, but it is was then shown in \cite{BER21a} that their emergence can be interpreted by the effect of frequency adaptation, and the Kuramoto model with inertia can be equivalently formulated as an adaptive dynamical network, see also~Sec.~\ref{sec:powerGrids}.

\subsection{Complete synchronization and master stability function}
\label{sec:CS}

Synchronization plays a crucial role in many applications where adaptive dynamical networks emerge. In brain networks this is important, for instance, under normal conditions in the context of cognition and learning~\cite{FEL11}, and under pathological conditions, such as Parkinson's disease~\cite{HAM07}, epilepsy~\cite{JIR13}, tinnitus~\cite{TAS12a}, schizophrenia, to name a few~\cite{UHL09}. In power grid networks, synchronization is essential for the stable operation~\cite{MOT13a,MEN14,TAH19}. It has been shown that adaptation rules such as spike-timing-dependent plasticity play an important role for achieving synchronization~\cite{KAR02a}. Rigorous conditions for the emergence of phase-locked states
in adaptive Kuramoto-like systems and for the complete oscillator death state in an adaptive Winfree model have been developed by Ha et. al.~\cite{HA16a,HA18,Ha2021InterplayOR}.

The methodology of the master stability function~\cite{Pecora1998} is one possibility for the analysis of synchronization phenomena. This method allows for separating dynamical from structural features for a given network. It drastically simplifies the problem by reducing the dimension and unifying the synchronization study for different networks. For adaptive networks, the master stability function was introduced in \cite{BER21b}, where the following class of $N$ adaptively coupled systems was considered
\begin{align}
	\dot{\bm{x}}_i&=f(\bm{x}_i)-{\sigma}\sum_{j=1}^N a_{ij}\kappa_{ij}g(\bm{x}_i,\bm{x}_j),\label{eq:adaptiveNW_x}\\
	\dot{\kappa}_{ij} &= -\epsilon\left(\kappa_{ij} + a_{ij} h(\bm{x}_i-\bm{x}_j)\right), \label{eq:adaptiveNW_kappa}
\end{align}
with  $f(\bm{x}_i)$ describing the local dynamics of each node, and $g(\bm{x}_i,\bm{x}_j)$ the coupling function. 
The coupling is weighted by the scalar variables $\kappa_{ij}$, which are adapted dynamically according to Eq.~(\ref{eq:adaptiveNW_kappa}) with the nonlinear adaptation function $h(\bm{x}_i-\bm{x}_j)$ depending on the difference of the corresponding dynamical variables. 
The base connectivity structure is given by the matrix elements ${a_{ij}\in\{0,1\}}$ of the $N\times N$ adjacency matrix $A$ which possesses a constant row sum $r$, i.e., $r=\sum_{j=1}^N a_{ij}$ for all $i=1,\dots,N$. 
The assumption of the constant row sum is necessary to allow for synchronization. 
The Laplacian matrix is $L=r\mathbb{I}_N -A$, where $\mathbb{I}_N$ is the $N$-dimensional identity matrix. The eigenvalues of $L$ are called Laplacian eigenvalues of the network. The parameter $\sigma >0$ defines the overall coupling input, and $\epsilon >0$ is a time-scale separation parameter. 
%In particular, if the adaptation is slower than the local dynamics, the parameter $\epsilon$ is small.
% We note that the phase space of system ~\eqref{eq:adaptiveNW_x}--\eqref{eq:adaptiveNW_kappa} is $N(d+N)$ dimensional.

Complete synchronization is defined as $\bm{x}_1=\bm{x}_2=\cdots=\bm{x}_N$. 
Denoting the synchronization state by $\bm{x}_i(t)=\bm{s}(t)$ and $\kappa_{ij}=\kappa_{ij}^s$, we obtain from Eqs.~\eqref{eq:adaptiveNW_x}--\eqref{eq:adaptiveNW_kappa} the following equations for $\bm{s}(t)$ and $\kappa_{ij}^s$
\begin{align}
	\dot{\bm{s}}&=f(\bm{s})+\sigma r h(0)g(\bm{s},\bm{s}),\label{eq:syncState_s}\\
	{\kappa}^{s}_{ij} &= -a_{ij} h(0). \label{eq:syncState_kappa}
\end{align}
In particular, we see that $\bm{s}(t)$ satisfies the dynamical equation (\ref{eq:syncState_s}), and $\kappa_{ij}^s$ are either $-h(0)$ or zero, if the corresponding link in the base connectivity structure exists ($a_{ij}=1$) or not ($a_{ij}=0$), respectively. 

In \cite{BER21b}, the stability problem for the synchronous state is reduced to the largest Lyapunov exponent $\Lambda(\mu)$, depending on a complex parameter $\mu$, for the following low-dimensional system 
\begin{align}
	\begin{split}
	\dot{\bm{\zeta}} &= \bigg(\mathrm{D}f(\bm{s})+\sigma r h(0)\big[\mathrm{D_1}g(\bm{s},\bm{s}) \\
	&\quad\quad + (1-\frac{\mu}{r}) \mathrm{D_2}g(\bm{s},\bm{s})\big]\bigg) \bm{\zeta} - {\sigma} g(\bm{s},\bm{s}) \kappa, 
	\end{split} \label{eq:adaptiveNW_MSF_zeta}\\ 
	\dot \kappa &=  -\epsilon\left(\mu \mathrm{D}h(0) \bm{\zeta} + \kappa\right)\label{eq:adaptiveNW_MSF_kappa}.
\end{align}
The function $\Lambda(\mu)$ is called master stability function.  Note that the first bracketed term in $\bm{\zeta}$ of~\eqref{eq:adaptiveNW_MSF_zeta} resembles the master stability approach for static networks, which, in this case, is equipped by an additional interaction representing the adaptation. Furthermore, the shape of the master stability function depends on the choice of $\sigma$ and $r$ explicitly. In case of diffusive coupling, i.e., $g(\bm{x},\bm{y})=g(\bm{x}-\bm{y})$, the master stability function can be expressed as $\Lambda(\sigma\mu)$ such that the shape of $\Lambda$ scales linearly with the coupling constant $\sigma$. This master stability approach has been further generalized for systems of coupled phase oscillators with heterogeneous adaptation rules~\cite{BER21f}.

Another example of the master stability function is given in \cite{SOR10a} for a system with the coupling weights $a_{ij} = \sigma_i {\overline a}_{ij}(t)$, where ${\overline a}_{ij}(t)$ is an externally given function of the coupling topology, and $\sigma_i$ an adaptively changing weight. The system considered in \cite{SOR10a} is a mixture of temporally evolving networks (see Sec.~\ref{sec:Temporally-evolving}) and adaptive networks. The coupling structure is defined by ${\overline a}_{ij}(t)$ and it is predefined, while the weights $\sigma_i$ are determined adaptively according to a certain control rule. We discuss it more specifically in Sec.~\ref{sec:control}.
%--------------------------------------------------

\subsection{Frequency clusters\label{sec:multicluster}}

Partial synchronization plays an important role in many real systems \cite{Scholl2021}. In the brain, partial synchrony is a common phenomenon allowing for coordinated activity across different brain regions and allows for complex cognitive processes.
Specific examples are coherent activity in motor cortex \cite{Baker1997} or coordinated reset stimulation \cite{POP15}, to name a few. 

Frequency clustering is one of the manifestation of partial synchrony and a typical phenomenon occurring in adaptive networks \cite{AOK15,KAS17,BER19a,BER19,BER21a,POP15,ROE19a,FEK20}. It is characterized by the emergence of groups of strongly coupled oscillators with the same average frequency within the groups. In other words, each group is frequency synchronized, but there is no frequency synchronization between the groups -- frequency clusters. At the same time, the phases of the oscillators in the frequency clusters need not be identical. 
Frequency clusters have been reported in dynamical networks of adaptive phase oscillators \cite{AOK11,AOK09,KAS17,BER19,BER19a}, Hodgkin-Huxley \cite{ROE19a}, Hindmarsh-Rose~\cite{CHA17a}, FitzHugh-Nagumo~\cite{VOC21}and Morris-Lecar \cite{POP15} neurons with spiking-time dependent plasticity, as well as in multiplexed networks \cite{KAS18}. 

We note that the frequency clusters that we report here are induced by the adaptivity, and they occur even in the case when the oscillators are identical. On the other hand, if the oscillators are not identical, e.g., two different groups with different frequencies, frequency clusters can emerge naturally without adaptation \cite{Ivanchenko,Montbrio2004,Strogatz1,Yue2020}.

\begin{figure}[t]
\includegraphics[width=\columnwidth]{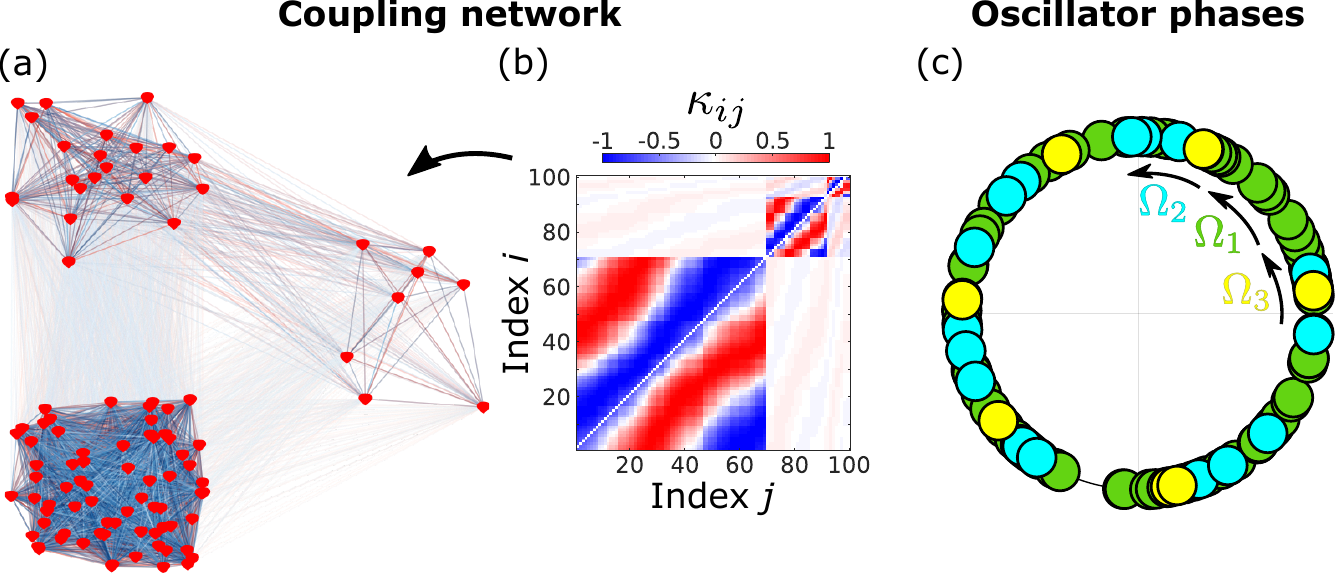}
\centering
\caption{Example of frequency clusters in a network of adaptively coupled phase oscillators \eqref{eq:AdaptiveKS_phi}--\eqref{eq:AdaptiveKS_kappa}. (a) Schematic representation of the emerging three clusters and (b) asymptotic coupling weights matrix.  (c) Oscillator's phases arranged over the circle. Colors denote different clusters. The oscillators from different clusters move with different frequency $\Omega_i$. Parameter values: $\varepsilon=0.01$, $\alpha=0.3\pi$, $\beta=0.23\pi$, $\omega=0$, and $N=100$. 
\label{fig:FC}}
\end{figure}

Figure~\ref{fig:FC} illustrates the frequency clusters observed in a network of adaptively coupled phase oscillators \eqref{eq:AdaptiveKS_phi}--\eqref{eq:AdaptiveKS_kappa}. The network is globally coupled with adaptive coupling weights. In the case shown, three clusters emerge self-consistently. As one can see from the coupling matrix in Figure~\ref{fig:FC}(b), the oscillators interact more strongly within the clusters while there is only a weak interaction between different clusters. One can show that the strength of the intercluster interaction is proportional to the time-scale splitting between the node and adaptation dynamics $\varepsilon$ in system \eqref{eq:AdaptiveKS_phi}--\eqref{eq:AdaptiveKS_kappa}, cf. \cite{KAS17}.

The most important feature of the frequency cluster is that all its elements move with the same (average) frequency. A structure of the mutual dynamics (phase lags)  within the cluster may be very different ranging from complete synchronization to anti-phase or so-called splay states. In \cite{BER19}, it was shown that only three types of clusters may appear in system \eqref{eq:AdaptiveKS_phi}--\eqref{eq:AdaptiveKS_kappa}: splay, antipodal (including in-phase and anti-phase), and douple-antipodal. In more realistic neuronal models \cite{POP15,CHA17a,ROE19a}, the observed clusters display complete synchronization between the elements. Beyond neuronal models, frequency cluster states have been reported for oscillating electrochemical systems~\cite{PAT21a}.

In order to understand the stability of frequency cluster states different approaches have been used. Utilizing slow adaptation, see e.g. Eq.~\eqref{eq:AdaptiveKS_phi}-\eqref{eq:AdaptiveKS_kappa} with small $\epsilon$, Berner et. al. described the stability in the asymptotic limit ($\epsilon\to 0$) ~\cite{BER19a,BER21c}. A more rigorous result on the stability of frequency clusters in Kuramoto-like systems has been developed by Feketa et.al.~\cite{feketa2021stability}. In their work, they provide a set of sufficient condition that guarantee the existence and stability of frequency cluster states.

In addition to the localized spatial structures, the emergence of modular and scale-free networks has been reported~\cite{ITO01a,ITO03,STA10b,GUT11,ASS11,YUA11,AOK12,WIN12,AOK15a,BOT12,BOT14,POP15,CHA17a,makarov2016emergence,avalos2012assortative} in networks with STDP. This fact underlines the potential importance of adaptive mechanism in the formation of connectivity structures as they have been experimentally found in brain networks~\cite{MEU10a,ASH19}. Furthermore, activity based adaptive rewiring has even been shown to enhance modularity~\cite{RUB09}.

\subsection{Solitary states \label{sec:solitary}}

A particular case of frequency cluster states (sometimes called multiclusters) are solitary states for which only one single element behaves differently compared with the behavior of the background group, i.e., the neighboring elements. These states have been found in diverse dynamical systems such as generalized Kuramoto-Sakaguchi models \cite{Maistrenko2014}, the Kuramoto model with inertia \cite{Jaros2015,BER21a}, the Stuart-Landau model, the FitzHugh-Nagumo model, systems of excitable units  and even experimental setups of coupled pendula \cite{KapitaniakKuzmaWojewodaEtAl2014}. Solitary states are considered as important states in the transition from coherent to incoherent dynamics \cite{Jaros2015}. In \cite{BER20c}, the emergence of solitary states in the presence of plastic coupling weights is investigated. The bifurcation scenarios in which solitary states are formed and (de)stabilized are studied.

\subsection{Recurrent synchronization}\label{sec:recurrent}

Recurrent synchronization is a macroscopic phenomenon in dynamical networks involving the recurrent switching between synchronous and asynchronous behavior. In \cite{Thiele2023}, a periodic switching between phase-locking and frequency clustering is reported. 
During recurrent synchronization, a macroscopic observable exhibits bursting behavior, whereas the individual nodes (neurons) are not required to burst at the microscopic level.
In the study \cite{Thiele2023}, the authors consider the nodes to have simple oscillatory dynamics. 
Therefore recurrent synchronization as a macroscopic effect contrasts bursting found in neuronal networks \cite{BEL11a,TAS12a,BEL08,GER14a}, where single neurons can have alternating periods of quiescence and fast spiking.

Other studies \cite{BLA99b,STO02, PAN12, SCH08, GAS20} have reported the occurrence of a seemingly similar phenomenon, called collective bursting, in neuronal networks. An important difference of \cite{BLA99b,STO02, PAN12, SCH08, GAS20} from \cite{Thiele2023} is that the collective bursting phenomenon is induced and observable on the microscopic level of individual neurons, whereas recurrent synchronization is not manifested by bursting on the microscopic level.
An adaptation-induced switching between phase-locking and periodic oscillation has been observed in~\cite{CIS20}, which is related to fold-homoclinic bursting~\cite{IZH00}.

The mechanism for recurrent synchronization  in \cite{Thiele2023} is based on a recurrent slow dynamics of adaptive coupling weights (hidden variables) between neuronal populations. When adaptation is heterogeneous (asymmetric), the hidden variables lead to a re-emergence of episodes of synchronization and desynchronization.
\cite{Thiele2023} reveals the importance of the following ingredients for the emergence of recurrent synchronization: slow adaptation, i.e., the timescale separation between the adaptation and the individual neuronal dynamics; asymmetry of adaptation rules; and recurrent (periodic/spiking) dynamics of the individual neurons. These results suggest that asymmetric adaptivity might play a fundamental role in the emergence and impairment of neuronal pattern generators, e.g., in Parkinsonian resting tremor.

\begin{figure}
    \centering
    \includegraphics[width=5cm]{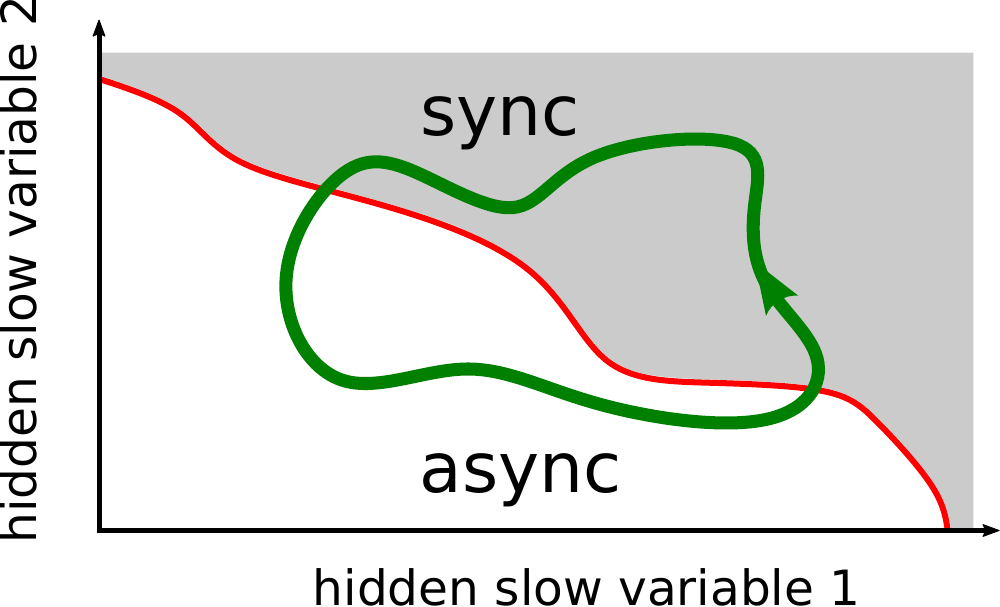}
    \caption{Schematic figure illustrating the emergence of recurrent synchronization. The hidden slow variables govern the recurrent transitions through the phases of synchronization and asynchrony of a fast network dynamics.}
    \label{fig:schema-recurrent}
\end{figure}

\subsection{Self-organized noise resistance, enhanced synchronization}
\label{sec:noise-resistance}

In \cite{POP13}, the authors show that adaptive dynamical networks can be more resilient to the desynchronizing effect of noise. This work considers spiking Hodgkin-Huxley neural populations with adaptive synaptic weights governed by spike timing-dependent plasticity (STDP). It is found that the mean synaptic coupling in such systems increases dynamically in response to the increase of the noise intensity, and there is an optimal noise level, where the amount of synaptic coupling gets maximal in a roverviewesonance-like manner. This constitutes a noise-induced self-organization of the synaptic connectivity~\cite{HES14,SHE15a,ZER21}, which effectively counteracts the desynchronizing impact of independent noise over a wide range of the noise intensity. 

The phenomenon of self-organized noise resistance was studied in more details in \cite{LUE16}, where the authors investigate a minimal model of two coupled oscillatory neurons in the presence of noise. First, they considered two coupled phase oscillator models 
\begin{eqnarray}
d\theta_1 & = & \left[ \omega_1 +w_1 g(\theta_2 - \theta_1)\right] dt + \sqrt{\mu}dW_1, \\
d\theta_2 & = & \left[ \omega_2 +w_2 g(\theta_1 - \theta_2)\right] dt + \sqrt{\mu}dW_2, 
\end{eqnarray}
where the natural fluctuations (noise) are represented as independent Wiener processes with the intensity $\sqrt{\mu}$, and the coupling weights $w_i(t)$ are adaptive accordingly to phase-difference-dependent plasticity (PDDP) rule 
\begin{equation}
\label{eq:PDDP}
\dot w_i (t) = P_i(\theta_2-\theta_1).
\end{equation}The authors apply an averaging procedure to derive specific functions $P_i$, that approximate the STDP rule \eqref{eq:STDP-update}--\ref{eq:STDP}. Thus,  \cite{LUE16} presents a connection between the discontinuous STDP adaptation rule and the continuous PDDP adaptation \eqref{eq:PDDP}. 

The work \cite{ZHO06f} shows that adaptivity can improve the synchronizability of networks of different topologies compared to corresponding networks with fixed structure. 
The authors of  \cite{ZHO06f} consider diffusively coupled identical oscillators of the form
\begin{equation}
    \dot x_i = F(x_i) + \sum_{j=1}^N G_{ij} \left( H(x_j) - H(x_i) \right),
\end{equation}
where $G_{ij}(t)=A_{ij}V_{i}(t)$ with $A_{ij}$ a binary adjacency matrix and $V_{i}$ weights that are controlled by the local synchronization properties of the nodes. More specifically, 
$$
\dot V_i = \gamma \frac{\Delta_i}{1+\Delta_i}, \quad 
\Delta_i = \left| 
H(x_i) - \frac{1}{k_i}\sum_j A_{ij}H(x_j)
\right|,\quad \gamma>0,
$$
where $k_i$ is the number of neighbors of the node $i$. The variable $\Delta_i$ measures the difference of the activity of node $i$ and the average activity of its neighbors. Such adaptation  leads to synchronization and thus suppression of the difference $\Delta_i$. Also, one can observe that the coupling weights increase in time ($\dot V_i > 0$) if the node $i$ is not synchronized with its neighbors. This latter fact gives a partial explanation for the particular mechanisms behind the enhancement of synchronization in this particular model. Moreover, the transition to synchronization appears to be hierarchical as the nodes of different degrees $k_i$ possesses different effective coupling strengths. 

The work \cite{Kim2018} reports how the stochastic burst synchronization can be either enhanced or depressed by the STDP in the presence of noise. More specifically, a Matthew effect in synaptic plasticity has been found due to a positive feedback process. Good burst synchronization (with higher bursting measure, see definition in \cite{Kim2018}) gets better via long-term potentiation of synaptic strengths, while bad burst  synchronization (with lower bursting measure) gets worse via long-term depression. As a result, a step-like rapid transition
to stochastic burst synchronization has been found to occur by changing the noise intensity, in contrast
to the relatively smooth transition in the absence of STDP.
For the individual nodes, the Izhikevich neuronal model was used \cite{Izhikevich2004}. 

\subsection{Multistability}
\label{sec:multistability}

The emergence of different collective states and synchronization patterns has been discussed in the above sections. Moreover, the co-emergence of the different states has been studied for various complex dynamical systems~\cite{PIS14}. This, so-called multistability, plays an important role in mathematics and physics. It is also crucial in a wide range of applications, e.g., in climate science or for the understanding of dynamic coordination in the brain~\cite{FEU18,KEL12b}. Various mechanisms have been described in complex dynamical systems that induce multistability. In fact, an adaptive network structure provides additional degrees of freedom for the co-stability of different dynamical states.

In models of phase oscillators with various forms of network adaptivity, multistability is a common phenomenon. The bistability of splay type (coherent) and other phase locked states has been found in~\cite{AOK09,AOK11}. In~\cite{SKA13a}, the authors observe the co-emergence of incoherence and synchronization. Beyond phase oscillator models, bistable dynamical regimes have been found in adaptive epidemic systems for healthy and endemic states~\cite{GRO06b} or in neuronal networks of Morris-Lecar bursting neurons with synaptic plasticity~\cite{POP15}. In the latter study~\cite{POP15}, the two co-stable states correspond to tinnitus related abnormal synchronization and a healthy conditions. The unlearning of the pathological state was analyzed by using coordinated reset stimuli~\cite{TAS03a,TAS12,TAS12a}. These findings have also triggered recent work on the treatment of Parkinson's disease using vibrotactile stimuli~\cite{PFE21}, where modelling systems have shown the same bistable dynamics~\cite{KRO20,KRO20a,KHA21,MAN21}.

In addition to bistable regimes, an adaptive network structure often leads to multistability of more than two collective states. These co-stable states may include different types of phase-locked states, e.g., splay and in-phase synchronous states~\cite{SEL02,PIC11a}, locked and incoherent states~\cite{CHA14a}, various forms of frequency cluster states~\cite{MAI07,BER19,BER19a} or chimera and solitary states~\cite{KAS17,BER21a,KAS21a,THA22}. Consequently, the resulting network structures for the co-emerging states take various forms, leading to a plethora of different asymptotic coupling structures~\cite{RAT21}. The robustness of multistability with respect to the variation of network structures has been analyzed in~\cite{KAS18a}. In~\cite{BER20}, the co-stability of different phase locked states for multiplex adaptive networks  has been analytically described using the multiplex decomposition method~\cite{BER21}. In addition, the authors found the appearance of additional phase-locked states induced by the multiplex and adaptive structure.

The multistable emergence of synchronized states and frequency clusters has also been found for coupled Hodgkin-Huxley systems with synaptic plasticity~\cite{LUE16,ROE19a}, where an analytic link between these models and adaptive phase oscillator models has also been established. Multistability of chaotic dynamics, periodic motion, and steady states have been found~\cite{AOK15a} on complex dynamical networks with dissipative processes between the nodes. Recently, multistability has been considered as an important feature for functional modelling approach of tumor and sepsis disease by using adaptive network models~\cite{SAW21b}.

\subsection{Explosive synchronization and hysteresis}
\label{sec:explosive}

In the previous sections, we have discussed the role of the adaptive network structure for the formation of synchronization patterns in homogeneous systems. For heterogeneous systems, where e.g. the individual nodes of a network represent dynamical units with different dynamical properties, the emergence of collective behaviour may depend on the interaction strength of the dynamical units. A classical example is the emergence of synchronization in populations of phase oscillators, where each oscillator possesses a different natural frequency drawn from a given frequency distribution. In these systems it has been shown that synchronization (coherence) emerges gradually with increasing interaction strength starting from asynchrony (incoherence) and ending in a completely synchronous state~\cite{KUR84}. From the statistical physics point of view, this transition from incoherence to coherence can be regarded as a second-order phase transition.

Another type of phase transition, known as first-order transition, has started attracting attention in the field of dynamical systems. This type of transition has been found to emerge in various dynamical systems, e.g., showing an abrupt jump from incoherence to coherence. The latter feature led also to the alternative notion of \emph{explosive synchronization}. Explosive synchronization has been found in phase oscillator systems on scale-free networks~\cite{GOM11a}. The mechanisms leading to an abrupt change in the synchronization of phase oscillators have been further investigated by looking at the synchronization state's basin of attraction~\cite{ZOU14} or considering the continuum limit~\cite{VLA15}.  Beyond phase oscillator networks, first-order transitions have been also been found in experimental systems of coupled relaxation oscillators~\cite{CAL20}. For reviews and recent developments on explosive synchronization and explosive percolation, we refer the reader to~\cite{BOC16,SOU19}. More recently, a universal mechanism has been unveiled that is responsible for the emergence of explosive transitions in a generic class of dynamical systems~\cite{KUE21}. In their work, Kuehn and Bick~\cite{KUE21} exemplified their mathematical findings further by showing the existence of an explosive transition in an adaptive epidemic network model. 

Explosive transitions and the related phenomenon of hysteresis have been reported for adaptive epidemic networks. In~\cite{GRO06b}, the authors found an abrupt transition from no infections to a very high number of infections, while increasing the probability of becoming infected. Also in dynamical systems of coupled phase oscillators, a first-order transition can be induced by applying adaptive coupling schemes. In~\cite{ZHA15a}, the coupling between the oscillators is scaled by the value of the order parameter, which in turn depends explicitly on the current state of the oscillators and leads to an explosive transition. Another type of adaptivity is introduced in~\cite{AVA18}, where the adaptation of the link between two nodes is governed by a differential form of the logistic map and driven by the correlation of the two nodes' states. The influence of the node correlations is controlled by a parameter called the correlation threshold which also determines the emergence of an explosive transition. Next to these differentiable forms of adaptation, in~\cite{eom2016concurrent} the enhancement of synchronization was studied in networks that, at each iteration time step, adapt their random Erdös-Renyi structure with respect to a connection probability that depends on the current state of the oscillators. This form of adaptation also induces a sudden transition to synchronization. Building on these findings, the emergence of first-order transitions has also been studied in other complex network structures such as multilayer networks. In~\cite{KAC20,KUM20a}, Kuramoto phase oscillator models on multiplex networks have been investigated. It was shown that in these setups, too, the different forms of adaptation of the interlayer coupling weights lead to explosive synchronization transitions.

\subsection{Heterogeneous nucleation in adaptive networks}
\label{sec:nucleation}

Besides explosive transitions, other types of first-order transitions have been reported for adaptive dynamical networks. In particular, in~\cite{FIA22} the authors considered a globally coupled adaptive network of phase oscillators \eqref{eq:AdaptiveKS_phi}-\eqref{eq:AdaptiveKS_kappa}, see also Sec.~\ref{sec:phaseOscModel}. This paper describes two qualitatively different transitions to synchronization induced by the interplay of an adaptive network structure and finite size inhomogeneities in the natural frequency distributions have been described: single-step and multi-step transitions. These transitions are a result of multistable multiclusters and explosive synchronization. In the multi-step transition, a single large cluster (nucleus) is formed around an inhomogeneity in the frequency distribution and grows seccesively until full synchronization is reached. In contrast, in the single-step synchronization transition, multiple equal-sized clusters (nuclei) are formed around multiple inhomogeneities, grow and coexist stably until they merge in an abrupt first-order transition to full synchronization for high coupling strengths. These transition phenomena are very similar to heterogeneous nucleation induced by local impurities known, e.g., from cloud formation~\cite{PRU10}, crystal growth~\cite{MUL01} or Ostwald ripening in equilibrium and nonequilibrium systems~\cite{SCH91}. This observation bridges between synchronization transitions in finite-size dynamical complex networks and thermodynamic phase transitions, where the finite-size induced inhomogeneities in the natural frequencies play the role of impurities.

In order to derive conditions for the transition scenarios, a mean-field approach for the considered adaptive dynamical network has been developed in~\cite{FIA22} using a collective coordinate ansatz~\cite{GOT15,HAN18b,SMI19,SMI20,SMI21}. To account for the various multistable multicluster states, the population of all oscillators is split up into subpopulations. The dynamics of the sub-populations, of the intra-population and of the inter-population coupling weights is then described by a few mean-field variables. In \cite{DUC22}, a similar approach utilizing the Ott-Antonson ansatz~\cite{OTT08} to describe the dynamics within the subpopulation has been introduced. A rigorous approach to mean-field description for adaptive dynamical networks can be found in Sec.~\ref{sec:mathmeth}.

\subsection{Chimera states\label{sec:chimeras}}

In this section, we briefly summarize some findings on chimera states in networks of adaptively coupled dynamical nodes. Commonly, chimera states are considered as symmetry broken states that consist of coherent (or synchronous) and incoherent (or asynchronous) parts. Hence, the network of nodes can be divided into these two groups. It is further often assumed that the two groups have some kind of "spatial" relationship between the nodes within each group. In particular, in the first article in which the notion "chimera state" was coined, the spatial relation was provided by a nonlocal coupling structure~\cite{ABR04}. Since then various kinds of chimera state in a variety of dynamical systems have been described~\cite{ZAk20} and even the concept of a chimera state has evolved over the years~\cite{HAU21a}. For more information and insights into the research on chimeras states, we refer the reader to the rich selection of review and perspective articles~\cite{MOT10,SCH16a,MAj18a,OME19c,PAR21a}. A review on the mathematics of chimera states can be found in~\cite{OME18a}.

One of the first works that focused on the phenomenon of chimera stets in adaptive networks is by Kasatkin et.al.~\cite{KAS17}. Here, the authors used a simple model of globally and adaptively coupled phase oscillators \eqref{eq:AdaptiveKS_phi}-\eqref{eq:AdaptiveKS_kappa}, see also Sec.~\ref{sec:phaseOscModel}, to show the emergence of chimera states~\footnote{Note that the authors called their states chimera-like states due to the missing spatial relationship between the nodes of the coherent and incoherent group.}. In fact, they have shown that the network splits up into two or more groups (clusters) of oscillators for which at least one group consists of frequency synchronized phase oscillators while another group stays incoherent. The groups size, further, has been shown to have hierarchical structure meaning that the number of oscillators in each group differs "strongly" from each other. It is further worth to mention that adaptive dynamics leads to intercluster coupling that vanishes on average, i.e., an effective decoupling between the clusters. Recently the notion of "strong chimeras" has been introduced, which are chimera states that "(i) are permanently stable, (ii) exhibit identically synchronized coherent domain, and (iii) do not co-occur with stable global synchronization"~\cite{ZHA21}. While the first condition would have to be proved, the numerics indicates that the chimera states from~\cite{KAS17} fall into this category of strong chimeras. In fact, for some regions in the analyzed parameter space, the presence of the incoherent cluster stabilizes the coherent cluster, while stable global synchronization is impossible. The stabilization of the coherent cluster could be possibly proved by an analysis as described in~\cite{BER19a,BER21c}.

The chimera states found in~\cite{KAS17} have been also analysed on multiplex networks which allowed for an investigation of the dynamical interaction of chimera states. In~\cite{KAS18}, Kasatkin and Nekorkin have shown the robustness of chimeras states against complexity of the network structure~\cite{KAS18}. Beyond that, in ~\cite{BER21b}, the emergence of chimera states in adaptive networks of phase oscillators and FitzHugh-Nagumo oscillators have been observed in an adaptivity induced desynchronization transition. Similar states as those found in~\cite{KAS17}, have been also observed in networks of adaptively coupled active rotators~\cite{THA22}, adaptively coupled FitzHugh-Nagumo oscillators~\cite{HOU19} and populations of bursting neurons equipped with burst-timing-dependent plasticity~\cite{WAN20g}. Using a mathematical slow-fast analysis, the emergence of chimera and breathing chimera states have been recently described for a two-layer system of phase oscillators with adaptive interlayer coupling~\cite{venegas2023stable}.

In systems of adaptively and pulse-coupled dynamical units, another type of chimera called itinerant chimera state have been described~\cite{KAS19}. This novel type of chimera states emerged in parameter regions between chimera states and coherent states. The phenomenon is characterized by a temporally and spontaneously changing organization of the coherent group. More precisely, at any point in time, the observed state can be split into a coherent and an incoherent group. However, the elements of these groups change with time in contrast to chimeras states where the elements stay the same. The elements in the coherent group may even change completely which distinguishes itinerant chimeras also from breathing chimeras. Itinerant chimeras have been also observed in adaptive networks of active rotators~\cite{THA22}.

To conclude this section, we would like to comment on the emergence of so-called "weak chimeras" in adaptive networks. Weak chimeras have been introduced by Ashwin and Burylko~\cite{ASH15} as a mathematically well-defined notion for a chimera state in coupled systems of indistinguishable, i.e., identical and interchangeable, phase oscillators. These states are simply characterized by the existence of three nodes, two of which have the same average frequency and the remaining node has a different average frequency. Therefore, a weak chimera state describes a symmetry broken state. See~\cite{ASH15} for more details. With regards to this definition, in fact, multicluster states on e.g. globally or non-locally coupled networks could be classified as weak chimera states, see Sec.~\ref{sec:multicluster} for details.

%\subsection{Chaotic dynamics}

%Type I and type II transitions

%multiscale properties (CK), meso and macro-scales…

%% file: parts/mathmeth.tex
A natural approach to improve the toolbox of mathematical techniques for adaptive network dynamics is to start by trying to extend existing methods from static networks. To understand the difficulties in this process, it is helpful to briefly review for each method the static setting and then discuss, where the challenges to an extension are. In fact, a mathematically rigorous theory for adaptive networks is still in its infancy at this point, so we can only sketch the first steps here focusing on a didatic introduction to the techniques that have already proven to be useful.

%%%%%%%%%%%%%%%%%%%%%%%%%%%%%%%%%%%%%%%%%%%%%%%%%%%%%%%%%%%%%%%%%%%%%%%%%%%%%
\subsection{Mean-Field: Vlasov-Fokker-Planck Equations}
\label{sec:VFPE}

The idea of a mean-field for interacting particle systems is deeply ingrained within the foundations of statistical physics~\cite{Boltzmann,Fokker,Jeans,Planck,Vlasov}. The idea is that a ``typical particle'' can be identified, which feels the interactions with other particles in a sufficiently uniform way to allow for a single equation to describe the system dynamics. As an illustration consider a set of ordinary differential equations (ODEs) in the form
\be
\label{eq:ips}
\frac{\txtd {\bm x}_i}{\txtd t} = \dot{{\bm x}}_i=f_i({\bm x}_i)+\frac1N\sum_{j=1}^N K({\bm x}_i,{\bm x}_j), \quad i\in\{1,2,\ldots,N\}=:[N],
\ee
where ${\bm x}_i={\bm x}_i(t)\in\cX$ describes the state of the $i$-th particle/node, $\cX$ is the node's phase space, $t\in \R$ is the time variable, $K$ is a given interaction kernel with $K({\bm x},{\bm x})=0$, and the individual dynamics $f_i({\bm x}_i)$ is often chosen to be very similar (or even identical) for each node. We have already seen several application examples, which are of a form identical or similar to~\eqref{eq:ips}. In fact, many famous models fall into the class~\eqref{eq:ips} including gas as well as plasma dynamics~\cite{Cercignani1}, the Kuramoto model~\cite{KUR84}, the Desai-Zwanzig model~\cite{DesaiZwanzig}, coupled van-der-Pol/FitzHugh-Nagumo systems~\cite{Winfree1,SomersKopell}, the continuous Hopfield equations~\cite{Hopfield}, the Hegselmann-Krause model~\cite{HegselmannKrause}, and the Cucker-Smale model~\cite{CuckerSmale1}, see also Eqs.~\eqref{eq:dynNetworkGen}, \eqref{eq:dynNetworkGenAdj}, \eqref{eq:dynNetwork}, and \eqref{eq:weightedDynNetwork}. 

To see mathematically, how a mean-field can arise, let us consider~\eqref{eq:ips} for the slightly simpler case $f_i\equiv0$ for all $i\in [N]$; for example, this situation arises in the Kuramoto model of identical oscillators in a rotating frame~\cite{Strogatz1}. If we could show that 
\benn
\frac1N\sum_{j=1}^N K({\bm x}_i,{\bm x}_j) \ra \int_{\cX} K({\bm x}_i,\tilde{{\bm x}})
~u(t,\tilde{{\bm x}})~\txtd \tilde{{\bm x}}\quad \text{as $N\ra \I$},
\eenn
for a density $u(t,{\bm x})$, then it is appealing to replace~\eqref{eq:ips} by a single evolution equation
\be
\label{eq:kin1}
\dot{{\bm x}} = \int_{\cX} K({\bm x},\tilde{{\bm x}})~u(t,\tilde{{\bm x}})~\txtd \tilde{{\bm x}}. 
\ee
In this context, the density $u(t,{\bm x})$ describes the density of particles found at time $t$ at position ${\bm x}$. One observes that~\eqref{eq:kin1} is the characteristic ODE of a partial differential equation (PDE)
\be
\label{eq:Vlasov1}
\partial_t u = -\nabla_{\bm x} \cdot (u V[u]),\qquad 
V[u](t,{\bm x}):=\int_{\cX} K({\bm x},\tilde{{\bm x}})~u(t,\tilde{{\bm x}})~\txtd \tilde{{\bm x}}.    
\ee 
The PDE~\eqref{eq:Vlasov1} is also known as the Vlasov equation~\cite{Golse}. One expects that for large $N$, we may use solutions of~\eqref{eq:Vlasov1} to approximate the original interacting particle system~\eqref{eq:ips}. To make this more precise, one considers the empirical measure
\benn
\delta^{N}(t):=\frac{1}{N}\sum_{i=1}^N \delta_{{\bm x}_i(t)},
\eenn   
where $\delta_{{\bm x}_i(t)}$ are Dirac measures at location ${\bm x}_i={\bm x}_i(t)$. Considering the time-dependent family of probability measures $\mu(t)$ given by $\mu(t)(\cA):=\int_\cA u(t,{\bm x})~\txtd {\bm x}$ for measurable subsets $\cA\subset \cX$, one can hope that for a fixed time $T>0$
\be
\label{eq:approx}
\lim_{N\ra \I}\sup_{t\in[0,T]}d(\delta^N(t),\mu(t))=0,
\ee
where a choice of the metric $d$ is required on the space of (probability) measures. There are several choices of comparison metric available~\cite{Dobrushin,Neunzert}. The question whether~\eqref{eq:approx} holds can be answered rigorously in the affirmative as long as $K$ is sufficiently regular and becomes more difficult for very singular kernels $K$. One might anticipate that it is possible to generalize the idea to derive a Vlasov equation in the context of (fixed) network dynamical systems of the form~\cite{Arenasetal,ROD16} 
\be
\label{eq:ips1}
\frac{\txtd {\bm x}_i}{\txtd t} = \dot{{\bm x}}_i=f_i({\bm x}_i)+\sum_{i=1}^N a^N_{ij}K({\bm x}_i,{\bm x}_j), \quad i\in\{1,2,\ldots,N\}=:[N],
\ee
where we write $A^N=(a^N_{ij})_{i,j=1}^N$ for the, possibly scaled, adjacency matrix, where the superscript emphasizes the dependence on the number of nodes $N$. Already the static network case~\eqref{eq:ips1} is very difficult as several questions arise: Since the scaling $1/N$ is just natural for the all-to-all or very dense networks, what is the correct scaling of $A^N$ as $N\ra \I$? What do we mean by a limit $A^N\ra A^\I$, i.e., what is the limit of a network of infinitely many nodes? What is the analog to the Vlasov equation for heterogeneous networks? If a Vlasov-type equation does exist, how is the network heterogeneity encoded in it? How do we prove approximation results between the solutions of the finite-dimensional ODEs and the Vlasov PDE now? A lot of recent progress has been made on these questions. The easiest case occurs if the adjacency matrix is sampled from a graphon $W:[0,1]\times [0,1]\ra \{0,1\}$~\cite{Lovasz}. One subdivides the unit interval into $N$ equally spaced sub-intervals $I_j$ and sets $(a^N_W)_{ij}=1$ if $W$ is $1$ at the center of the square $I_i\times I_j$ and $(a^N_W)_{ij}=0$ otherwise. Therefore, $W$ effectively encodes all links and can be viewed as the limiting operator of the adjacency matrix that can act (spatially) on densities
\be
\label{eq:graphon}
\lim_{N\ra \I} A^N_W\ra A^\I_W,\quad (A^\I_W u)(t,{\bm x}):=\int_0^1 W({\bm x},{\bm y})u(t,{\bm y})~\txtd {\bm y}.
\ee 
In the case of graphons for static network dynamics, a lot of detailed results exist~\cite{ChibaMedvedev,KaliuzhnyiVerbovetskyiMedvedev1}. For example, one obtains a Vlasov-type equation for~\eqref{eq:ips1} and $f_i\equiv 0$ as 
\be
\label{eq:VlasGraphon}
\partial_t u = - \nabla_{\bm x}(uV_W[u]),\quad V_W[u](t,{\bm x},y):=\int_\cX \int_0^1 
K({\bm x},\tilde{{\bm x}}) W(y,\tilde{y}) u(t,\tilde{{\bm x}},\tilde{y})~\txtd \tilde{y}~\txtd \tilde{{\bm x}},
\ee
where now the density $u=u(t,{\bm x},y)$ also depends on the heterogeneity via the variable $y\in[0,1]$, which encodes the position of a node in the network. Of course, one could try to average over the heterogeneity coordinate in some form to obtain a single mean-field PDE, which captures the average properties of the network over all nodes. There are now already several studies exploiting mean-field graphon Vlasov limit PDEs to study stability, bifurcations, and related topics~\cite{BickSclosa,ChibaMedvedev1,ChibaMedvedevMizuhara}. Indeed, all the usual tools for PDEs become applicable once one exchanges the large finite-dimensional ODE system for a family of Vlasov equations, which is much smaller in many applications. Unfortunately, there are only relatively few networks, relative to the space of all networks, which admit a graphon limit~\cite{BackhauszSzegedy}. This problem has also recently been tackled and a solution is to view graph limits more abstractly via general linear operator limits $A^\I$, so-called graphops~\cite{BackhauszSzegedy}, and/or via their associated fiber measures $\{\mu^\I_y\}_{y\in[0,1]}$. In particular, viewing $A^\I$ as an operator limit is natural since each finite-dimensional adjacency matrix $A^N$ is a linear operator already (albeit on growing spaces). Hence, the Vlasov equation in this more general setting is formally given by~\cite{KuehnGraphops}
\be
\label{eq:VlasGraphop1}
\partial_t u = - \nabla_{\bm x}(uV_A[u]),\quad V_A[u](t,{\bm x},y)=\int_\cX  
K({\bm x},\tilde{{\bm x}}) (A^\I u)(t,\tilde{{\bm x}},y)~\txtd \tilde{{\bm x}},
\ee
where instead of a graphon acting via a convolution we more abstractly have the action of the graphop  
$A^\I$ acting as an operator. Alternatively, we can use the associated fiber measures representing $A^\I$ to re-write \eqref{eq:VlasGraphop1} as 
\be
\label{eq:VlasGraphop2}
\partial_t u = - \partial_{\bm x}(uV_\mu[u]),\quad V_\mu[u](t,{\bm x},y)=\int_\cX \int_0^1 
K({\bm x},\tilde{{\bm x}}) u(t,\tilde{{\bm x}},\tilde{y})~\txtd \mu_y(\tilde{y})~\txtd \tilde{{\bm x}}.
\ee
Several rigorous results exist proving that \eqref{eq:VlasGraphop1}-\eqref{eq:VlasGraphop2} are good approximations to the dynamics on large networks~\cite{GkogkasKuehn,KuehnXu}. Yet, static networks are only a first step, and for adaptive networks a typical class of ODEs is
\be
\label{eq:ips1adapt}
\begin{aligned}
\dot{{\bm x}}_i&=f_i({\bm x}_i)+\frac1N\sum_{j=1}^N a^N_{ij}K({\bm x}_i,{\bm x}_j),\\
(\dot{a}^N_{ij})&=g_{ij}({\bm x},A^N),
\end{aligned}
\ee
where the dynamics of the edges is given by the vector field $g=g_{ij}$, which may potentially depend on all node values ${\bm x}=({\bm x}_1,{\bm x}_2,\ldots,{\bm x}_N)^\top$ and the entire adjacency matrix $A^N$. In fact, we have already seen many examples of the form~\eqref{eq:ips1adapt} arising in applications. We remark that in the substantially simpler case of time-dependent networks~\cite{AyiDuteil,Duteil} we have that $g_{ij}$ is independent of ${\bm x}$, so that we have a non-autonomous variant of~\eqref{eq:ips1adapt}. In this time-dependent case, one anticipates under reasonable conditions that~\eqref{eq:VlasGraphop1} and \eqref{eq:VlasGraphop2} are still valid replacing $A^\I$ by a time-dependent operator $A^\I(t)$ and $\mu_y$ by time-dependent measures $\mu_y(t)$. As we have emphasized several times described in all the applications above, adaptive networks go beyond this, and then one has to deal with the limit of the ODEs $(\dot{a}_{ij}^N)=g_{ij}({\bm x},A^N)$. For general vector fields $g_{ij}$, this is beyond current mathematical techniques. A first starting point has been obtained in~\cite{GKO22} considering evolution equations for the links of the form 
\be
\label{eq:locrule}
\dot{a}_{ij}=-\epsilon(a_{ij}+h({\bm x}_j-{\bm x}_i)),
\ee
where $h$ is a given coupling kernel and $\epsilon>0$ controls the scale separation; cf.~Section~\ref{sec:multiscale}. The rule~\eqref{eq:locrule} is local and linear in the link weights. In this case, one can again prove the validity of a Vlasov equation~\cite{GKO22}. The idea is to use the variation-of-constants formula for~\eqref{eq:locrule}
\benn
a_{ij}(t)=\txte^{-\epsilon t}a_{ij}(0)+\int_0^t \txte^{-\epsilon(t-s)}h({\bm x}_j(s)-{\bm x}_i(s))~\txtd s
\eenn
and then insert this formula into~\eqref{eq:ips1adapt}. This yields a nonlocal-in-time generalized Kuramoto-type model. Under suitable conditions on $h$, one can then establish a generalized Vlasov equation. Yet, the full adaptive case is mathematically still unresolved at this point.\medskip

A directly related class of mean-field problems arise if the original ODEs are replaced by stochastic ordinary differential equations (SODEs), e.g., given by
\be
\label{eq:ips1SDE}
\dot{{\bm x}}_i=f_i({\bm x}_i)+\sum_{j=1}^N a^N_{ij}K({\bm x}_i,{\bm x}_j)+\sigma\xi_i,
\ee
where $\{\xi_i=\xi_i(t)\}_{i=1}^N$ are independent vectors of white noises and $\sigma>0$ is a constant controlling the noise level. For the classical case of all-to-all coupling, it is well-understood that one obtains a modification of the Vlasov equation, sometimes called Vlasov-Fokker-Planck equation (VFPE), which contains a Laplacian $\Delta_x$ arising due to the noise~\cite{Risken,Frank,Pavliotis1}. Therefore, one conjectures that for heterogeneous networks, the VFPE for~\eqref{eq:ips1SDE} should be  
\be
\label{eq:VlasGraphop1SDE}
\partial_t u = \frac{\sigma^2}{2}\Delta_{\bm x}u - \nabla_{\bm x}(uV_A[u]),\quad V_A[u](t,{\bm x},y)=\int_\cX  
K({\bm x},\tilde{{\bm x}}) (A^\I u)(t,\tilde{{\bm x}},y)~\txtd \tilde{{\bm x}}.
\ee
For all-to-all coupled systems, many mathematical techniques exist to derive~\eqref{eq:VlasGraphop1SDE}, see e.g.~\cite{ChaintronDiez}. However, even the theory of VFPEs for static networks is still being developed at this point and the case of adaptive networks is mostly open although the technique mentioned above via the variation-of-constants formula evidently generalizes quite directly. 

%%%%%%%%%%%%%%%%%%%%%%%%%%%%%%%%%%%%%%%%%%%%%%%%%%%%%%%%%%%%%%%%%%%%%%%%%%%%%
\subsection{Mean-Field: Moment Equations}
\label{sec:moments}

Instead of studying a detailed mesoscopic evolution equation as in Section~\ref{sec:VFPE}, one might be content with capturing important macroscopic observables. One option to derive such observables are the classical moments of a probability distribution. Indeed, if we consider the VFPE~\eqref{eq:VlasGraphop1SDE}, then one can define the $p$-th moments and the $p$-th centered moments as
\be
\label{eq:moments}
m_k(t,y):=\int_\cX u(t,{\bm x},y)^p~\txtd {\bm x},\qquad \mathfrak{m}_k(t,y):=\int_\cX [u(t,{\bm x},y)-m_0(t,y)]^p~\txtd {\bm x}.
\ee 
As before, averaging over the heterogeneity variable $y$ would lead to even more reduced moments. Once certain moments are chosen, then one can differentiate~\eqref{eq:moments} with respect to $t$ and derive differential equations~\cite{Socha}. Yet, for adaptive networks, using the definitions~\eqref{eq:moments} is oftentimes physically not very informative as one is primarily interested in the interplay between node and link dynamics. 

Hence, it can be beneficial to design/select observables particularly well-suited for adaptive networks~\cite{GRO09,KissMillerSimon,KuehnMC}. Let us illustrate this approach first in the context of susceptible-infected-susceptible (SIS) dynamics on a static network with infection rate $\beta$ and recovery rate $\gamma$, which we already encountered in Section~\ref{sec:epidemics}. Let $v_i(t)$ for $i\in[N]$ denote the random variable having a value $U\in\{S,I\}$ at time $t$ at node $i$. The expected values are defined by
\benn
[U](t):=\sum_{i=1}^N \P(v_i(t)=U)
\eenn
so that $[S](t)$ and $[I](t)$ are deterministic processes denoting the average number of the susceptible and infected populations respectively. Then we define the expectations for the proportions of the links as well
\benn
[UV](t):=\sum_{i=1}^N\sum_{j=1}^N a^N_{ij}\P(v_i(t)=U,v_j(t)=V),~\qquad U,V\in\{S,I\},
\eenn
where we recall that $(a^N_{ij})_{i,j=1}^N$ is the adjacency matrix. Note carefully that here links between two susceptibles contribute twice to the $[SS]$ count, so there are hidden combinatorial factors\footnote{In fact, definition/conventions of the combinatorial factors seem to differ across the literature. Hence, it is always advisable to check the conventions used in a certain source.}. Of course, similar words-of-warning and generalized definitions apply for triplets
\benn
[UVW](t):=\sum_{k=1}^N\sum_{i=1}^N\sum_{j=1}^N a_{ij}a_{jk}\P(v_i(t)=U,v_j(t)=V,v_k(t)=W),
\eenn
for $U,V,W\in\{S,I\}$, and so on for higher-moments. The idea of a mean-field approximation is now based upon tracking averages. For an SIS epidemic with the infection rate $\beta$ and recovery rate $\gamma$ on an undirected unweighted graph one may prove~\cite{KissMillerSimon} that
\be
\label{eq:ODESIS1}
\begin{array}{lcl}
\dot{[S]} &=& \gamma [I]-\beta[SI],\\
\dot{[I]} &=& \beta[SI]-\gamma[I].\\
\end{array}
\ee
The result is intuitive: infections take place proportional to the average number of $SI$-links multiplied by the infection rate $\beta$, while recovery takes place proportional to average 
infected population multiplied by the recovery rate $\gamma$. Of course, we could even omit one equation by using the conservation property $[S]+[I]=N$.

The proof of~\eqref{eq:ODESIS1} uses the exact master equation for SIS dynamics. Let $x^k_j(t)$ be the probability of being in state 
\benn
\cS^k_j\qquad \text{for $j\in\{1,2,\ldots,c_k\}$ where }
c_k:=\left(\begin{array}{c}N\\ k\end{array}\right)
\eenn
and we can interpret $c_k$ as the number of possible states having $k$ infected nodes. We can then group the probabilities in a (column)-vector   
\benn
x^k:=(x_1^k,x^k_2,\ldots,x^k_{c_k})^\top.
\eenn
Then a relatively straightforward calculation~\cite{KissMillerSimon} shows that the master equation on a general undirected and unweighted graph $\cG_N$ for the SIS model takes the form
\be
\label{eq:masterSIS}
\frac{\txtd}{\txtd t}(x^k)=A^kx^{k-1}+B^kx^k+C^kx^{k-1},\qquad k\in\{0,1,\ldots,N\}
\ee
with $A^0$ and $C^N$ being zero matrices and directly computable matrices $A^k$, $B^k$ and $C^k$. Note that the structural form of~\eqref{eq:masterSIS} arises because in a single step of infection or recovery, the state of the dynamics only changes by one infected node. To see a link between the master equation~\eqref{eq:masterSIS} and \eqref{eq:ODESIS1}, we observe that
\benn
[I](t)=\sum_{k=0}^N \sum_{j=1}^{c_k} kx^k_j(t),\quad 
[S](t)=\sum_{k=0}^N \sum_{j=1}^{c_k} (N-k)x^k_j(t)
\eenn 
as well as 
\benn
[SI](t)=\sum_{k=0}^N\sum_{j=1}^{c_k} N_{SI}(\cS^k_j)x_j^k(t),
\eenn
where $N_{SI}(\cS^k_j)$ is the number of edges between infected and susceptible vertices in the state $\cS^k_j$. We introduce the helpful auxiliary notation 
\benn
e_k:=(1,1,\ldots,1)\in\R^{1\times c_k}
\eenn
to express summations in a more compact form. The strategy is now clear: differentiate the definitions of $[I](t)$ and $[S](t)$, and then simplify until $[I]$ and $[SI]$ appear. We start with $[I](t)$ and
obtain
\beann
\dot{[I]}&=& \sum_{k=0}^N ke_k \frac{\txtd}{\txtd t}(x^k)=\sum_{k=0}^N ke_k (A^kx^{k-1}+B^kx^k + C^kx^{k+1})\\
&=& \sum_{k=0}^{N-1} (k+1)e_{k+1} A^{k+1}x^{k}
+ \sum_{k=0}^N ke_k B^kx^{k}
+ \sum_{k=1}^N (k-1)e_{k-1} C^{k-1}x^{k}\\
&=& \sum_{k=0}^N\left( (k+1)e_{k+1} A^{k+1}+ke_k B^k+(k-1)e_{k-1} C^{k-1}\right) x^{k}.
\eeann
To simplify this expression further, we claim that 
\be
\label{eq:claimintOP}
ke_{k+1} A^{k+1}+ke_k B^k+ke_{k-1} C^{k-1}=0\qquad \forall k\in\{0,1,\ldots,N\}.
\ee
To see this, consider one finds after some algebraic manipulation that
\benn
-(e_{k+1}A^{k+1})_i-(e_{k-1}C^{k-1})_i=B_{ii}^k=(e_kB^k)_i
\eenn
which holds for all $i\in[c_k]$ so \eqref{eq:claimintOP} is indeed true. This
simplifies our derivative computation to yield
\benn
\dot{[I]}=\sum_{k=0}^N\left( e_{k+1} A^{k+1}-e_{k-1} C^{k-1}\right) x^{k}=
\beta[SI]-\gamma[I],
\eenn
where we have used a few further relatively simple algebraic manipulations. The calculation for $\dot{[S]}$ is almost completely analogous. 

For adaptive networks, these types of calculations for master equations are still valid, just that more terms appear due to the adaptation rule. In particular, the structure of~\eqref{eq:masterSIS} has to also track changes in the links. As a concrete example, let us re-consider the SIS model with adaptive re-wiring considered in Section~\ref{sec:epidemics}. The adaptation rule is to re-wire $SI$-links at rate $w$ to $SS$-links~\cite{GRO06b}, i.e., individuals apply social distancing by breaking links to infected nodes and re-wiring them to susceptible nodes; see also~\cite{HorstmeyerKuehnThurner,HorstmeyerKuehnThurner1,ShawSchwartz,ShawSchwartz1}. Following the procedure via master equations, or even directly writing down an ODE system analogous to \eqref{eq:ODESIS1}, one obtains
\be
\label{eq:ODESISThilo}
\begin{array}{lcl}
\dot{[S]} &=& \gamma [I]-\beta[SI],\\
\dot{[II]} &=& \beta[SI]+\beta [ISI]-\gamma [II],\\
\dot{[SS]} &=& \beta[SI]+w [SI]-\beta [SSI],\\
\end{array}
\ee
where the total number of links is conserved. So $[SI]$ can be written in terms of $[II]$ and $[SS]$ and where some care has to be taken regarding combinatorial pre-factors depending upon, which definition of the network motifs is used.

The key problem for the static case \eqref{eq:ODESIS1} and the adaptive case \eqref{eq:ODESISThilo} is that the ODEs are not closed, i.e., they actually form an infinite chain of equations, which is not really simpler than tracking the full network dynamics and/or VFPEs. A classical approach to circumvent this problem is to use moment closure methods, which aim to express higher moments as functions of lower moments; for a detailed literature review regarding moment closure see~\cite{KuehnMC}. In the context of~\eqref{eq:ODESIS1} and~\eqref{eq:ODESISThilo}, moment closure is the search for mappings $H_{1,2}$ that satisfy
\be
\label{eq:momentclosure}    
[SI]\approx H_1([S],[I]), \qquad ([SSI],[ISI])\approx H_2([S],[I],[SS],[SI],[II]).
\ee
The main issue in trying to find~\eqref{eq:momentclosure} is that the approximation ideally should hold for all possible initial conditions, all parameter ranges, and over all times. The same issue appears in all moment closure schemes. As expected, no general rigorous solution to this problem has been found. The current pre-dominant strategy is to use physical principles upon which to postulate a closure mechanism. For example, one frequently used closure scheme~\cite{BrauerCastillo-Chavez,BrauervandenDriesscheWu} for static SIS epidemic models such as \eqref{eq:ODESIS1} is
\benn
[SI]\approx [S][I],
\eenn
which is exact for the complete network but generally fails for heterogeneous complex networks. For \eqref{eq:ODESISThilo}, a commonly used closure scheme is the pair approximation~\cite{KeelingEames,Rand,Tayloretal}
\benn
[UVW]\approx \frac{[UV][VW]}{[V]},\qquad \text{for $U,V,W\in\{S,I\}$.}
\eenn
The pair approximation turns out to be quite good when compared to direct numerical network simulations for many heterogeneous complex networks. Therefore, it also has been used frequently in adaptive epidemic dynamics and other classes of adaptive network models; see~\cite{KissMillerSimon} and references therein.

In addition to searching for good closures, e.g., based upon entropy principles~\cite{Levermore,RaghibHillDieckmann,Rogers}, an alternative approach to improve the mean field approximation via moments, is to change the set of observables. In contrast to the moment equations outlined above, which are also called homogeneous, taking observables that account directly for the node degree are called heterogeneous moment equations. A wide variety of approaches to heterogeneous moment equations and selecting associated closures exists, see e.g.~\cite{BoehmeGross,DemirelVazquezBoehmeGross,Gleeson,Gleeson1,KissMillerSimon,marceau2010adaptive,Morettietal,PorterGleeson,PuglieseCastellano,VazquezEguiluz}. To illustrate the idea, let us consider the adaptive voter model~\cite{HolmeNewman,KimuraHayakawa,NardiniKozmaBarrat,vazquez2008generic}. The classical voter model~\cite{Liggett} considers nodes with two opinions, say $A$ and $B$. At each time step a link $XY$ with $X,Y\in \{A,B\}$ is selected. If $X\neq Y$ the link is active with one node chosen at random adopting the neighbor's opinion and if $X=Y$ nothing happens. The adaptive voter model includes social segregation. For the selected link $XY$, there is now a probability $p$ of re-wiring an active link to an inactive one and with $1-p$ the usual process happens. The homogeneous moment equations are 
\be
\label{eq:voter1}
\begin{array}{lcl}
\dot{[A]} &=& 0,\\
\dot{[AA]} &=& \frac12[AB]+\frac{(1-p)}{2}\left(2[ABA]-[AAB]\right),\\
\dot{[BB]} &=& \frac12[AB]+\frac{(1-p)}{2}\left(2[BAB]-[ABB]\right),\\
\end{array}
\ee
with the obvious conservation laws determining $[B]$ and $[AB]$. One can then apply closure schemes, such as pair approximation, to~\eqref{eq:voter1}. Instead of homogeneous moments, such as $[A]$, $[B]$, $[AB]$, and so on, one may consider heterogeneous observables involving the degree of the nodes. For example, the heterogeneous pair approximation~\cite{PuglieseCastellano,vazquez2008generic} considers the active link densities $[AB]_{k,k'}$ between nodes of degree $k$ and $k'$. After quite a lengthy derivation counting all possible changes in these new observables, one can exactly write down moment equations~\cite{DemirelVazquezBoehmeGross}, which are structurally of the form
\be
\label{eq:vechet}
\dot{[AB]}_{k,j} = F_{k,j}\left(\{[A_lB_{l'}]:l,l'\leq\bar{l}\},\{Q_l:l<\bar{l}\};p\right) ,
\ee
where $\bar{l}$ is the maximal degree, $[A_lB_{l'}]$ are densities between nodes of type $A$ with degree $l$ to nodes of type $B$ with degree $l'$, $Q_l$ is the excess degree distribution, and we have an equation for each pair $(k,j)$. So the heterogeneous pair approximation~\eqref{eq:vechet} gives a vector field $F$ depending on densities of nodes between all possible degrees. Observe that~\eqref{eq:vechet} is a closed - albeit very large - system of ODEs, where one has already implicitly made an approximation of the correlations on the third-order level. Although the heterogeneous pair approximation tends to be more accurate in many situations, the immediate drawback is visible in comparing the analytical tractability of~\eqref{eq:voter1} to the structurally very complex ODEs~\eqref{eq:vechet}. This trade-off also occurs for all other known heterogeneous moment equations. For example, another approach are active neighborhood / approximate master equation techniques~\cite{Gleeson1,marceau2010adaptive,Noeletal}, where one selects observables $X_{k,m}$ of nodes with degree $k$ in state $X$ and $m$ neighbors in state $Y\neq X$ for $X,Y\in\{A,B\}$ and $m\in\{0,1,\ldots,k\}$. Then one obtains ODEs~\cite{DemirelVazquezBoehmeGross}, which are structurally of the form
\be
\label{eq:ana}
\dot{A}_{k,m} = \tilde{F}_{k,m}\left(\{A_{l,n}:l\leq k+1,n\leq m+1\};p\right), 
\ee     
where the system defined by the vector field $\tilde{F}$ can be closed again for a fixed maximal degree. Interestingly, the ODEs~\eqref{eq:ana} turn out to contain various sums such as
\benn
\sum_{k,m} A_{k,m},\quad \sum_{k,m} B_{k,m}, \quad \sum_{k,m} (k-n)A_{k,m},\quad \ldots,
\eenn
that can be interpreted as moments. Using these moments one can re-write the active neighborhood / approximate master equation ansatz~\eqref{eq:ana} as a PDE~\cite{SilkDemirelHomerGross}. To make this transition, one considers yet another observable $Q(t,x,y):=\sum_{k,m}A_{k,m}x^k y^m$, which yields structurally a PDE of the form 
\be
\label{eq:GrossPDE}
\partial_t Q = g_0(x,\bar{Q}_1;p)Q  + g_1(x,y,\bar{Q}_2;p)\cdot \nabla Q  , 
\ee
where $g_{0,1}$ are explicit maps and $\bar{Q}_{1,2}$ are nonlocal terms depending on derivatives of up to $j$-th order evaluated at a particular normalization point. Observe carefully that the terms appearing in~\eqref{eq:GrossPDE} are not completely unexpected, as one can draw clear similarities to Vlasov(-Fokker-Planck) mean-field limit PDEs for network dynamics as described above. Indeed, Vlasov equations for heterogeneous networks are first-order transport-type PDEs with nonlocal terms accounting for the heterogeneity. For Vlasov PDEs one can also use moments to go back to an infinite ODE system of observables, which demonstrates a very nice consistency of ideas needed in mathematical mean-field approaches to tackle adaptive network dynamics.  

%%%%%%%%%%%%%%%%%%%%%%%%%%%%%%%%%%%%%%%%%%%%%%%%%%%%%%%%%%%%%%%%%%%%%%%%%%%%%
\subsection{Continuum Limit}
\label{sec:continuum}

The mean field limit approaches in Sections~\ref{sec:VFPE}-\ref{sec:moments} are based upon a probabilistic viewpoint from kinetic theory that tracks the density of a typical node or suitable macroscopic observables. An alternative approach is to keep each individual node $v_n$ with phase space $\cX$, associate each node location with a geometric position in a space $\cY$, and then construct a limiting differential equation on $\cY$ with values in $\cX$. The most classical example is the model
\be
\label{eq:heat}
\frac{\txtd v_i}{\txtd t} = v_{i+1}-2v_i+v_{i-1},\quad, v_{-N}=v_N,\qquad v_i=v_i(t)\in\R=\cX,~i\in\{-N,\ldots,N\}.
\ee 
Taking $\cY=[-1,1]$, placing $v_i$ at spatial location $x_i=i/N$, scaling time as $t\mapsto t/N^2$, and taking $N\ra +\I$ yields in the continuum limit the diffusion/heat equation 
\be
\label{eq:heat1}
\partial_t v = \partial_x^2 v,\qquad v=v(t,x),~v(-1)=v(1),~x\in[-1,1].
\ee
Of course, one may view~\eqref{eq:heat} as a network dynamical system, albeit a very structured system with very local coupling (on a spatial circular domain). For static dynamical systems on complex networks, one naturally obtains continuum limits, which are integro-differential equations defined via a graph limit operator~\cite{KaliuzhnyiVerbovetskyiMedvedev,KUE19,Medvedev3,Medvedev2}. Let us illustrate this approach already more directly for adaptive network dynamical systems~\cite{GkogkasKuehnXu} of the form
\be
\label{eq:ips1adapt2}
\begin{aligned}
\dot{v}_i&=f(v_i)+\frac1N\sum_{j=1}^N (a^N_{ij})K(v_i,v_j),\\
(a_{ij}^N)'&=-\epsilon(a^N_{ij}+h(v_i,v_j)),
\end{aligned}
\ee
which we have already studied several times, e.g., when we have discussed a VFPE for the simpler case when $h$ only depends upon the difference $v_j-v_i$ (see also Eqs.~\eqref{eq:APO_phi}-\eqref{eq:APO_kappa}, \eqref{eq:AdaptiveKS_phi}-\eqref{eq:AdaptiveKS_kappa}). As before, let us assume that the graph limit of the adjacency matrices $(a^N_{ij})_{i,j=1}^N=A^N\ra A^\I$ as $N\ra \I$ exists. Let us fix a reference space $\cY=[0,1]$ for the nodes endowed with the Lebesgue measure $\mu_\cY$. Suppose $A^\I$ is a bounded, self-adjoint, and positivity-preserving operator $A^\I:L^\I([0,1],\R)\ra L^1([0,1],\R)$. This assumption just means that $A^\I$ behaves like an adjacency matrix in the infinite limit, and one also refers to $A^\I$ as a graphop~\cite{BackhauszSzegedy}. If we want to represent also sparse graphs in the continuum limit, then using only graphops defined via graphons as in~\eqref{eq:graphon} is not sufficient. Hence, we consider more general graphops, which can be represented via the Riesz representation theorem as
\benn
(A^\I g)(y)=\int_0^1 g(z)~\txtd \eta^y(z), 
\eenn  
where $\{\eta^y\}_{y\in [0,1]}$ are called fiber measures. Each $\eta^y$ describes the links connected to $y$. Then the continuum limit of~\eqref{eq:ips1adapt2} is given by~\cite{GkogkasKuehnXu}
\bea
\partial_t v(t,y)&=&f(v(t,y))+\int_0^1 K(v(t,z),v(t,y))~\txtd \eta^z_{t}(y),\label{eq:contlim1}\\
\partial_t \eta^y_t(z)&=&-\epsilon \eta^y_{t}(z)-\epsilon h(v(t,y),v(t,z)) \mu_\cY(z)\label{eq:contlim2}.
\eea
The first equation~\eqref{eq:contlim1} is relatively self-evident as it is the direct analog of our diffusion/heat equation construction, where the integral term captures the nonlocal coupling. These types of continuum limit integral equation appear in quite a number of contexts and they become easier, when $\eta^z_{t}(y)$ also has a density $W$ so that $\txtd\eta^z_{t}(y)=W(y,z)~\txtd z$ leading back to the graphon case as $W$ has then the interpretation of a graphon. The second equation~\eqref{eq:contlim2} arising due to adaptivity is more delicate as it is formally written as a differential equation of measures and has to be interpreted weakly via test functions. Proving the existence of a continuum limit mathematically seems to be slightly easier than working with VFPEs but effectively both types of equations are deeply linked as can be gathered from the discussion revolving around the characteristic equation~\eqref{eq:kin1} in Section~\ref{sec:VFPE}.   

%%%%%%%%%%%%%%%%%%%%%%%%%%%%%%%%%%%%%%%%%%%%%%%%%%%%%%%%%%%%%%%%%%%%%%%%%%%%%
\subsection{Multiscale Decomposition}
\label{sec:multiscale}

We have already developed mathematical techniques for adaptive network models of the form~\eqref{eq:ips1adapt2}, but have not exploited the time scale separation parameter $\epsilon$ between fast node dynamics and the slow link dynamics~\cite{KuehnBook}. This scale separation occurs quite often, e.g., in neuroscience and machine learning, where the speed of information propagation is much faster than the scale of synaptic plasticity or weight learning. Of course, in other applications, the time scale separation could potentially be reversed. Therefore, the general theory of fast-slow dynamical systems applies~\cite{Fenichel4,Jones,Kaper,KuehnBook,Wechselberger4}, which covers the case of general ODEs such as
\be
\label{eq:fsmain}
\begin{array}{lcr}
\dot{\bm x}&=& f({\bm x},{\bm y},\epsilon),\\
\dot{\bm y}&=& \epsilon g({\bm x},{\bm y},\epsilon),
\end{array}
\ee
where $f:\R^m\times \R^n\times \R\ra \R^m$ and $g:\R^m\times \R^n\times \R\ra \R^n$ are sufficiently smooth maps. In the context of adaptive networks, our previous notation entails that~\eqref{eq:fsmain} is the case of fast node dynamics and slow link dynamics but as discussed in Section~\ref{sec:classification}, the reverse case also occurs and then one would just have to swap the labels of the variables. Taking the limit $\epsilon\ra 0$ in~\eqref{eq:fsmain} yields the fast subsystem
\be
\label{eq:fs1}
\begin{array}{lcl}
\dot{\bm x}&=& f({\bm x},{\bm y},0),\\
\dot{\bm y}&=& 0,
\end{array}
\ee
so that only the fast variables ${\bm x}$ are dynamic, while the slow variables ${\bm y}$ become parameters. The steady states of~\eqref{eq:fsmain} form the critical manifold
\be
\label{eq:C0}
\cC_0:=\{({\bm x},{\bm y})\in\R^{m+n}:f({\bm x},{\bm y},0)=0\}.
\ee
In addition to the fast subsystem dynamics that takes place in $\R^m$ over a ${\bm y}$-parametric family of spaces, one can induce a slow subsystem dynamics on $\cC_0$. Re-scaling $s:=\epsilon t$ in~\eqref{eq:fsmain} and taking the limit $\epsilon\ra 0$ we get
\be
\label{eq:fs2}
\begin{array}{lcl}
0&=& f({\bm x},{\bm y},0),\\
{\bm y}'&=& g({\bm x},{\bm y},0),
\end{array}
\ee  
where prime denotes differentiation with respect to $s$. The differential-algebraic equation~\eqref{eq:fs2} can be viewed as an ODE with phase space $\cC_0$. The main regularity assumption of multiple time scale dynamics is to start from submanifolds $\cS_0\subseteq\cC_0$, which are normally hyperbolic, i.e., if for any $p\in\cS_0$, the fast subsystem linearization $\txtD_{\bm x}f(p,0)\in\R^{m\times m}$ has no eigenvalues with a zero real part. Then the Fenichel-Tikhonov Theorem~\cite{Fenichel1,Fenichel4,Tikhonov} implies that there exists a perturbed slow manifold $\cC_\epsilon$ on which the dynamics is conjugate to the slow subsystem on $\cS_0$, and $\cC_\epsilon$ has the same stability properties with respect to the fast subsystem as $\cS_0$. These results do directly apply to adaptive network dynamics of the form~\eqref{eq:ips1adapt2}, for an example with fast link dynamics and slow node dynamics, see~\cite{Haetal}. However, one problem frequently appears that calculating~\eqref{eq:C0} explicitly (or even numerically) for large finite-size networks of size $N\gg 1$ can be difficult, as it amounts to solving a large system of nonlinear equations $f({\bm x},{\bm y},0)=0$. Furthermore, even if we were able to characterize $\cC_0$ easily by knowing all the steady states of the static network, the adaptivity can add up to $m=N^2$ slow variables if the link weights scalar-valued, so the slow flow \eqref{eq:fs2} is potentially even more complicated. Hence, alternative strategies are required.

One option is to start with small adaptive networks focusing just on the behavior of small motifs, which links to the theory in Section~\ref{sec:moments}. As an example, let us consider adaptive consensus dynamics on a triangle motif~\cite{JardonKuehn1}
\be
\label{eq:consensus}
\begin{array}{lcl}
\dot{\bm x}&=&L({\bm x},{\bm y},\epsilon){\bm x},\\
\dot{\bm y}&=&\epsilon g({\bm x},{\bm y},\epsilon),
\end{array}\qquad L({\bm x},{\bm y},\epsilon)=
\left(
\begin{array}{ccc}
w+1 & w & -1\\
-w & w+1 & -1\\
-1 & -1 & 2\\
\end{array}
\right)
\ee  
where ${\bm x}\in\R^3$, ${\bm y}\in\R$, and the weight $w=w({\bm x}_1,{\bm x}_2,{\bm y},\epsilon)$ in the Laplacian $L$ is a smooth function of its arguments. Although the fast dynamics has a nice structure and we only consider one slow variable to adapt the link dynamics, using geometric methods from multiple time scale dynamics is already challenging~\cite{DumortierRoussarie,KruSzm4}. Under the assumption of an affine linear mapping for $w$, using a conservation law to eliminate one fast variable, eliminating another fast variable by a globally stable fast direction and applying suitable coordinate transformations, one can reduce \eqref{eq:consensus} to the following planar fast-slow system
\be
\label{eq:fsconsensus}
\begin{aligned}
\dot{X}_3=& -(2\tilde{W}(X_3,Y)+1)X_3,\\
\dot{Y}=& \epsilon G(X_3,Y,\epsilon).
\end{aligned}
\ee    
The key point to observe is that already the subset $\cS_0=\{(X_3,Y)\in\R^2:X_3=0\}$ is not normally hyperbolic in general as $-(2\tilde{W}(0,Y)+1)$ may be zero for sufficiently generic adaptive $Y$-dynamics. Hence, even in the simplest case, we cannot hope for completely normally hyperbolic multiscale adaptive network dynamics. Analyzing singularities, where normal hyperbolicity is lost turns out to be much more complicated. There is a multitude of asymptotic~\cite{BenderOrszag,KevorkianCole,DeJagerFuru,MisRoz}, geometric~\cite{DumortierRoussarie,JardonKuehn,KruSzm3,JonesKopell}, and/or numerical techniques~\cite{DesrochesKrauskopfOsinga2,GearKaperKevrikidisZagaris,GuckenheimerKuehn2,KaperKaper} to carry this out; for a detailed list of references of multiple time scale technqiues we refer to~\cite{KuehnBook}. A geometric desingularization (or blow-up) analysis proves~\cite{JardonKuehn1} that already semi-linear consensus dynamics~\eqref{eq:consensus} on a triangle motif can have highly sensitive orbits called canards~\cite{BenoitCallotDienerDiener,Eckhaus}. 

Instead of going to small motifs, we can also take the limit as $N\ra \I$ in the spirit of Sections~\ref{sec:VFPE} and \ref{sec:continuum}. Then one hopes that the mathematical situation improves for fast-slow systems in infinite dimensions. As an example, consider the continuum limit \eqref{eq:contlim1}-\eqref{eq:contlim2}, which we can (very formally!) view as an evolution equation of the form
\be
\label{eq:clfs}
\begin{array}{lcl}
\partial_t v(t,y)&=&\mathfrak{F}(v(t,y),\eta^z_{t}(y)),\\
\partial_t \eta^y_t(z)&=&\epsilon \mathfrak{G}(v(t,y),\eta^y_t(z)).
\end{array}
\ee  
Evidently, the mathematical interpretation is already delicate as one might want to use a Banach space as a phase space for~\eqref{eq:clfs}, but the measures $\eta^z_{t}$ describing the graph limit are often best treated on metric spaces. In the case of Banach spaces, there has been recently a development of multiple time scale dynamics in the infinite-dimensional setting~\cite{EngelKuehn1,HummelKuehn,EngelHummelKuehn}, which has the clear potential to be general enough to eventually also apply to multiscale network limit systems such as~\eqref{eq:clfs}.

Lastly, we briefly comment on the use of multiscale techniques for adaptive network dynamics when the node and edge dynamics is not explicitly given by ODEs. For ODEs, it is often slightly easier to identify the multiscale structure and also to apply existing ODE techniques. For models, which at least partially, use discrete-time updates, the first question is how one identifies scale separation in the first place. As a first example, let us consider the Jain-Krishna adaptive network model~\cite{JainKrishna2,JainKrishna}. The model considers a set of continuous-time autocatalytic ODEs~\cite{Kauffman} together with an adaptive discrete-time directed network structure. At each node $j$, one considers the ODEs 
\be
\label{eq:JK}
\frac{\txtd {\bm x}_i}{\txtd t}=:
\dot{\bm x}_i=(A^\top {\bm x})_i-{\bm x}_i\sum_{k=1}^d (A^\top {\bm x})_k,\qquad i\in\{1,2,\ldots,N\},
\ee
where the adjacency matrix $A=A[0]$ is usually chosen as a random Erd\H{o}s-Renyi graph and one also imposes two constraints
\be
\label{eq:JKc}
\sum_{k=1}^N {\bm x}_k=1 \qquad \text{and}\qquad {\bm x}_k\geq 0\quad \forall k,
\ee
which are just mass and non-negativity conservation. Adaptivity arises by trying to model evolutionary elimination of non-successful species/nodes. Suppose the ODEs~\eqref{eq:JK} have converged (sufficiently close) to a steady state $x_*$ for the adjacency matrix $A=A[0]$. Then define the set of indices $\cJ_*:=\left\{j\in\{1,2,\ldots,d\}:{\bm x}_j=\min_k {\bm x}_k\right\}$, which just picks out the species with minimum steady state value. Then we select some $j_*$ at random from $\cJ_*$ and re-sample $a_{ij_*}$ and $a_{j_*i}$ for $i\neq j$ from a probability distribution, which yields a new matrix 
$A=A[1]$; the process can now be repeated. In this context, there is an implicit time scale separation made in the model, as we have assumed that the adaptivity step only occurs after we have fully, or at least approximately, converged to the steady state. This means that the node dynamics is effectively fast in comparison to a slow adaptive dynamics of the topology. This scale separation can then still be exploited in the mathematical analysis by treating the node and link dynamics as separate singular limit processes~\cite{KUE19a}. A similar principle to identify and exploit scale separation also works for purely discrete-time systems. As an example, let us consider the Bornholdt-Rohlf model~\cite{BornholdtRohlf,MEI09a} for self-organized criticality on directed and signed networks. Suppose nodes have states $\{{\bm x}_i(t)\}_{i=1}^N\in\{\pm 1\}$ and links have values $A=\{a_{ij}(t)\}_{i,j=1}^N\in\{-1,0,+1\}$ without loops. The initial network $t=0$ is usually random Erd\H{o}s-Renyi graph with uniformly distributed random node values. The node dynamics is by a parallel update given as
\be
\label{eq:BR_node1}
{\bm x}_i(t+1)=\left\{\begin{array}{ll}
\text{sgn}([A(t){\bm x}(t)]_i(t)) & \text{ if $[A(t){\bm x}(t)]_i(t)\neq 0$,}\\
{\bm x}_i(t) & \text{ if $[A(t){\bm x}(t)]_i(t)= 0$.}\\
\end{array}
\right.
\ee   
After $T_v$ node dynamics steps \eqref{eq:BR_node1} have finished, one measures the average activity over the last $T_a:=\lfloor T_{v}/2\rfloor$ steps
\be
\label{eq:BR_act}
\cA_i:=\frac{1}{T_{v}-T_a}\left[\sum_{t=T_a}^{T_{v}}{\bm x}_i(t)\right].
\ee
Nodes with $|\cA_i|=1$ indicate dynamically frozen states, while $|\cA_i|<1$ are active nodes. The adaptivity of the model is introduced by trying to activate frozen nodes and reduce activity in highly dynamic nodes. Choose a site $i$ at random and calculate $\cA_i$ using \eqref{eq:BR_act}. Let $0\leq\delta\leq 1$ be a parameter; if $|\cA_i|>1-\delta$ then $i$ receives a new link $a_{ij}$, with a randomly chosen link weight in $\{\pm 1\}$, from a node $j$ chosen at random. If $|\cA_i|\leq 1-\delta$ then one of the existing nonzero links is deleted. After this topological update the dynamics switches again to $T_v$ node dynamics steps \eqref{eq:BR_node1}. This model also has a natural time-scale separation given by the time scale $T_v$ which is usually very large so that the node dynamics is fast,  while the link dynamics is slow. One can again exploit this scale separation principle to understand self-organized criticality using low-dimensional fast-slow dynamical systems~\cite{KuehnNetworks}. 

%%%%%%%%%%%%%%%%%%%%%%%%%%%%%%%%%%%%%%%%%%%%%%%%%%%%%%%%%%%%%%%%%%%%%%%%%%%%%
\subsection{Bifurcations \& Patterns}
\label{sec:bifs}

Bifurcation theory, among the many other classical tools from nonlinear dynamics, can essentially be carried over to systems involving network dynamics directly, and this has been utilized to find many of the effects discussed in Section~\ref{sec:dynPhenomena}. Yet, for adaptive network dynamics, there has been one additional important paradigm that has evolved in recent years, where bifurcation theory~\cite{GH,Kuznetsov,Strogatz} has played a key role. Suppose we have, e.g. via the methods in Sections~\ref{sec:VFPE}-\ref{sec:multiscale}, derived a tractable system of differential equations or iterated maps for dynamical variables ${\bm X}$ with several system parameters $P$ remaining. For simplicity, let us consider the case of ODEs
\be
\label{eq:ODE}
\dot{\bm X}=F({\bm X},P),\qquad {\bm X}={\bm X}(t)\in\R^d.
\ee
As outlined in the beginning of Section~\ref{sec:mathmeth}, one often wants to compare the dynamics of adaptive networks with their static counterparts. Hence, in many problems one can identify a single parameter $q$ controlling the strength of the adaptive dynamics. For example, in~\eqref{eq:ips1adapt2} we would take the slow adaptation rate $q=\epsilon$, while for the epidemic SIS model~\eqref{eq:ODESISThilo}, we would take the re-wiring parameter $q=w$. Once this identification has been made, it has become common to analyze bifurcation scenarios varying two parameters $P=(p,q)$, where $p$ is a primary bifurcation parameter already present in the static network node dynamics. For example, in~\eqref{eq:ips1adapt2} we could take $p$ as a strength for the kernel coupling via $K$, while for the epidemic SIS model~\eqref{eq:ODESISThilo} one may take the infection rate $p=\beta$. Via this strategy, one can compare the differences in bifurcation diagrams with the main bifurcation parameter $p$ for non adaptivity ($q=0$) with those for the adaptive case ($q\neq 0$). In general, it has been found, that adaptivity can shift bifurcation points~\cite{Zschaleretal,HorstmeyerKuehnThurner1,LUE16}, change stability of patterns \cite{BER19}, generate new patterns~\cite{ZschalerTraulsenGross,KuehnZschalerGross}, alter the criticality of transitions~\cite{GRO06b}, and induce myriad novel effects in bifurcation diagrams such as new early-warning signs for bifurcations~\cite{HorstmeyerKuehnThurner1}. In fact, apart from the main physical consequences, the key mathematical observation is that co-dimension two bifurcation theory is the natural framework for adaptive networks as long as one wants to compare static to adaptive networks. This viewpoint has been recently used in~\cite{KUE21} to explain that the transition from non-explosive (second-order, supercritical, soft) transitions to explosive (first-order, subcritical, hard) ones cannot only be induced by adaptivity but is a generic effect once static network dynamics is expanded to include more complicated network dynamics such as multiplex or higher-order coupling.   

%% file: parts/conclusion.tex
%\begin{itemize}
%    \item BA network are scale free but the construction algorithm is not adaptive, adaptive networks may produce scale free structures themselves
%\end{itemize}

In this review, we have introduced the modeling approach of adaptive dynamical networks and illuminated various of its capabilities. As the notion of "adaptivity" has been used in several contexts in the literature, we have elaborated on the distinction of the class of models considered in this review from other model classes considered in the vast literature on adaptive mechanisms. In this regard, we have provided several ideas how to classify adaptive dynamical networks starting from various features of such systems. The major part of this review is concerned with the applications of adaptive dynamical networks. We have devoted individual sections to show applications to neural and physiological systems, artificial intelligence, control schemes, power grids as well as to social, epidemiological and transport networks. Building on the models that have been introduced in these sections, we have given an overview on the plethora of dynamical phenomena arising in such systems. In the last section of this review, we show several state-of-the-art mathematical tools that can be used to analyse adaptive dynamical systems.

In summary, adaptive dynamical systems are omnipresent in nature and technology and can show various dynamical phenomena. Here, we have reported on the current status of these fascinating class of dynamical models which capabilities for understanding real-world dynamical systems is far from exhausted and several are indeed in its infancy. In particular, also for the future research there a numerous possibilities to study these systems, their dynamics and to develop analytic methods to describe them.

More precisely, we have discussed in Sec.~\ref{sec:classification} the distinction between continuous and event-based network adaptation. For some event-based rules, in particular in systems of coupled oscillators, exist continuous adaption rules that provide a very good approximation and are analytically more tractable. These approximations have been successfully used to understand the dynamics in systems where the adaptation is governed by spike timing-dependent plasticity~\cite{LUE16,DUC22}. It is, however, an open question as to what the approximation of event-based adaption by a continuous rule could be used for other adaptive dynamical networks as well.

Another classification feature of adaptation rules, namely the adaptation rate, see Sec.~\ref{sec:classification}, has become of interest. In particular, recently an explicit splitting of time scales has been utilized to derive conditions for the emergence of several complex dynamical states in (rather small) adaptive dynamical networks~\cite{juttner2022complex,Thiele2023,venegas2023stable}. How to generalize these findings to larger or more complex systems is still elusive. In addition, small systems, in particular systems of two coupled oscillators, have been used to describe the emergence of chaos~\cite{juttner2022complex,emelianova2019intersection,emelianova2020third,emelianova2021emergence} or recurrent synchronization~\cite{Thiele2023} in adaptive dynamical networks or to explain the interplay of an adaptive network structure with noise~\cite{BAC18a,BAC18b,BAC20,FRA20} or delay~\cite{madadi2022delay}. Also here the question naturally arises, how the observed phenomena and derived findings transition to larger systems.

Another emerging research direction with respect to adaptive dynamical networks, concerns the study of mean-field theories. Mean-field approaches are powerful in describing the dynamics for large populations of interacting systems, see e.g.~\cite{BIC20} for a review on mean-field approaches for phase oscillator models. These approaches have been developed for a plethora of dynamical systems, however, for adaptive dynamical networks only recently first results have been achieved~\cite{GkogkasKuehnXu,GKO22,DUC22,FIA22} and therefor many problems are still unsolved.

Generalizing dynamical networks to more complex network structures such as multilayer networks, hypernetworks, higher-order networks or simplicial complexes is an interesting research perspective for adaptive dynamical networks since the interplay of the different "types of complexity" invoked by adaptivity and a complex network structure is little investigated. For mulilayer networks~\cite{pitsik2018inter,KAS18,BER20,KAC20,FRO21a} as well as simplicial complexes~\cite{kachhvah2022hebbian,kachhvah2022first,rajwani2023tiered} for results have been obtained but there is much more to understand to obtain a comprehensive picture of the interplay.. In more general terms, adaptive dynamical networks provide a test bed to study the relation between function and structure of a dynamical network. In this regard e.g. finding good estimates for the network topology from the resulting dynamics (data) is an ongoing research question~\cite{timme2007revealing,timme2014revealing,peixoto2019network} which has been only little studied from the adaptive dynamical network perspective~\cite{YU06a}.

Adaptivity in a general sense is a widespread feature of technological as well as natural systems and many research has been devoted to understand its impact on the emergent dynamics~\cite{sawicki2023perspectives}. Throughout this review, we have highlighted the power of adaptive dynamical network to model realistic dynamical systems by capturing their temporal changes of the interaction structure. We believe that adaptive dynamical networks can serve as a new generic modeling paradigm for complex dynamical systems, that is fundamental to understand the complex interplay of structure and function.

% \textbf{TODO:} 
% \begin{itemize}
%     \item ASH19 - can a system with dead zones interpreted as an adaptive network with very fast adaptation
%     \item mean-field induced adaptation Skardal, Torcini, Olmi, Taher, SOR09a
% \end{itemize}